\documentclass[a4paper]{article}

\usepackage{graphicx}
\usepackage{float}
\usepackage{amsmath}
\usepackage{amsfonts}

\title{Gravitational lensing and ghost images in the regular Bardeen no-horizon spacetimes}
\author{Jan Schee and Zden\v{e}k Stuchl\'{i}k\\
	          \emph{Institute of Physics, Faculty of Philosophy \& Science,}\\ 
						\emph{Silesian University in Opava,} \\
						\emph{Bezru\v{c}ovo n\'{a}m\v{e}st\'{i} 13, CZ-74601 Opava,}\\
						\emph{Czech Republic}\\
						email: zdenek.stuchlik@fpf.slu.cz, jan.schee@fpf.slu.cz}
\date{}

\newcommand{\beq}{\begin{equation}}
\newcommand{\eeq}{\end{equation}}
\newcommand{\bea}{\begin{eqnarray}}
\newcommand{\eea}{\end{eqnarray}}
\newcommand{\diff}{\mathrm{d}}

\begin{document}
\maketitle
\begin{abstract}
We study deflection of light rays and gravitational lensing in the regular Bardeen no-horizon spacetimes. Flatness of these spacetimes in the central region implies existence of interesting optical effects related to photons crossing the gravitational field of the no-horizon spacetimes with low impact parameters. These effects occur due to existence of a critical impact parameter giving maximal deflection of light rays in the Bardeen no-horizon spacetimes. We give the critical impact parameter in dependence on the specific charge of the spacetimes, and discuss "ghost" direct and indirect images of Keplerian discs, generated by photons with low impact parameters. The ghost direct images can occur only for large inclination angles of distant observers, while ghost indirect images can occur also for small inclination angles. We determine the range of the frequency shift of photons generating the ghost images and determine distribution of the frequency shift across these images. We compare them to those of the standard direct images of the Keplerian discs. The difference of the ranges of the frequency shift on the ghost and direct images could serve as a quantitative measure of the Bardeen no-horizon spacetimes. The regions of the Keplerian discs giving the ghost images are determined in dependence on the specific charge of the no-horizon spacetimes. For comparison we construct direct and indirect (ordinary and ghost) images of Keplerian discs around Reissner-N{\"{o}}rdstr{\"{o}}m naked singularities demonstrating a clear qualitative difference to the ghost direct images in the regular Bardeen no-horizon spacetimes. The optical effects related to the low impact parameter photons thus give clear signature of the regular Bardeen no-horizon spacetimes, as no similar phenomena could occur in the black hole or naked singularity spacetimes. Similar direct ghost images have to occur in any regular no-horizon spacetimes having nearly flat central region.
\end{abstract}

\section*{Introduction}
Black holes governed by the standard Einstein general relativity contain a physical singularity with diverging Riemann tensor components and predictability breakdown, where quantum gravity is expected to enter the play that should be able to overcome this internal defect of the general relativity. However, families of regular black hole solutions have been found that eliminate the physical singularity from the spacetimes having an event horizon. Such regular solutions describe even "no-horizon" spacetimes, if parameters of these solutions are properly chosen. Of course, these are not vacuum solutions of the Einstein equations, but contain necessarily a properly chosen additional field, or modified gravity, and the energy conditions related to the existence of physical singularities \cite{Haw-Elli:1973:LargeScaleStructure:} are then violated. 

Regular black hole solution containing a magnetic charge as a source parameter has been proposed by Bardeen \cite{Bar:1968:GR5Tbilisi:}, but for modified gravity; their magnetic charge parameter can be related to a non-linear electrodynamics as shown by \cite{AyB-Gar:2000:PhysLetB:}. A regular black hole solution of combined Einstein gravitational equations and non-linear electrodynamic equations has been introduced by Ayon-Beato and Garcia \cite{AyB-Gar:1998:PhysRevLet:,AyB-Gar:1999:PhysLetB:,AyB-Gar:1999:GenRelGrav:}. Both the Bardeen and the Ayon-Beato-Garcia (ABG) solutions are characterized by gravitational mass parameter $m$ and charge parameter $g$. Their geodesic structure is governed by the dimensionless ratio $g/m$. A different approach to the regular black hole solutions has been applied by Hayward \cite{Hay:2006:PhysRevLet:}. Modification of the mass function in the Bardeen and Hayward solutions and inclusion of the cosmological constant can be found in the new solutions of Neves and Saa \cite{Nev-Saa:2014:arXiv:1402.2694:}. Rotating regular black hole solutions were introduced in \cite{Mod-Nic:2010:PHYSR4:,Bam-Mod:2013:PhysLet:,Tos-Abd-Ahm-Stu:2014:PHYSR4:,Mus:2014:PHYSR4:}. For properly chosen charge parameter $g/m$, such solutions allow for existence of fully regular spacetime, without an event horizon. We call such solutions "no-horizon" spacetimes. Some of their properties were discussed in \cite{Pat-Jos:2012:PHYSR4:}. 

Properties of the geodesic motion in the field of regular black holes have been discussed in \cite{Gar-Hac-Kun-Lam:2013:arXiv:1306.2549:,Pat-Jos:2012:PHYSR4:,Eir-Sen:2013:PHYSR4:,Tos-Abd-Ahm-Stu:2014:PHYSR4:}. A detailed discussion of the circular geodesics of the regular Bardeen and ABG black hole and no-horizon spacetimes and its implication to simple optical phenomena as the silhouette shape and extension, or the profiled spectral lines generated by Keplerian rings constituted from test particles following stable circular geodesics has been recently presented in \cite{Stu-Sche:2015:IJMPD:}. It has been demonstrated that the geodesic structure of the regular black holes outside the horizon is similar to those of the Schwarzschild or Reissner-N\"{o}rdstr\"{o}m (RN) black hole spacetimes, but under the inner horizon, no circular geodesics can exist. The geodesic structure of the no-horizon spacetimes is similar to those of the naked singularity spacetimes of the RN type \cite{Stu-Hle:2002:ActaPhysSlov:,Pug-Que-Ruf:2011:PHYSR4:}, or the Kehagias-Sfetsos (KS) type \cite{Keh-Sfe:2009:PhysLetB:,Stu-Sche:2014:CLAQG:,Stu-Sche-Abd:2014:PHYSR4:} that represents an asymptotically flat solution of the modified Ho\v{r}ava quantum gravity \cite{Hor:2009:PHYSR4:,Hor:2009:PHYSRL:}. In all of these no-horizon and naked singularity spacetimes, an "antigravity" sphere exists consisting of static particles at a stable equilibrium position, given by the so called "static" radius. The "antigravity" sphere can be surrounded by a Keplerian disc or toroidal configuration \cite{Stu-Sche:2014:CLAQG:}. 

In the Bardeen and ABG spacetimes, character of the Keplerian (geodetical) discs strongly depends on the magnitude of the specific charge. If $g/m$ is close to the value corresponding to the extreme black-hole state, two Keplerian discs exist above the static radius, and even two photon circular orbits exist in the near-extreme states, the inner one being stable representing the outer edge of the inner Keplerian disc, while the outer disc is limited by the innermost stable circular geodesic. As an exceptional phenomenon, not occurring in the naked singularity or the Bardeen no-horizon spacetimes, an internal Keplerian disc can occur under the antigravity sphere in the ABG no-horizon spacetimes with $g/m > 2$ \cite{Stu-Sche:2015:IJMPD:}. 

However, there is a significant difference of the character of the optical phenomena in the RN or KS naked singularity spacetimes as compared to those occurring in the regular Bardeen and ABG no-horizon spacetimes, related to the fact that near their centre at $r=0$ the regular spacetimes are close to the de Sitter spacetime \cite{Stu-Sche:2015:IJMPD:}. In the regular no-horizon spacetimes the deflection of light rays is not a monotonic function of the impact parameter, but is has a maximum for a critical impact parameter depending on the specific charge of the Bardeen and ABG no-horizon spacetimes. Therefore, there should be a weak lensing of distant objects both for large and small impact parameters. Moreover, "ghost" direct images of some parts of the Keplerian discs orbiting in the no-horizon spacetimes occur due to the effect of non-monotonicity of the dependence of the deflection angle on the impact parameter of photons. We study all the optical phenomena in the case of the relatively simple Bardeen spacetime, in order to obtain a clear picture of the "ghost" phenomena and compare them to the case of the ghost images created by Keplerian discs in the field of Reissner-N\"{o}rdstr\"{o}m (RN) naked singularities. In next paper we plan to study these phenomena also for the ABG spacetimes where the situation is more complex due to the possible existence of inner Keplerian discs located in the spacetimes with $g/m > 2$ under their static radius. An important information is expected also from the study of the frequency shift distribution of the Keplerian disc images (including the ghost images) and the profiled spectral lines that give relevant complementary information on the optical phenomena \cite{Sche-Stu:2008:IJMPD:,Sche-Stu:2008:GenRelGrav:,Stu-Sche:2010:CLAQG:,Sche-Stu:2013:JCAP:,Stu-Sche:2013:CLAQG:}. 

The electromagnetic field related to the spherically symmetric regular Bardeen and ABG spacetimes \cite{AyB-Gar:1998:PhysRevLet:,AyB-Gar:1999:PhysLetB:,AyB-Gar:1999:GenRelGrav:} is irrelevant for our study, since we consider the geometry properties only. We assume that at $r=0$, the source of the (non-linear) electromagnetic field of the Bardeen or ABG background is located, where purely radial trajectories of test particles and photons terminate, similarly to the case of the central singular points in the spherically symmetric naked singularity spacetimes. 

\section{Regular Bardeen no-horizon spacetimes}
The regular Bardeen black-hole or no-horizon spacetimes are characterized in the standard spherical coordinates and the geometric units (c=G=1) by the line element
\beq
	\diff s^2=-f(r)\diff t^2+\frac{1}{f(r)}\diff t^2+r^2(\diff\theta^2+\sin^2\theta\diff\phi^2) ,\label{interval}
\eeq
where the "lapse" $f(r)$ function depends only on the radial coordinate, the gravitational mass parameter $m$, and the charge parameter $g$ and takes the form 
\beq
	f(r)=1 - \frac{2 m r^2}{(g^2 + r^2)^{3/2}}.
\eeq
The Bardeen spacetimes are constructed to be regular everywhere, i.e., the components of the Riemann tensor, and the Ricci scalar are finite at all $r\ge 0$ \cite{AyB-Gar:1999:GenRelGrav:}, but these spacetimes are not Ricci flat. The loci of the Bardeen black hole horizons are determined by the relation \cite{Stu-Sche:2015:IJMPD:} 
\beq
		g^6 + (3g^2 - 4m^2)r^4 + 3g^4r^2 + r^6 = 0 . 
\label{pseudosingularity}		
\eeq
This equation allows maximally for two real solutions at $r \geq 0$ corresponding to the black hole horizons \cite{Stu-Sche:2015:IJMPD:}. If real and positive solutions of the equation (\ref{pseudosingularity}) do not exist, the spacetime is fully regular, having no event horizon. We call it "no-horizon" spacetime. The critical value of the dimensionless parameter $g/m$ separating the black-hole and the "no-horizon" Bardeen spacetimes reads \footnote{In the well known RN spacetimes, the critical specific charge separating black holes and naked singularities reads $Q/M=1$ \cite{Mis-Tho-Whe:1973:Gravitation:}.}  
\beq
		(g/m)_{NoH} = 0.7698 .
\eeq
In the "no-horizon" Bardeen spacetimes the metric is regular at all radii $r \geq 0$. We assume $r=0$ to be the site of the self-gravitating charged source of the spacetime; test particle or photon trajectories terminate at $r=0$. 

\section{Geodesics of the Bardeen spacetimes}

In the spherically symmetric spacetimes, the geodesic motion is restricted to the central planes. For a single particle (photon) we can choose the central plane to be the equatorial plane, with $\theta = \pi/2$. Keplerian discs are usually assumed to be located in the equatorial plane of the spacetime. However, photons radiated by matter of the equatorial Keplerian discs that follows circular geodesics are then moving in non-equatorial central planes if they have to reach distant observers at any given inclination angle $\theta_o$.

The spherically symmetric spacetimes posses two Killing vector fields, $\partial/\partial t$ and $\partial/\partial\phi$, implying existence of two constants of motion, energy $E\equiv -p_t$ and axial angular momentum $L\equiv p_\phi$. For the motion in the non-equatorial planes, there exists an additional constant of motion $Q^2$, governing the "latitudinal" angular momentum defined by the relation $p_\theta^2\equiv Q^2-L^2\cot^2\theta$. \footnote{Since the latitudinal motion constant has to be positive in the spherically symmetric spacetimes, contrary to the Carter constant in the Kerr spacetimes that can be also negative \cite{Car:1973:BlaHol:}, we denote it as $Q^2$.} For massive particles, the constants of motion can be related to the constant rest mass (energy) $\mu > 0$, and are then specific energy and specific angular momenta of the motion. For photons, there is $\mu=0$. The equations of the geodesic motion can be written in the integrated form, giving components of the four-velocity (four-momentum) of the particle (photon) in the form 
\bea
	p^t&=&\frac{E}{f(r)},\\
	(p^r)^2 &=&E^2-f(r)\left(\kappa+\frac{L^2+Q^2}{r^2}\right),\\
	(p^\theta)^2&=&\frac{1}{r^4}\left(Q^2-L^2\cot^2\theta\right),\\
	p^\phi&=&\frac{L}{r^2\sin^2\theta}
\eea
where $\kappa=0$ for photons and $\kappa=1$ for massive particles. For further analysis it is convenient to define an effective potential of the motion by the relation 
\beq
	V_{eff}=f(r)\left(\kappa+\frac{L^2+Q}{r^2}\right).
\eeq

\subsection{Keplerian orbits}

The Keplerian orbits of matter in the equatorial Keplerian discs \cite{Nov-Tho:1973:BlaHol:,Pag-Tho:1974:ApJ:} are represented by the circular geodesics of the spacetime under consideration. In the equatorial plane of spherically symmetric spacetimes there is $Q^2=0$, and the circular geodesics are determined by an effective potential related to the specific angular momentum $L_c$ that takes a simple form
\beq
            V_{eff} = f(r)\left(1+\frac{L_c^2}{r^2}\right). 
\eeq
The radial motion is then governed by the equation for the radial component of the 4-velocity 
\beq
           (u^r)^2 = E^2 - V_{eff} . 
\eeq
For the photon motion, $\mu = 0$, the effective potential can be related to the impact parameter $b=E/L$ and reads  
\beq
           V_{ph} = f(r)\left(\frac{b^2}{r^2}\right). 
\eeq
Properties of the effective potential and circular geodesics outside the event horizon of the Bardeen black hole spacetimes are similar to those occurring in the Schwarzschild spacetime, while in the no-horizon spacetimes they are similar to those occurring in RN naked singularity spacetimes \cite{Stu-Hle:2002:ActaPhysSlov:,Pug-Que-Ruf:2011:PHYSR4:}, or in the KS naked singularity spacetimes \cite{Vie-etal:2014:PHYSR4:,Stu-Sche-Abd:2014:PHYSR4:,Stu-Sche:2014:CLAQG:}. The circular geodesics of the Bardeen no-horizon spacetimes were studied in detail \cite{Stu-Sche:2015:IJMPD:}, here we briefly summarize the results. 

At a given radius $r$ in the Bardeen spacetime, the specific angular momentum $L_c$ and the specific covariant energy $E_c$ of the circular geodesics are determined by the conditions $V_{eff}=0$ and $dV_{eff}/dr = 0$ that imply the relations \cite{Stu-Sche:2015:IJMPD:} 
\bea
	L^2_c&=&\frac{m r^4(r^2-2g^2))}{(r^2+g^2)^{5/2}-3m r^2},\\
	E^2_c&=&\frac{\left[1-2m r^2/(r^2+g^2)^{3/2}\right]^2(r^2+g^2)^{5/2}}{(r^2+g^2)^{5/2}-3mr^2}
\eea 

\begin{figure}[H]
\begin{center}
\includegraphics[scale=0.5]{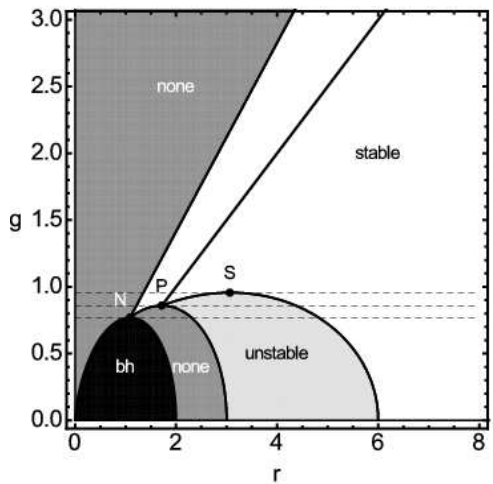}
\caption{Loci of circular geodesic orbits in the Bardeen spacetimes. The white regions correspond to stable orbits, while the light gray regions represent unstable orbits. In the dark gray regions are no circular orbits. In the Bardeen spacetimes we define three characteristic points N (No-horizon), P (photon) and S (stable) (see the text for details): $N=(1.09,0.76988)$, and $P=(1.7179, 0.85865)$, and $S=(3.07,0.95629)$. Boundary of the black region determines loci of the horizons. Boundary of the outer dark gray region determines the photon circular orbits. The boundary of the inner dark gray region finishing at the point $N$ represents the static radius $r_{stat}$. Boundary of the light gray region determines the marginally stable circular geodesics. The black line finishing at the point $P$ represents $r_{\Omega}$. In this figure and all the following figures we put for simplicity $m=1$; and the charge parameter and the radius are thus dimensionless quantities expressed in units of $m$. \label{fig1}}
\end{center}
\end{figure}

In figure \ref{fig1} we give curves separating the $r/m - g/m$ space into regions corresponding to stable circular orbits and unstable circular orbits respectively, and we also give radii of photon circular orbits, horizons, and the static radius representing the so called antigravity sphere. Radial profiles of the specific energy and specific angular momentum of the circular geodesics, and the behavior of the effective potential can be found in \cite{Stu-Sche:2015:IJMPD:}

In the Bardeen no-horizon spacetimes an "antigravity" effect of the geometry is demonstrated by vanishing of the angular momentum $L_c = 0$ at the so called static radius where test particles can remain in stable equilibrium position, similarly to the case of the KS naked singularity spacetimes of the modified Ho\v{r}ava gravity \cite{Stu-Sche:2014:CLAQG:,Vie-etal:2014:PHYSR4:}, or the RN naked sigularity spacetimes \cite{Stu-Hle:2002:ActaPhysSlov:,Pug-Que-Ruf:2011:PHYSR4:}. The static radius is located at  
\beq
	r_{stat} = \sqrt{2} g
\eeq
and no circular geodesics are possible under the stable static radius. Character of the Keplerian discs located above the static radius strongly depends on the charge parameter $g/m$ of the Bardeen no-horizon spacetimes governing the existence of unstable circular geodesics and existence of the photon circular geodesics \cite{Stu-Sche:2015:IJMPD:}. 

The photon circular geodesics are determined by the divergences of the relations for the specific energy and angular momentum $E_c$ and $L_c$. Their radii are determined by the relation
\beq
	(r^2+g^2)^{5/2}-3m r^4=0.
\eeq
The photon circular geodesics exist in the spacetimes with parameter $g/m$ smaller that the critical charge parameter $(g/m)_{P}$ given by 
\beq
		(g/m)_{P} = 0.85865 .
\eeq
In the black hole spacetimes, one unstable photon circular geodesic exists. In the no-horizon spacetimes with $(g/m)_{NoH} < g/m < (g/m)_{P}$, two photon circular geodesics are allowed, the outer one being unstable relative to radial perturbations, and the inner one being stable -- see Figure 1. In vicinity of the stable photon circular geodesic, just above the stable static radius, trapped photons can occur. 

The inner edge of the Keplerian discs is located at the innermost stable circular orbit (ISCO) determined by the condition $\frac{\diff^2 V}{\diff r^2}=0$. Under the ISCO, unstable circular geodesics exist being terminated at the unstable photon circular orbit. In the no-horizon spacetimes, the existence of unstable circular geodesics is allowed in the spacetimes with the charge parameter smaller that the critical charge $(g/m)_{S}$ given by 
\beq
		(g/m)_{S} = 0.95629 .
\eeq
Then two regions of stable circular geodesics can exist above the stable static radius, and a second marginally stable orbit, an outermost stable circular orbit (OSCO), exists along with the ISCO -- see Figure 1. 

\subsection{Classification of the Bardeen spacetimes}

We give a short review of the classification of the regular Bardeen spacetimes due to the properties of the circular geodesics governing the Keplerian accretion discs, details can be found in \cite{Stu-Sche:2015:IJMPD:}. The classification is reflected in Figure 1. 

\subsubsection{Black holes}

The event horizons exist in the spacetimes with the charge parameter in the interval  
\beq
     0 < g/m < (g/m)_{NoH} .
\eeq
No circular orbits exist under the inner horizon. Above the outer horizon, the character of the circular orbits follows those of the standard Schwarzschild geometry. 

\subsubsection{No-horizon spacetimes admitting photon circular orbits}

The photon circular geodesics occur for the charge parameter in the interval 
\beq
      (g/m)_{NoH} < g/m < (g/m)_{P} .
\eeq
Two regions of circular geodesics exist above the stable static radius. The outer one ranges, as in the black-hole spacetimes, from infinity down to the unstable photon circular geodesic.  The stable circular orbits exist down to the ISCO, under which unstable circular orbits exist. The inner region consists of stable circular geodesics that range from the stable photon circular geodesic down to the static radius allowing for stable equilibrium positions of test particles. 

\subsubsection{No-horizon spacetimes admitting unstable circular orbits}

The no-horizon spacetimes with the charge parameter in the interval  
\beq
      (g/m)_{P} < g/m < (g/m)_{S} 
\eeq
allow no photon circular geodesics, but unstable circular geodesics can exist. The circular geodesics extend from infinity down to the stable static radius. Two regions of stable circular geodesics are separated by a region of unstable circular geodesics. The outer region of stable orbits extends from infinity down to the ISCO, the inner region of the stable orbits extends from the OSCO down to the stable static radius. 

\subsubsection{No-horizon spacetimes admitting only stable circular orbits}

The no-horizon spacetimes with the charge parameter in the interval  
\beq
       (g/m)_{S} < g/m .
\eeq 
allow only stable circular geodesics that extend from infinity down to the static radius. No unstable circular geodesics are possible. No circular geodesics are allowed under the static radius in all three classes of the no-horizon Bardeen spacetimes. \footnote{Note that in the ABG spacetimes with $g/m > 2$ circular geodesics can exist under the stable static radius, being limited from above by an unstable static radius \cite{Stu-Sche:2015:IJMPD:}.}

We have to stress that character of the circular geodesic motion in the Bardeen no-horizon spacetimes is the same as in the KS naked singularity spacetimes \cite{Stu-Sche:2014:CLAQG:,Stu-Sche-Abd:2014:PHYSR4:,Vie-etal:2014:PHYSR4:}, or in the RN naked singularity spacetimes \cite{Stu-Hle:2002:ActaPhysSlov:,Pug-Que-Ruf:2011:PHYSR4:}. On the other hand, significant differences occur in the Kerr naked singularity spacetimes, especially in the case of near-extreme Kerr spacetimes \cite{deF:1974:ASTRA:,Stu:1980:BAC:,Stu:1981:BAC:,Vir-Elli:2002:PHYSR4:,Tak-Har:2010:CLAQG:,Stu-Sche:2010:CLAQG:,Stu-Hle-Tru:2011:CLAQG:,Pat-Jos:2011:CLAQG:,Stu-Sche:2012a:CLAQG:,Stu-Sche:2012b:CLAQG:,Stu-Sche:2013:CLAQG:}. 

\subsection{Angular velocity of Keplerian orbits}

Of special interest is existence of an alteration of the gradient of the radial profile of the angular frequency of circular geodesics that can have a crucial impact on the standard mechanism of the Keplerian accretion, as the local maximum of the angular velocity profile defines an effective edge of the Keplerian accretion disc with accretion governed by viscosity effects, as the Keplerian accretion stops its functioning for vanishing of the angular velocity gradient \cite{Stu-Sche:2014:CLAQG:,Vie-etal:2014:PHYSR4:}. The angular frequency of the circular geodesics is given by the formula 
\begin{equation}
\Omega_{c}^2=\frac{m(r^2-2g^2)}{(r^2+g^2)^{5/2}}.
\end{equation}
The extrema of the function $\Omega_c^2(r;g,m)$ are located along the curves $r_{\Omega}(g,m)$ determined by the condition $\diff\Omega^2_c/\diff r=0$ that for the Bardeen no-horizon spacetimes implies 
\beq
		r_{\Omega} = 2g;
\eeq
The function $r_{\Omega}$ is illustrated in Figure 1. The vanishing and change of the sign of the gradient of the Keplerian angular velocity radial profile has another important consequence for the Keplerian accretion discs, as the standard accretion governed by the MRI instability requires decreasing of the Keplerian frequency with increasing radius. Therefore, the radius $r_{\Omega}$ can be considered as the inner edge of the standard Keplerian discs. Possible scenarios of the subsequent accretion, located in regions under the radius $r_{\Omega}$, are discussed in \cite{Stu-Sche:2014:CLAQG:} and will not be repeated here. We shall assume here that in the Bardeen spacetimes with $g/m > (g/m)_{S}$, the Keplerian discs have their inner radius at $r_{\Omega}$. 

\subsection{Photon motion}

For a general photon motion, not confined to the equatorial plane where the Keplerian disc location is assumed, the radial component of the photon 4-momentum reads  
\begin{equation}
\left[P^{r}\right]^{2}=E^{2}-f(r)\left(\frac{L^{2}+Q^2}{r^{2}}\right).\label{eq_radial}
\end{equation}
The trajectories of photons are independent of energy, therefore, it is convenient to relate the effective potential of the photon motion to the impact parameters 
\begin{equation}
                l = \frac{L}{E} , q^2 = \frac{Q^2}{E^2}
\end{equation}
For the photon motion, it is further convenient to use the coordinates 
\begin{equation}
       u=\frac{1}{r} ,
\end{equation}
\begin{equation}
       \mathfrak{m}=\cos\theta ,
\end{equation}
and to re-parameterize the radial motion equation by $Ew\rightarrow w$, using simultaneously also the constants of motion in the form of impact parameters. The equations of the radial and the latitudinal motion expressed in terms of the affine parameter $w$ then take the form \cite{Stu-Sche:2014:CLAQG:}
\begin{equation}
\frac{du}{dw}=\pm u^{2}\sqrt{1-\widetilde{f}(u)\left(l^{2}+q^{2}\right)u^{2}}
\end{equation}
where  
\begin{equation}
\tilde{f}(u)=f(1/u)
\end{equation}
and
\begin{equation}
\frac{d\mathfrak{m}}{dw}=\pm u^{2}\sqrt{q^{2}-(l^{2}+q^{2})\mathfrak{m}^{2}} 
\end{equation}
that can be properly integrated when photons radiated by Keplerian discs are considered \cite{Rau-Bla:1994:ApJ:,Fan-etal:1997:PASJ:,Sche-Stu:2008:IJMPD:,Sche-Stu:2008:GenRelGrav:,Stu-Sche:2010:CLAQG:,Kra:2011:CLAQG:,Kra:2014:GRG:}. 

\section{Deflection of light rays}

We first study deflection of light in the Bardeen no-horizon spacetimes concentrating on the interplay of  photons with low and high values of the impact parameter that implies a variety of interesting consequences in optical phenomena. 

\subsection{Special character of the regular spacetimes imprinted in deflection of light} 

The Bardeen spacetimes are regular at the central region $r \sim 0$, being nearly flat there,  
\begin{equation}
	r\rightarrow 0 \quad \Rightarrow \quad g_{tt}\rightarrow -1 \textrm{ and } g_{rr}\rightarrow 1 .
\end{equation}
To be exact, they have de Sitter character at $r \sim 0$ \cite{Stu-Sche:2015:IJMPD:}. Therefore, contrary to the case of naked singularity spacetimes, the gravitational effects are weak at the central region of the regular Bardeen no-horizon spacetimes. This property has an important effect on the deflection of light, giving similar deflection for photons with high and low values of the impact parameter, and implying extraordinary consequences for images of the Keplerian discs. The most interesting seems to be 
creation of ghost images in the central region of the Bardeen no-horizon spacetimes considered here, but it can be relevant also for profiled spectral lines generated by Keplerian discs, light curves of hots spots on the discs,or for weak lensing of distant objects that will be considered in future papers.  

Of course, similar optical phenomena arise in all the regular no-horizon spacetimes having a near-flat central region, e.g., the regular ABG no-horizon spacetimes.

The deflection of light will be studied for simplicity in the equatorial plane in order to obtain a clean demonstration of the phenomenon of dependence of the deflection angle on the magnitude of the impact parameter. The deflection of light is then governed by the equation
\begin{equation}
	\phi=\phi_0+\int^{u_e}_{u_t}\frac{l\diff u}{\sqrt{1-f(u)l^2 u^2}}+\int^{u_o}_{u_t}\frac{l\diff u}{\sqrt{1-f(u)l^2 u^2}}
\end{equation}
where $u_e, u_o, u_t$ denote the radial coordinate of the emitter, observer, and the turning point of the radial photon motion, and $l$ is the impact parameter. 

We illustrate the origin of the phenomenon of critical impact parameter giving maximal deflection angle of photon trajectories in Figures 2 and 3 where light rays are constructed in two characteristic Bardeen spacetimes for typical values of the impact parameter $l$. We can see that if the deflection angle is considered as a function of the impact parameter $l$, the deflection is very weak for large values of $l$ and photon trajectories at large minimal distances from the centre $r=0$, and for low values of $l$ and trajectories at minimal distance in vicinity of $r=0$; there is a maximal value of the deflection angle that occurs for a critical value of the impact parameter $l_c$. In the Bardeen no-horizon spacetimes with $(g/m)_{NoH}< g/m < (g/m)_{P}$, allowing for existence of the photon circular geodesics, trajectories winding up to the unstable photon geodesic exist (Figure 2), while no winding trajectories exist in the Bardeen spacetimes with $g/m > (g/m)_P$ (Figure 3). In both cases, trajectories with low impact parameters are only  slightly deflected in the cetral region, being well observable by distant observers because of non-existence of an event horizon. 

\begin{figure}[H]
\begin{center}
	\begin{tabular}{cc}
	\includegraphics[scale=0.51]{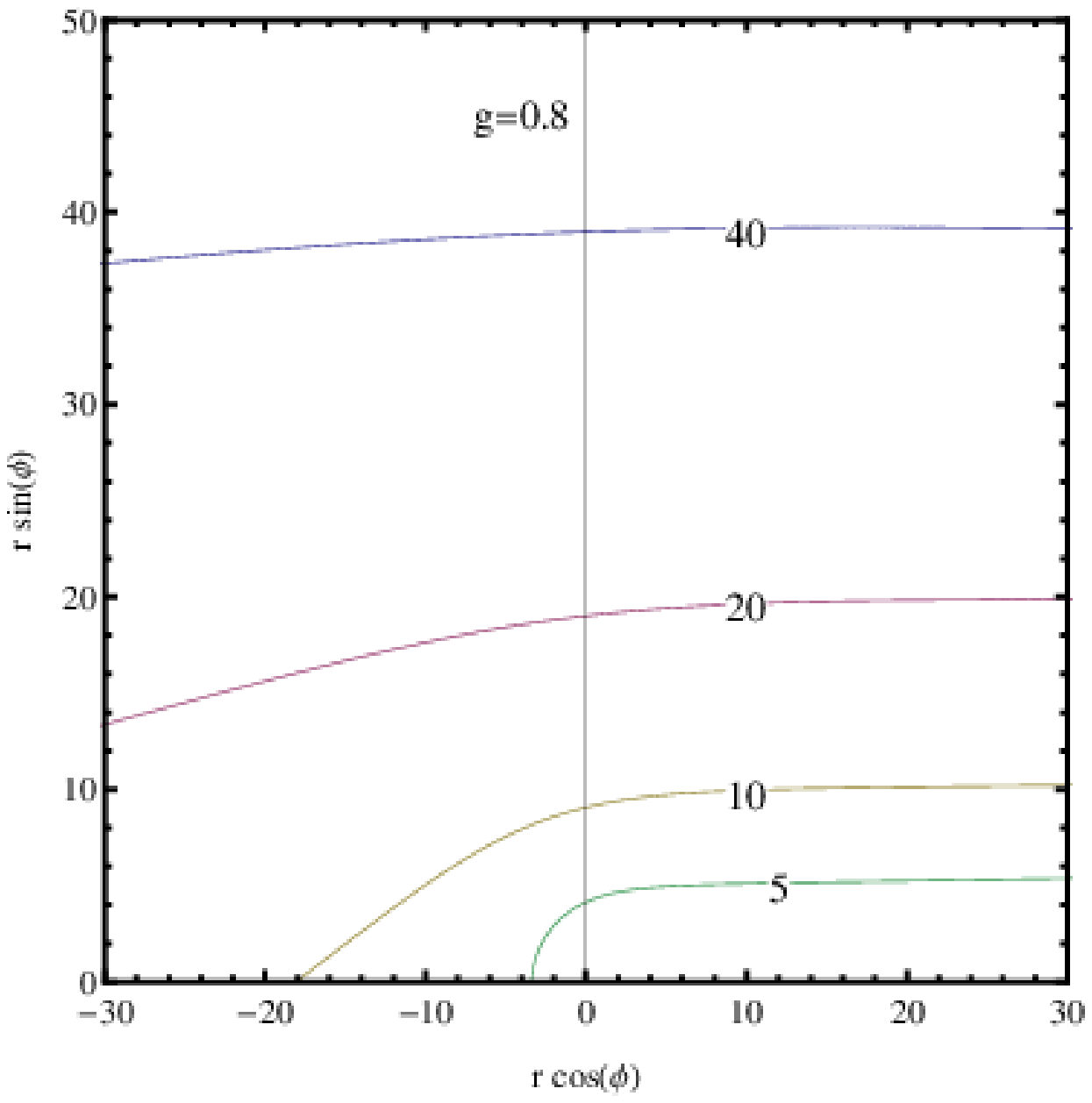}&\includegraphics[scale=0.51]{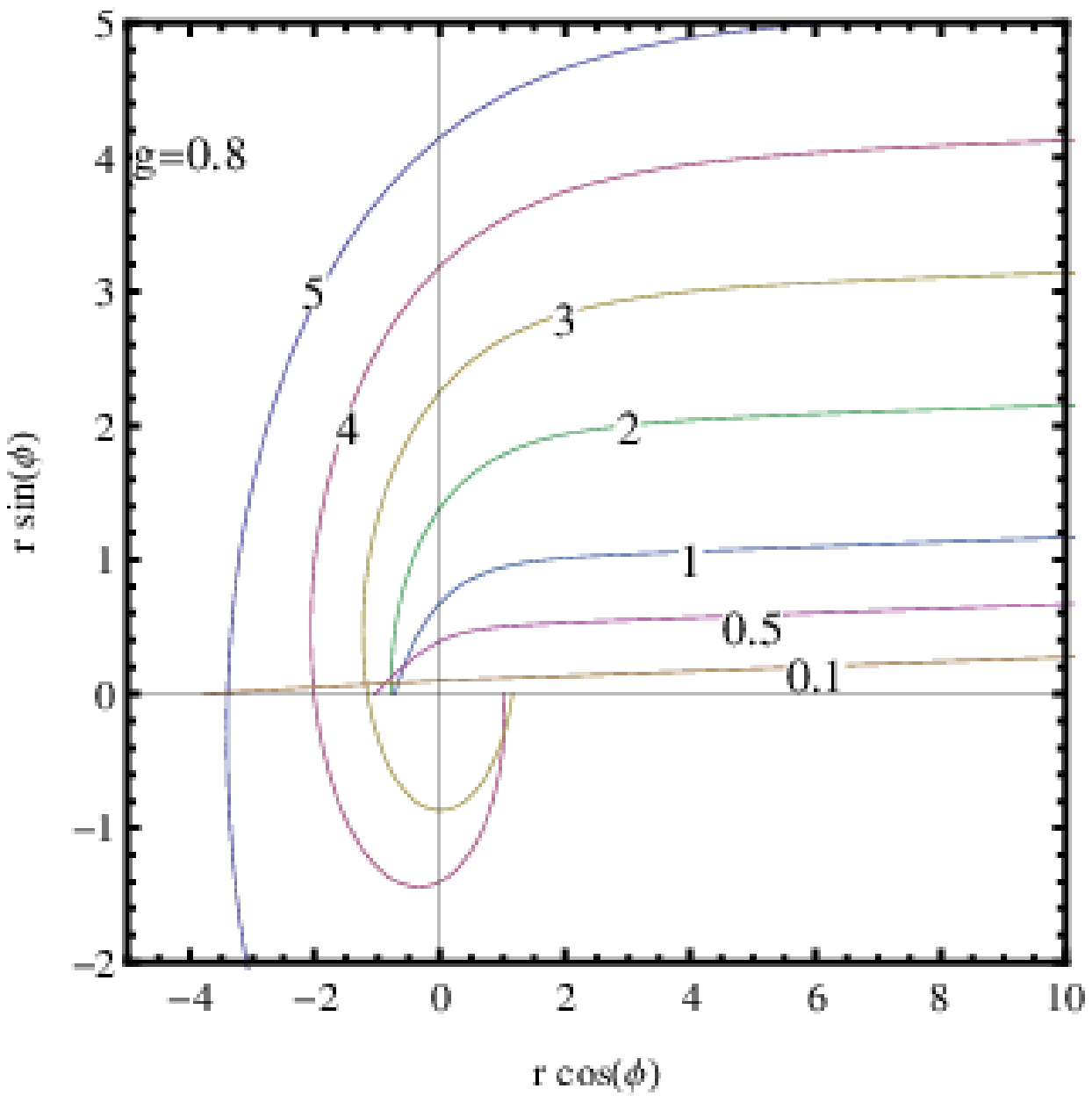}
	\end{tabular}
	\caption{Illustration of the light rays bending in the Bardeen no horizon regular spacetime with magnetic charge parameter $g=0.8$ for large (left) and small (right) impact parameters. The maximal deflection angle diverges due to the existence of the photon circular orbit.}
\end{center}
\end{figure}

\begin{figure}[H]
\begin{center}
	\begin{tabular}{cc}
	\includegraphics[scale=0.3]{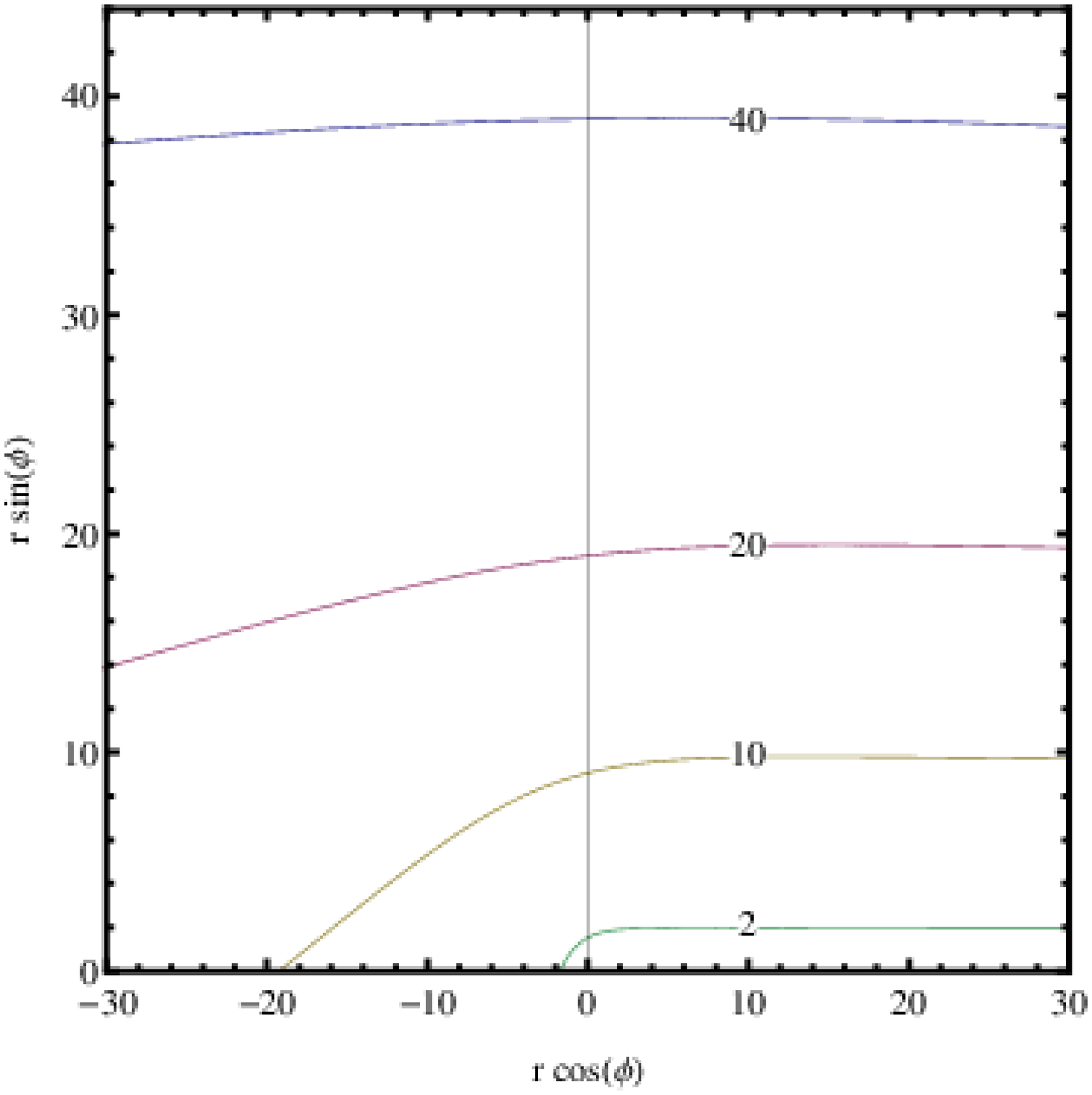}&\includegraphics[scale=0.3]{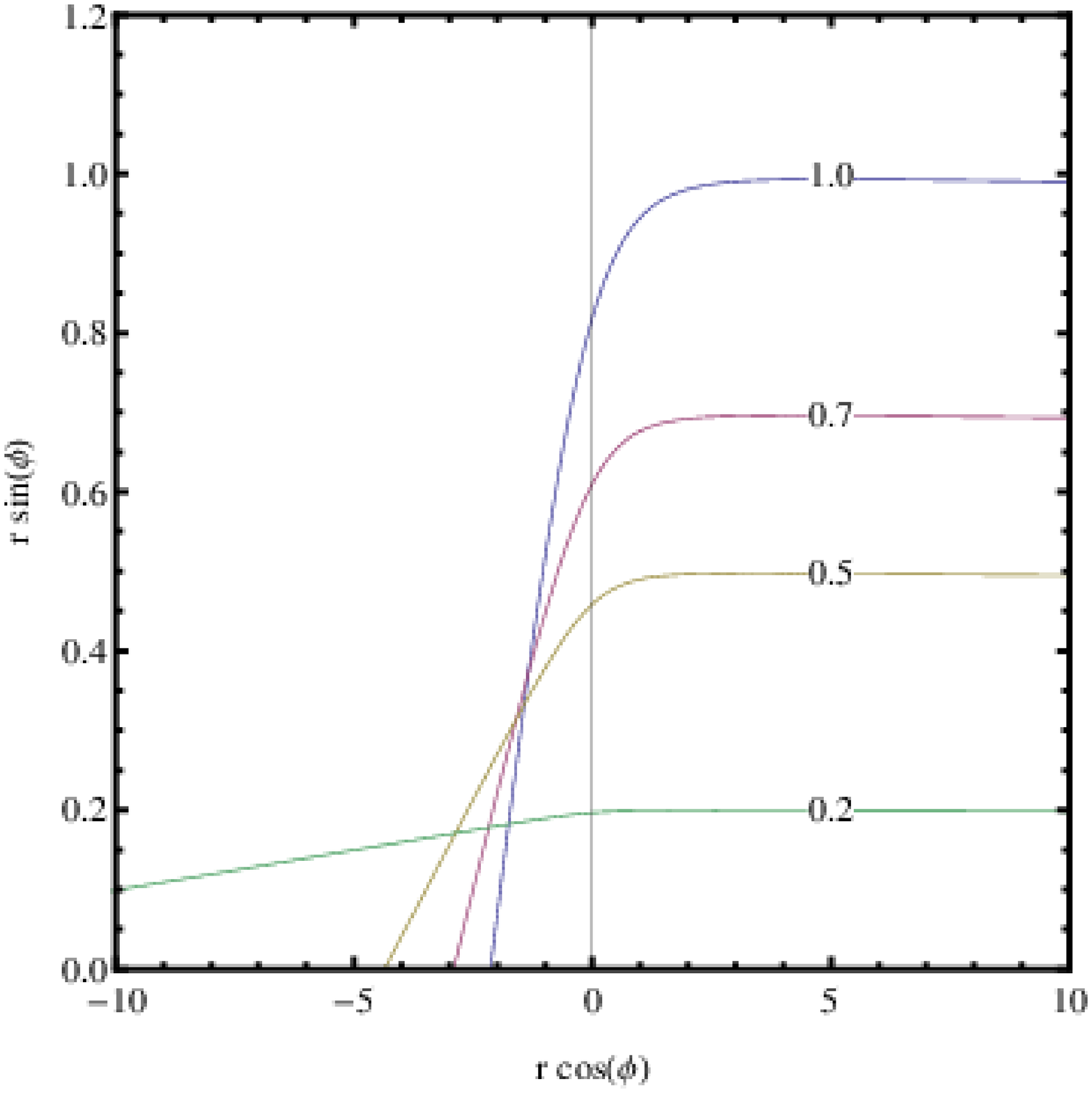}
	\end{tabular}
	\caption{Illustration of the light rays bending in the Bardeen no horizon regular spacetime with magnetic charge parameter $g/m=1.3$ for large (left) and small (right) impact parameters. No photon circular orbit exists in this Bardeen spacetime and the maximal deflection angle is finite.}
\end{center}
\end{figure}

\subsection{Deflection angle} 

The peculiar properties of the regular Bardeen no-horizon spacetimes can be seen from the behaviour of deflection angle defined by the formula
\begin{equation}
	\delta=2\psi_\infty-\pi,
\end{equation}
where is
\begin{equation}
	\psi_\infty=\int_{u_t}^{0}\frac{l\diff u}{\sqrt{1-f(u)l^2u^2}}
\end{equation}
and $\psi$ denotes the azimuthal angle measured in the central plane where motion of a particular photon occurs. We have constructed the deflection angle as function of the impact parameter $l$ for any spacetime characterized by the charge parameter $g/m$. The functions $\delta\equiv\delta(l;g)$ are represented for four representative values of the magnetic charge parameter $g/m=1.0$, $1.5$, $2.0$, and $2.5$ (see Figure 4). 
For any value of $g/m > (g/m)_{NoH}$ a local maximum of the function $\delta\equiv\delta(l;g)$ exists, giving thus the maximal deflection angle $\delta_{max}$ corresponding to the critical value of the impact parameter $l_c$. For $l<l_c$ the bending of light is decreasing because of decreasing gravitational effects of the spacetime. 

One can clearly see that with increasing value of the charge parameter $g/m$ the value of maximal deflection angle $\delta_{max}$ is decreasing. This kind of behavior is expected since the metric lapse function $f(r)$ is getting closer to value $1$ as value of the charge parameter $g/m$ is increasing and radius $r$ is fixed. On the other hand, with decreasing parameter $g/m$, the maximal deflection angle increases. As expected, the maximal deflection angle diverges for $g/m \to (g/m)_P$, and it remains divergent for all the Bardeen spacetimes with $(g/m)_{NoH}< g/m < (g/m)_{P}$, allowing for existence of photon circular geodesics. The divergence corresponds to the unstable (outer) photon circular geodesic at $r=r_{phu}$. \footnote{Trapped photons in vicinity of the stable (inner) photon circular orbit are irrelevant for the phenomena related to distant observers \cite{Stu-Sche:2015:IJMPD:}.} We give dependence of the maximal deflection angle $\alpha_{max}$, the corresponding critical impact parameter $l_c$, and the related turning point $r_t$ of the trajectory corresponding to the maximal deflection angle on the charge parameter $g/m$ in Figure 4. For $g/m \to (g/m)_P$, the critical impact parameter $l_c \to l_{phu}$ and the turning radius $r_t \to r_{phu}$. 
\begin{figure}[H]
\begin{center}
\begin{tabular}{cc}
	\includegraphics[scale=0.6]{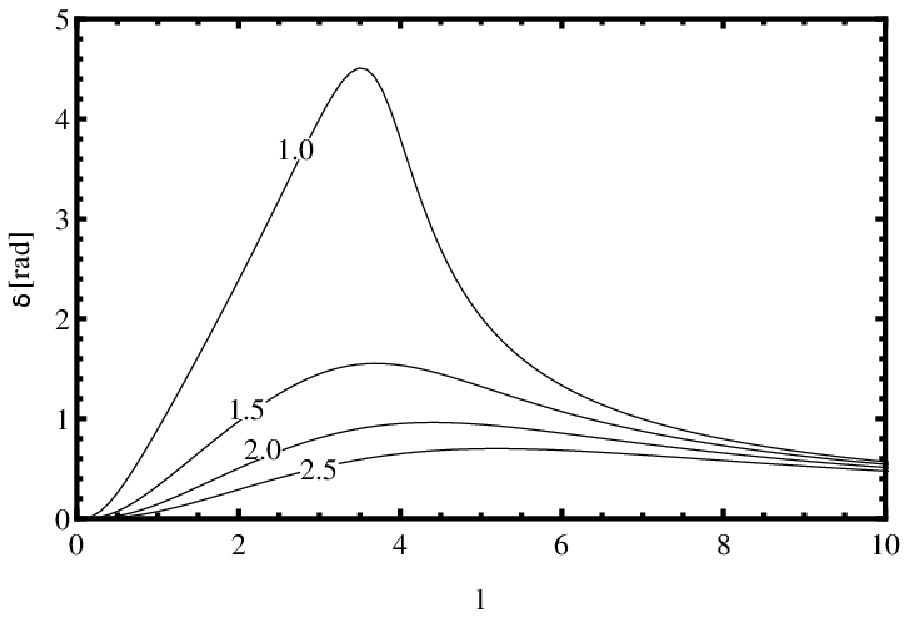}&\includegraphics[scale=0.6]{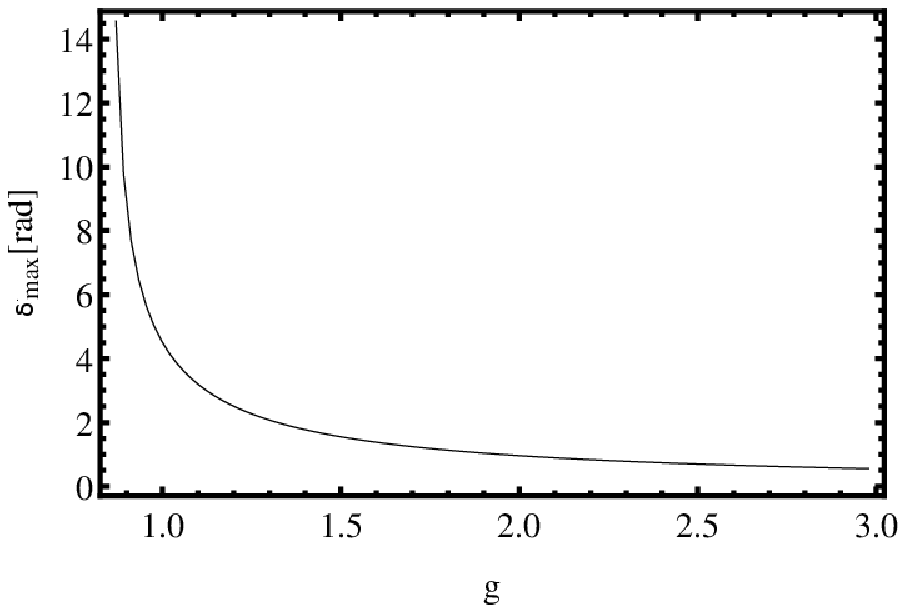}\\
	\includegraphics[scale=0.6]{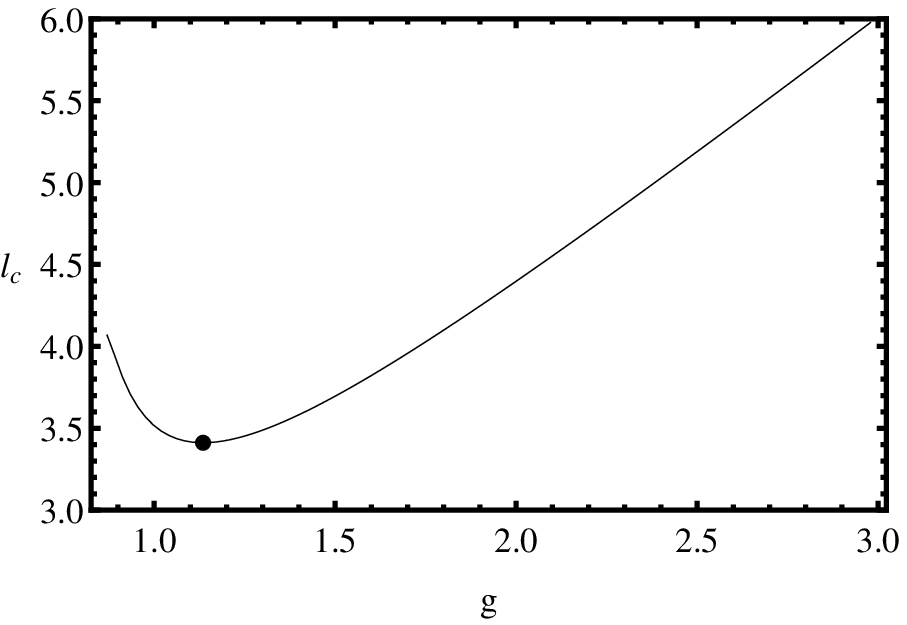}&\includegraphics[scale=0.6]{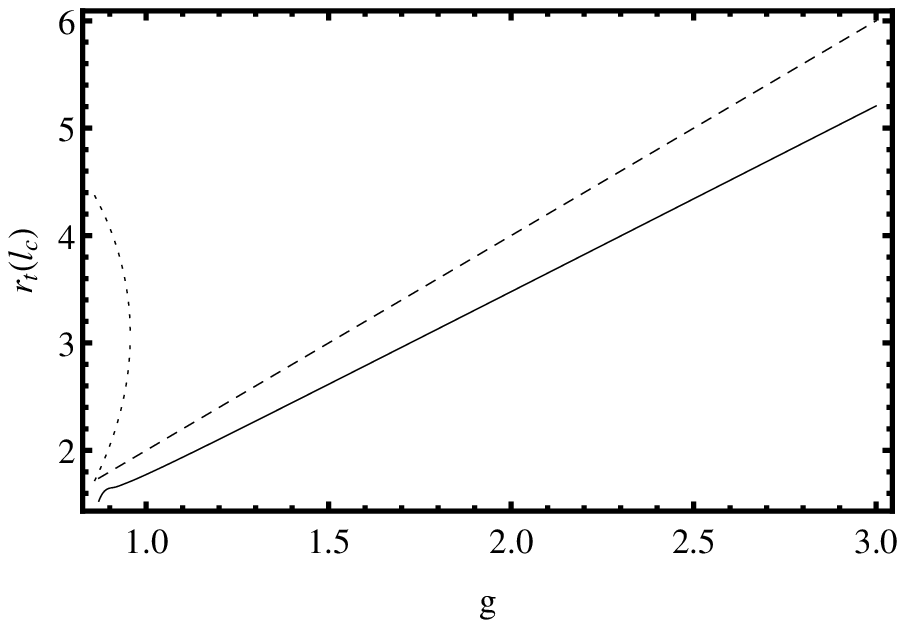}\\
\end{tabular}
	\caption{Illustration of the behavior of the deflection angle $\delta(l,g)$ (\emph{top, left}) is presented as function of the impact parameter $l$ for four representative values of the magnetic charge parameter $g/m=1.0$, $1.5$, $2.0$, and $2.5$.  The plot of the maxima $\delta_{max}(g)$ (\emph{top, right}) of the deflection angle $\delta$ as function of the magnetic charge parameter $g/m\ge (g/m)_{P}\simeq 0.85865$ (the situation where no photon orbits are present; if photon orbits are present then the maximal deflection is, of course, infinite). The plot of the critical impact parameter $l_c(g)$ (\emph{bottom, left}) corresponding to the $\delta_{max}(g)$ with the minimum of the curve $l_c(g)$ located at $[g/m,l]=[1.13518, 3.4115]$, are presented with the plot of the radial turning point $r_t(l_c)$ (\emph{bottom, right}) corresponding to the critical impact parameter $l_c$, that is plotted along with the radius $r_{\Omega}(g)$ of vanishing of the angular velocity gradient (dashed) and the radius of the marginally stable orbits $r_{ms}(g)$ (dotted).}
\end{center}
\end{figure} 
One can see that the function $\delta_{max}(g)$ is monotonically increasing with the magnetic charge $g/m$ decreasing to $(g/m)_P$ where it diverges. The radius of the turning point of the radial motion of the photon with critical impact parameter $l_c$ decreases with the magnetic charge parameter decreasing to $(g/m)_P$. In the Bardeen no-horizon spacetimes with $g/m < (g/m)_P$, the maximal deflection angle $\delta_{max}$ diverges and it corresponds to the unstable (outer) photon circular orbit with $l_c = l_{phu}$; the turning point of the radial motion is related to the radius of the unstable circular photon orbit. For $g/m > (g/m)_P$, the critical impact parameter function $l_{c}(g)$ has a local minimum $l_{c(min)}(g/m \sim 1.1) \sim 3.4$. 

\subsection{Classification of images}

In a spherically symmetric spacetime a radiating source following a (circular) geodesic is confined to a central plane. Inclination angle of an observer receiving the radiation is related to the axis perpendicular to the central plane. Keplerian (geometrically thin) discs are assumed to be located in the equatorial plane of the spacetime. Classification of images of the radiating source is governed by the character of the motion of photons creating the images; namely it is given by the number of half-turns of the photon trajectory around the centre of the spacetime. 

The direct (primary) images are created by photons going directly to the observer, without crossing the equatorial plane. The indirect (secondary) images are created by photons crossing once the equatorial plane. (In order to map all the optical phenomena, here we assume that the Keplerian discs are optically thin enabling thus such a crossing of the equatorial plane.) 

In the spacetimes allowing for existence of unstable photon circular orbit with the impact parameter $l_{ph}$, infinite number of images can be created near the region corresponding to $l=l_{ph}$. The $n-th$ order image is related to photons with impact parameter $l \sim l_{ph}$ that cross the equatorial plane $(n-1)$ times. In the black hole spacetimes, such images can be created only by photons with impact parameter slightly higher than $l_{ph}$, while for the no-horizon or naked singularity spacetimes, also photons with impact parameter slightly lower than $l_{ph}$ enter the play. 

In all naked singularity spacetimes photons with small impact parameter are reflected by the central repulsive barrier, while in all the regular no-horizon spacetimes such photons can cross the central region. Such photons create the so called ghost images, additional to both the direct and indirect images. If such spacetimes admit existence of circular photon orbits, the ghost images are related to photons with $l<l_{ph}$. We shall concentrate our study in determining properties of the ghost images, mapping differences between the regular no-horizon spacetimes and the naked singularity spacetimes. 

\section{Direct ghost images of Keplerian discs}

The effect of the maximal deflection angle implies that for a radiating source orbiting at the equatorial plane in strong gravity region of a Bardeen no-horizon spacetime, a distant observer located at a large inclination angle can receive photons of significantly different impact parameters that correspond to two images of the radiating point. This effect is fully governed by the dependence of the emission radius of photons received by a distant observer at a given inclination angle as function of their impact parameter. We shall see that even three photons with different impact parameters emitted from a given source at a fixed radius at the strong gravity region can reach the distant observers at large inclination angles. In such situations, the photon with large impact parameter corresponds to the standard direct image, while the photon (photons) with small impact parameter correspond to the ghost image. For images of the Keplerian discs, it is necessary to have the region generating the ghost images to be located above the inner edge of the Keplerian disc. In general situations, the inner edge corresponds to the ISCO, but in the no-horizon spacetimes with $g/m > (g/m)_P$, the radius of the vanishing of the angular velocity radial profile gradient should represent the inner edge. 

Therefore, we study appearance of the innermost parts of the Keplerian discs orbiting the regular Bardeen no-horizon spacetimes of all three classes concentrating on the behavior of the standard direct images and the ghost images related to the low impact parameter photons. We give their optical appearance reflecting their shape distortions due to the gravitational lensing, and the frequency shift of the emitted radiation due to the gravitational and Doppler effects. We use the ray-tracing method developed for the black hole case in \cite{Fan-etal:1997:PASJ:,Sche-Stu:2008:IJMPD:} and for the naked singularity case in \cite{Stu-Sche:2010:CLAQG:,Stu-Sche:2012a:CLAQG:,Sche-Stu:2013:JCAP:}. We demonstrate the combined gravitational and Doppler shifts by a simple map assuming the Keplerian discs radiating locally at a fixed frequency corresponding, e.g., to a Fe X-ray line. In order to have a complete picture of the phenomenon of the ghost images, we give also explicitly the range of the frequency shift of the standard direct images and the ghost direct images. 

Imaging of the Keplerian discs observed by distant observers at arbitrary inclination angles means that we have to consider the role of photons with both large and small impact parameters. The photons with large impact parameters give the standard direct image of the Keplerian discs and they reflect the whole area of the discs, while the photons with small impact parameters give ghost images reflecting usually a small part of the Keplerian discs, namely the regions located just behind the centre of the no-horizon spacetime. The ghost images can exist for large enough inclination angles of the observer, being located inside the standard direct images and separated from them. However, for very large inclination angles the ghost images can be joined to the direct images or they can coincide with them. 
 
The disc appearance can be relevant for sources close enough when the current observational technique enables a detailed study of the innermost parts of the accretion structures just at close vicinity of the black hole horizon or the innermost parts of the no-horizon spacetimes, as can be expected in near future for the Sgr A* source \cite{Doe-etal:2009:APJ:}. \footnote{The profiled spectral lines generated by the Keplerian discs, or the light curves of hot spots on the surface of orbiting Keplerian discs, are relevant and well measurable also for much more distant sources when efficient X-ray satellite observatories as LOFT \cite{Fer-etal:2012:ExpA:} will be used. These will be studied in future works.}

\subsection{Frequency shift}

We assume the Keplerian discs composed from isotropically radiating particles following the circular geodesics in the equatorial plane. The frequency shift of radiation emitted by a point source moving along such a circular orbit is given in the standard way by 
\begin{equation}
1+z = \frac{k_{\mu}U^{\mu}|_{o}}{k_{\mu}U^{\mu}|_{e}}.
\end{equation}
For the emitter we use notation (e), while for the observer we use notation (o). For circular orbits, the emitter four-velocity has only the time and axial non-zero components 
\begin{equation}
U^{\mu}=\left[U^{t},0,0,U^{\phi}\right].
\end{equation}
For the static observers at infinity, the frequency shift formula $1+z$ reads
\begin{equation}
1+z=\frac{1}{U_{e}^{t}(1-l\Omega)}  
\end{equation}
\begin{equation}
\frac{1}{U_{e}^{t}} = (f(r) - r_{e}^2 \Omega^2)^{1/2} ,  
\end{equation}
where $l$ is the impact parameter of the photon, and $\Omega$ is the angular velocity of the emitter, here assumed to be the Keplerian angular velocity $\Omega_c$.  
 
\subsection{Construction of the direct images and direct ghost images}

Existence of the ghost images in the Bardeen no-horizon spacetimes is directly related to the existence of the maximal deflection angle implying the possibility to have two photons radiated from a point-like source at radius $r_e$ close to the centre at $r=0$ with significantly different impact parameters that can reach a distant observer, as illustrated in Figure 5.   
\begin{figure}[H]
\begin{center}
	\includegraphics[scale=0.6]{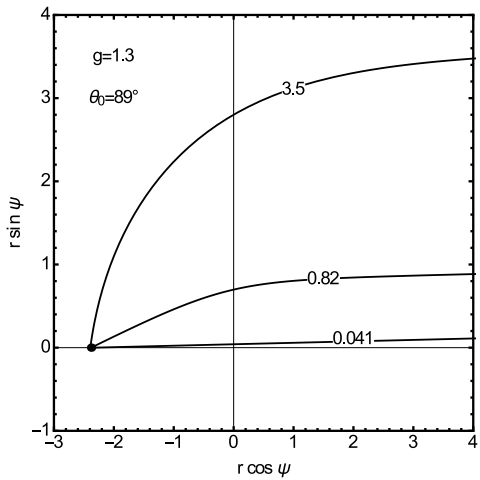}
	\caption{We illustrate origin of the ghost image in the case of the no-horizon spacetime with magnetic charge parameter $g=1.3$. The ghost image arises due to fact that in close vicinity of the center there is $g_{tt}\rightarrow -1$ and $g_{rr}\rightarrow 1$, therefore the light rays starting at a point at the radius $r_e$ with sufficiently small impact parameter $l<l_c$ are only slightly bended creating the ghost image, while the light rays starting at the same point with $l>l_c$ are bended more strongly, creating the direct image in the observer sky. The ghost image has to be related to a restricted region located behind the central object and the inclination angle of the observer has to be sufficiently large for creation of this image. Here the situation is illustrated for very large inclination angle $\theta_o = 89^{o}$ and the three photons with impact parameters allowing to reach the observer while radiated at $r_e = 2.5$.}
\end{center}
\end{figure}
Both direct and ghost images are considered to be primary images as they are generated by photons having no full turning about the central source. In the Bardeen no-horizon spacetimes with $g/m > (g/m)_P$ only the primary images can be created, while in the spacetimes with $(g/m)_{NoH} < g/m < (g/m)_P$, also the secondary and higher-order images can be created due to the existence of the unstable photon circular geodesic. Here we restrict our attention to the primary images. 

Construction of the primary images is reflected in Figure 6 where the motion is illustrated for the central plane $r - \theta$ containing the observer located at $\theta_o$. We have to calculate the integrals along the trajectories of the light given in three different ways. There are trajectories having no turning points at both the radial and latitudinal motion (type I), trajectories having the turning points of both the radial and latitudinal motion (type II), and trajectories having a turning point of the latitudinal motion, but no turning point of the radial motion (type III). Note that the orbits of the Type III and I have to start between the spacetime centre and the observer, while orbits of the Type II have to start behind the centre. 
The integrals are given by the formulae 
\begin{eqnarray}
	\int^{rt}_{r_o}\frac{\diff r}{\sqrt{R}}+\int^{r_t}_{r_e}\frac{\diff r}{\sqrt{R}}&=&\int^{\theta_+}_{\theta _o}\frac{\diff \theta}{\sqrt{W}}+\int^{\theta_+}_{\pi/2}\frac{\diff \theta}{\sqrt{W}},\\
	\int^{r_e}_{r_o}\frac{\diff r}{\sqrt{R}}&=&\int^{\theta_+}_{\theta_o}\frac{\diff \theta}{\sqrt{W}}+\int^{\theta_+}_{\pi/2}\frac{\diff \theta}{\sqrt{W}},\\
	\int^{r_e}_{r_o}\frac{\diff r}{\sqrt{R}}&=&\int^{\pi/2}_{\theta_o}\frac{\diff \theta}{\sqrt{W}}.
\end{eqnarray}

\begin{figure}[H]
	\begin{center}
	\includegraphics[scale=0.4]{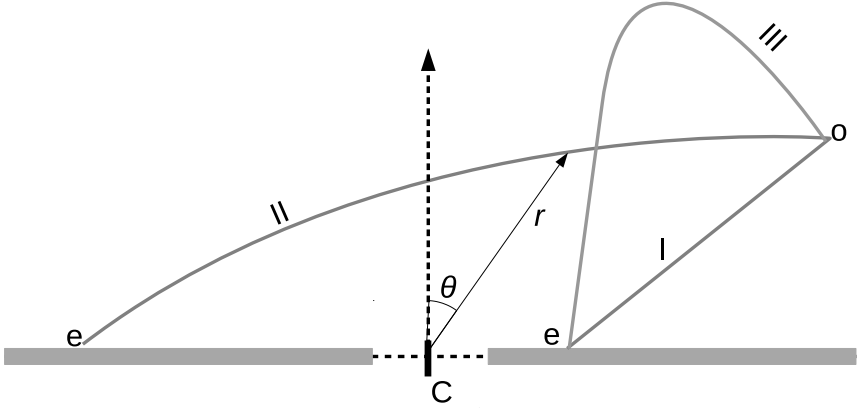}
	\caption{Schema of the Keplerian disc primary images construction.}
	\end{center}
\end{figure}

In the distant observer sky, the images of the source are determined by the angles related in the standard way to the impact parameters $l$ and $q$ of radiated photons \cite{Bar:1973:BlaHol:,Cun-Bar:1975:ApJ:}
\begin{eqnarray}
\alpha&=&-\frac{l}{\sin\theta_o},\\
\beta^2&=&q^2-l^2\cot^2\theta_o.
\end{eqnarray}
For imaging, we consider the innermost regions of the Keplerian discs limited by the inner edge at 
\begin{equation}
	r_{in}=r_{\Omega}=2g
\end{equation}
in the Bardeen no-horizon spacetimes with $g/m > (g/m)_P$, and at 
\begin{equation}
	r_{in}=r_{ISCO}
\end{equation}
in the Bardeen no-horizon spacetimes with $(g/m)_{NoH} < g/m < (g/m)_S$. The outer edge of the radiating region of the Keplerian disc is always assumed at 
\begin{equation}
	r_{out}=20m.
\end{equation}
The points of the images are given by the angles $\alpha$ and $\beta$ governed by the impact parameters $l,q$ that are determined for given coordinates $r_e, \phi_e$ of the radiating pointlike sources in the equatorial Keplerian disc ($\theta=\pi/2$) and the observer coordinates $r_o, \theta_o, \phi_o=0$ by integrating the equations of the photon motion (see, e.g., \cite{Bao-Stu:1992:ApJ:}). For each point of the image, given by the pairs ($\alpha$ - $\beta$), ($l-q$), the frequency shift of the imaged pointlike source is given by the frequency shift formula with the relevant values of the impact parameters $l,q$ and the positions of the source and the observer.

\subsection{Appearance of the Keplerian discs}

The results of the construction of the primary images, i.e., both the direct and ghost images of the innermost parts of the Keplerian discs are represented in Figures 6-8 for all the classes of the regular Bardeen no-horizon spacetimes. The frequency shift mapping of both the direct and ghost images is reflected by the color code. In all the constructed images we present the map of the relative frequency shift 
\begin{equation}
                    g*=\frac{z-z_{min}}{z_{max}-z_{min}}
\end{equation}
related to the range $(z_{max}-z_{min})$ giving the extension of the frequency shift in all the considered cases of the appearance of the Keplerian discs. We also give the extremal values of the frequency shift $(z_{max}$ and $z_{min})$. It should be stressed that the frequency range could serve as a strong tool for determining the spacetime parameters, as demonstrated in the case of the Kerr naked singularities \cite{Stu-Sche:2010:CLAQG:}. 

It should be stressed that the ghost images occur only for large inclination angles ($\theta_o > 80^\circ$). This is quite natural result, as the photons with low impact parameter have to move very close to the origin of coordinates and experience only very small deflection. 

The direct images correspond to the whole radiating region of the Keplerian discs, while the ghost images are created only for a strictly limited region of the radiating discs that is located just behind the gravitating source from the point of view of the distant observer. 

There exist two different relations of the standard direct images and the ghost images. First, the ghost images can be completely separated from the direct images, second, the ghost images can merge to some extent with the direct images. In a given Bardeen no-horizon spacetime the convergence of the ghost and direct images occurs for some critical inclination angle of the distant observer, or inversely, for a fixed inclination angle of the observer, the ghost image is touching the standard direct image for a critical value of the charge parameter of the Bardeen spacetime. 

We can see that for a given inclination angle of the observer, the extension of the ghost image increases with increasing value of the charge parameter of the spacetime, while for a fixed Bardeen spacetime, the extension of the ghost image increases with the inclination angle. No ghost images have been found in the case of the Bardeen no-horizon spacetimes allowing for existence of the photon circular orbits since in such spacetimes the inner edge of the Keplerian disc is located above the region allowing for creation of the ghost images. In the case of the internal Keplerian orbits located under the stable (inner) circular photon orbits the ghost images could be created, but existence of such Keplerian discs is astrophysically unrealistic. 

The joining of the ghost images to the direct images occurs for very large inclination angles that slightly decrease with increasing charge parameter of the spacetime. 

The frequency shift of the direct images of Keplerian discs demonstrates an interesting dependence on the inclination angle -- $z_{max}$ increases with increasing $\theta_{o}$ in a fixed spacetime up to $\theta_{o} \sim 85^{\circ}$, but it decreases for larger inclination angles. The maximum $z_{max}$ decreases with increasing charge parameter $g$ of the Bardeen spacetime, while $z_{min}$ decreases with increasing $g$ up to $g \sim 1$, and increases for $g > 1$. The frequency range of the direct ghost image is substantially restricted in comparison with the frequency range of the standard direct image, as the ghost radiation is related to a restricted region of the disc lying just behind the source centre moving transversely relative to the observer. 

\begin{figure}[H]
	\begin{center}
	\begin{tabular}{c}
		\includegraphics[scale=0.75]{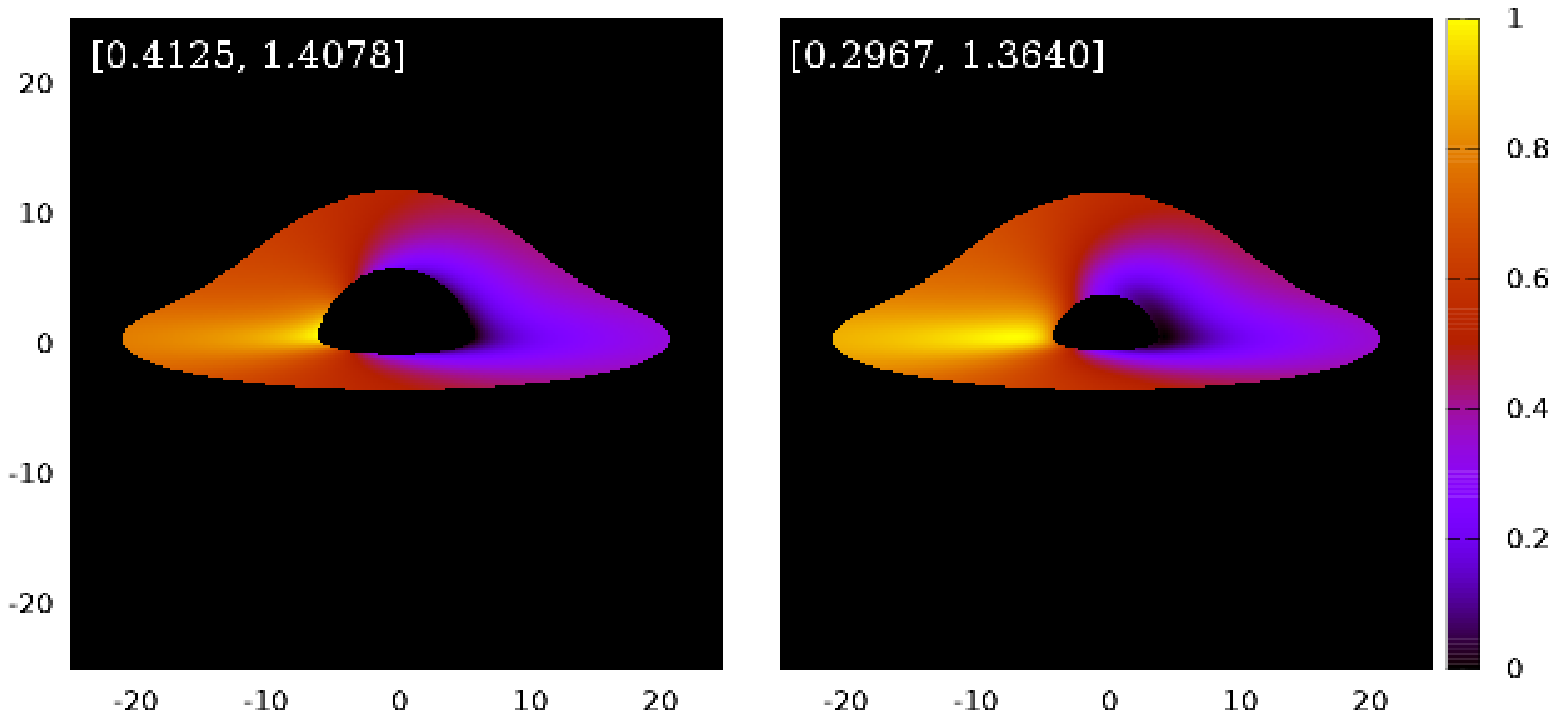}\\
		\includegraphics[scale=0.75]{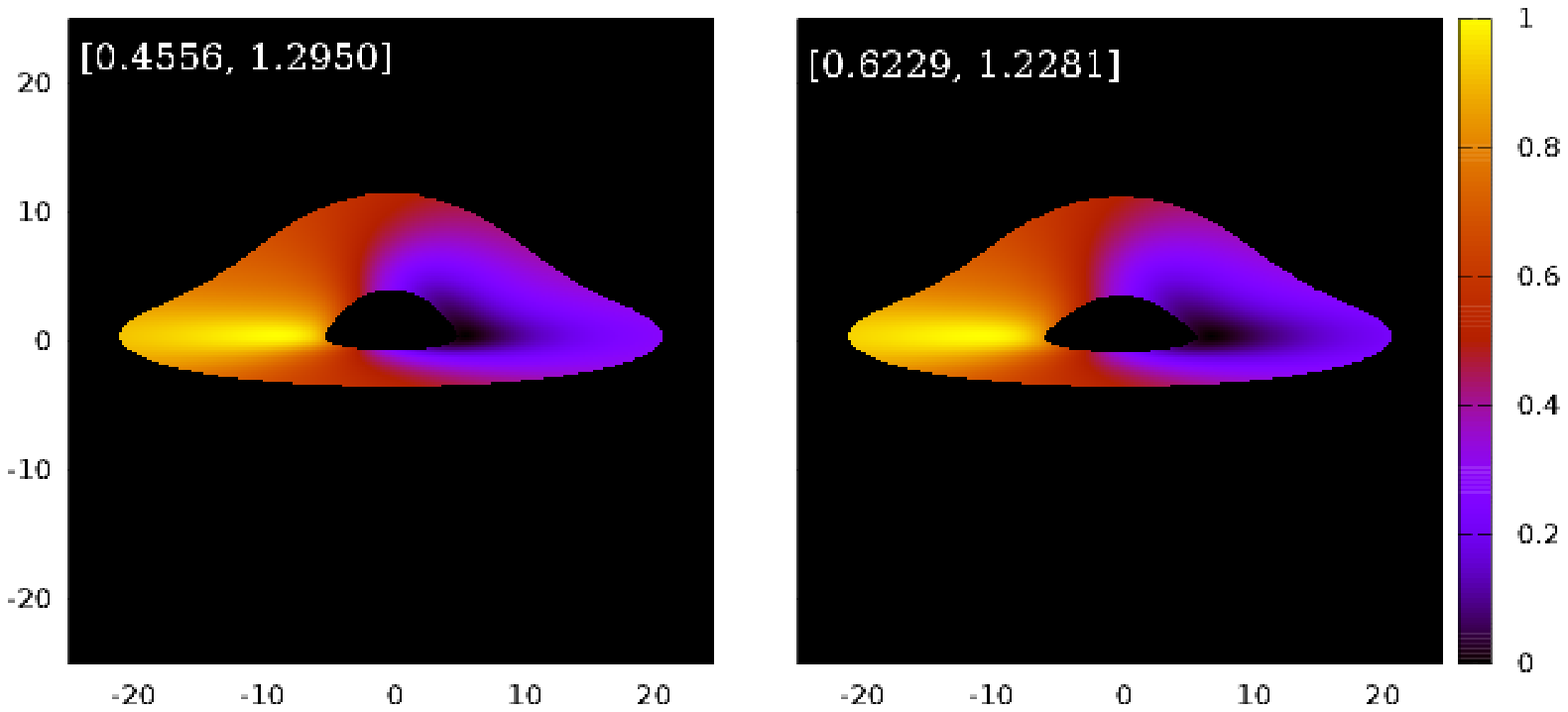}
	\end{tabular}
	\caption{Map of the frequency shift in the Keplerian disc direct images for the charge parameter $g/m=0.8$, $1.1$, $1.5$ , and $2.5$ (from top left to right bottom). The observer inclination is set to $\theta_o=80^\circ$. No direct ghost images occur for this inclination angle. The rotation of the disc is assumed anticlockwise here and in all the following figures giving the images of Keplerian discs. Here and in the following figures, we give in all the constructed images the extremal values of the frequency shift across the discs $(z_{max}$ and $z_{min})$.}
	\end{center}
\end{figure}

\begin{figure}[H]
	\begin{center}
	\begin{tabular}{c}
			\includegraphics[scale=0.75]{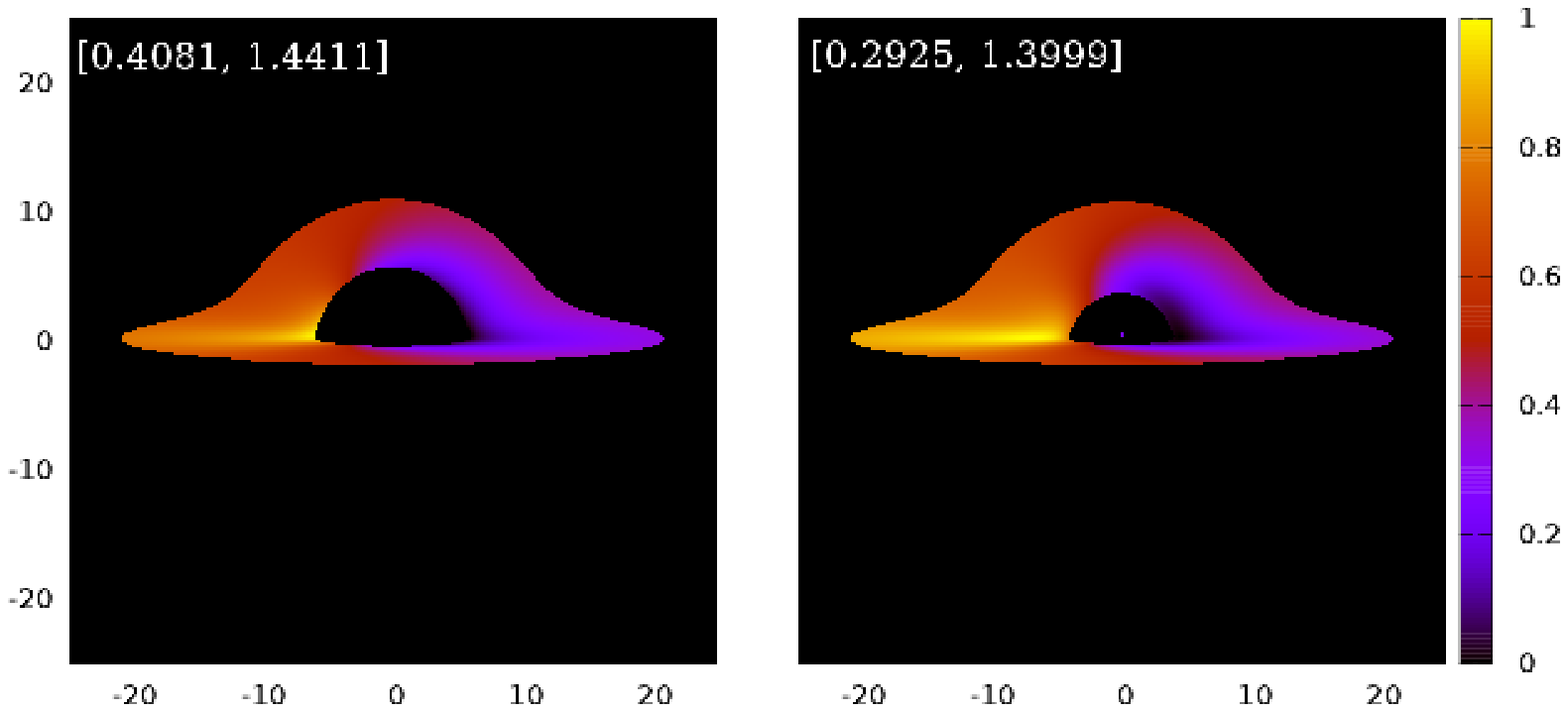}\\
			\includegraphics[scale=0.75]{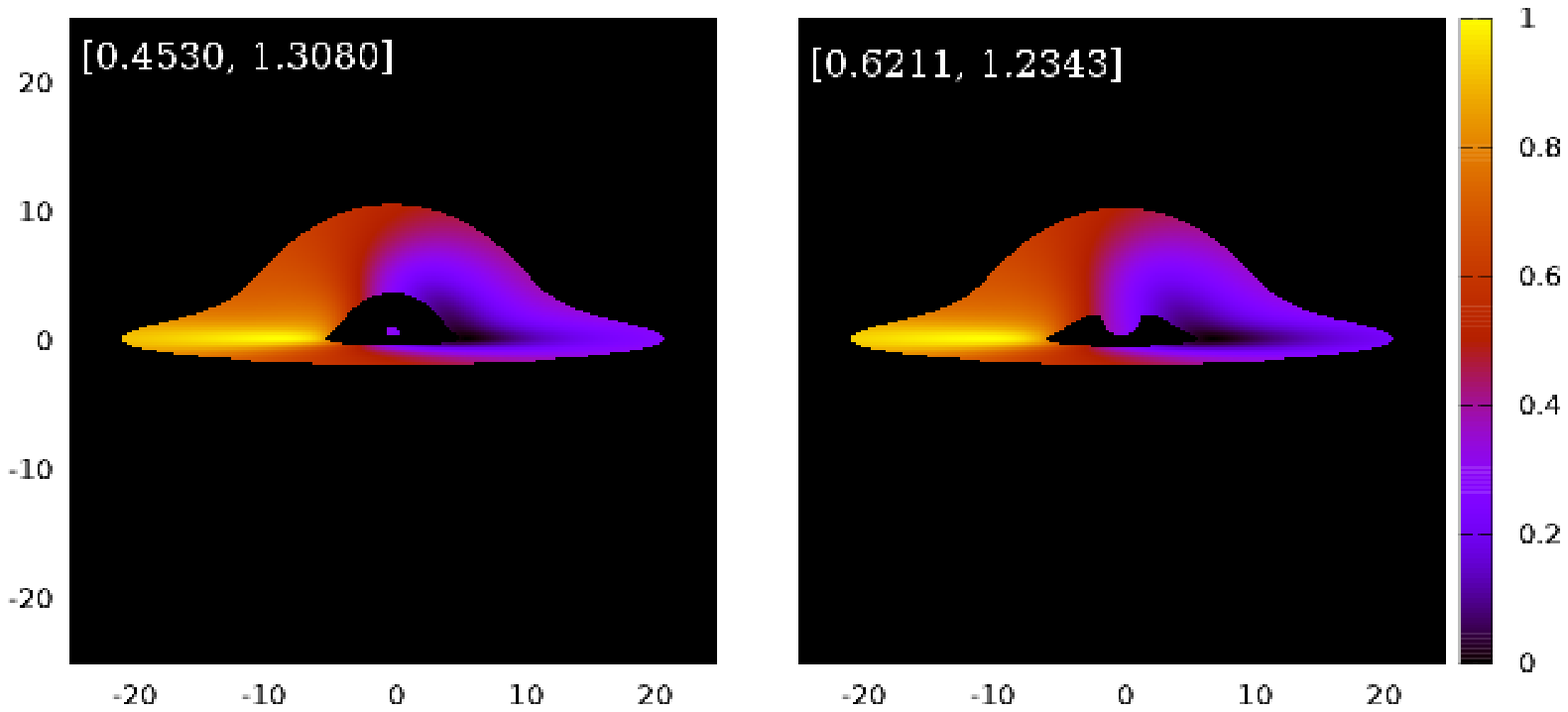}
		\end{tabular}
	\caption{Map of the frequency shift in the Keplerian disc direct images for the charge parameter $g/m=0.8$, $1.1$, $1.5$, and $2.5$ (from top left to right bottom). The observer inclination is set to $\theta_o=85^\circ$. Direct ghost images occur for the charge large enough -- $g/m=1.5, 2.5$.}
	\end{center}
\end{figure}

\begin{figure}[H]
	\begin{center}
\begin{tabular}{c}
			\includegraphics[scale=0.75]{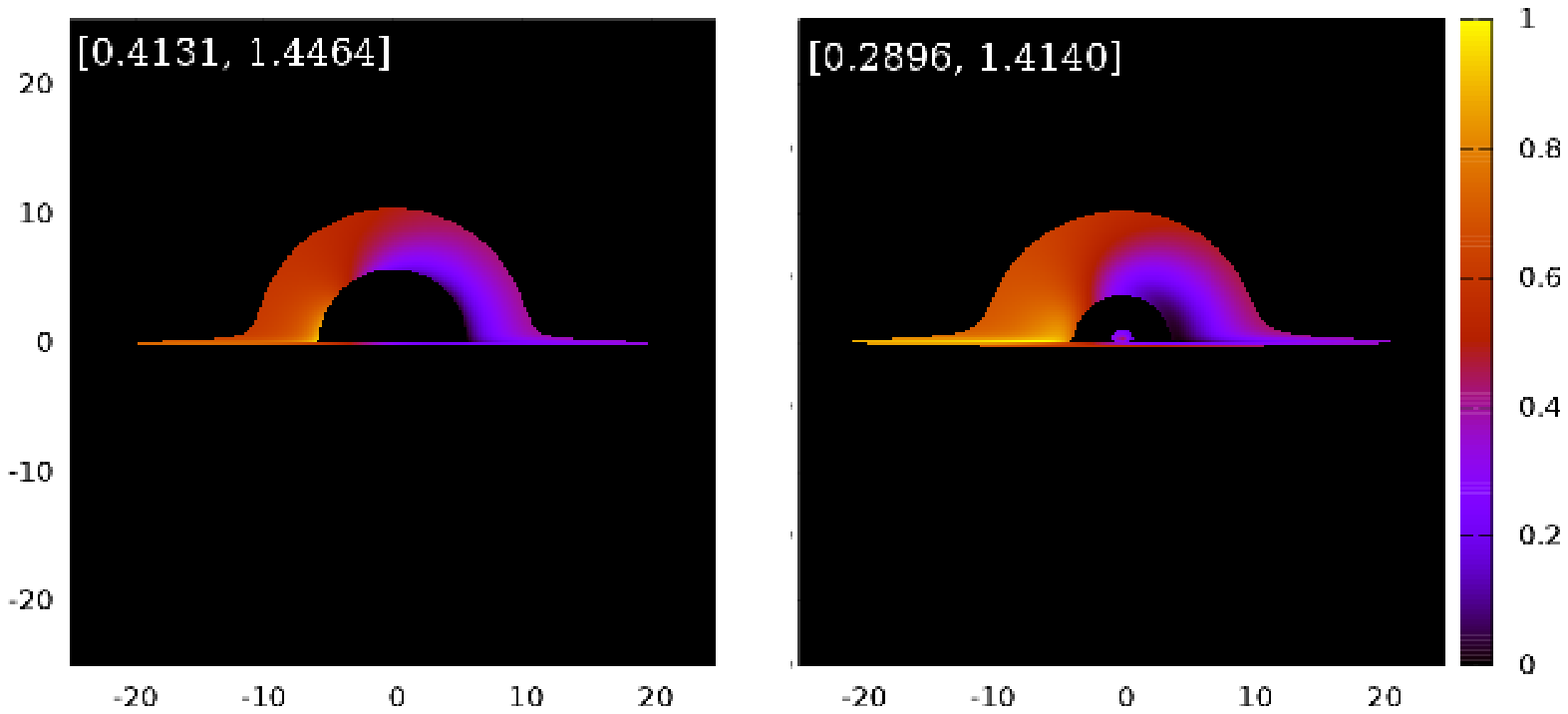}\\
			\includegraphics[scale=0.75]{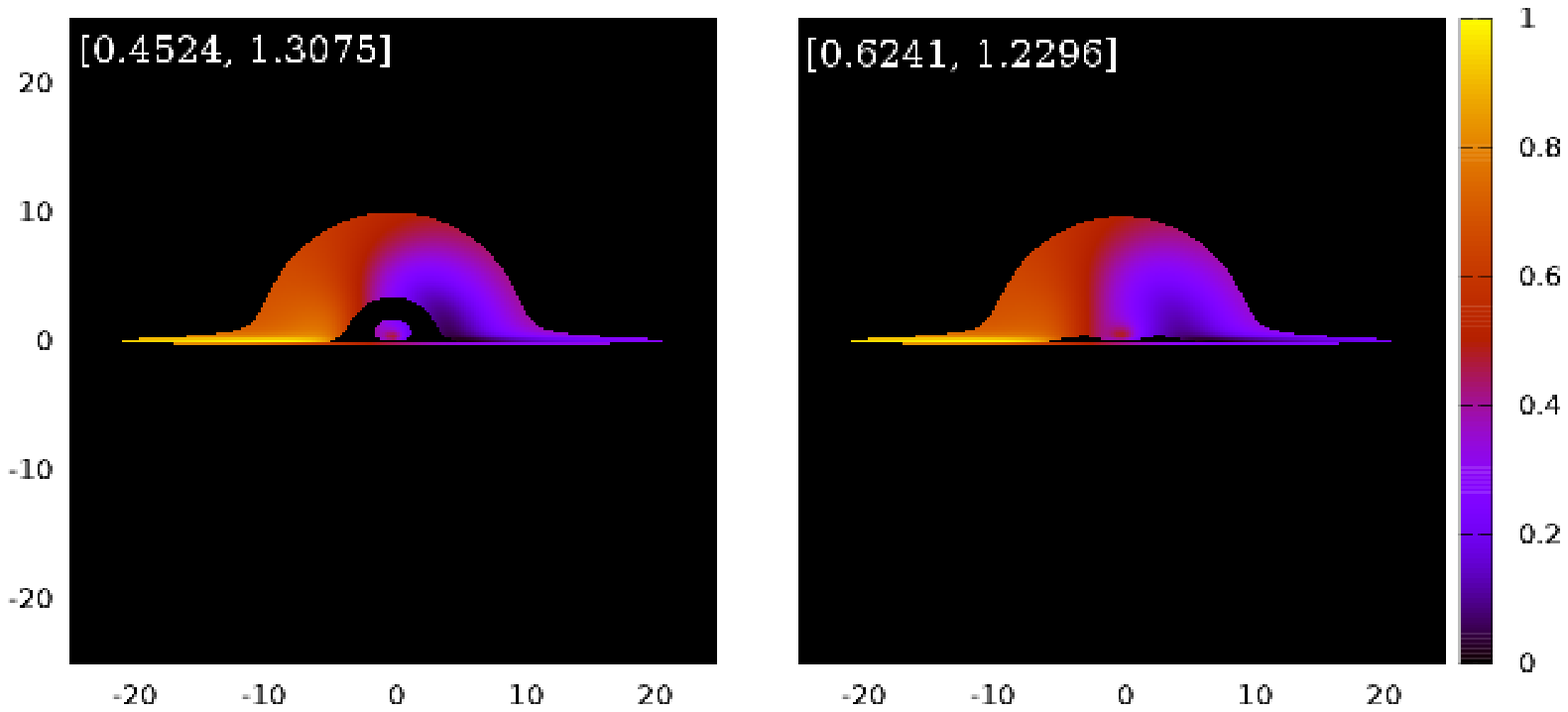}
		\end{tabular}
	\caption{Map of the frequency shift in the Keplerian disc direct images for the charge parameter $g/m=0.8$, $1.1$, $1.5$, and $2.5$ (from top left to right bottom). The observer inclination is set to $\theta_o=89^\circ$. Direct ghost images occur except the case of $g/m=0.8$.}
	\end{center}
\end{figure}

\section{Analysis of the direct ghost images} 

It is instructive and useful to discuss the relation of the ghost and standard direct images and the regions of the radiating Keplerian discs that are responsible for the creation of the ghost images. 

\subsection{Relation of the ghost and standard direct images}

From the images of the Keplerian discs presented in Figures 7-9 we can see that the ghost images are of two kinds. First, they are \emph{separated} from the standard direct image, second they are \emph{merged} with the direct image. We also see that the ghost images occur only for high inclination angles $\theta_o > 80^{o}$. In order to understand the origin of the ghost image phenomenon, we consider for simplicity only the motion in the $r - \theta$ plane when the photon trajectory is characterized by a single impact parameter $l$.

To understand the origin of two kinds of ghost images and the dependence of the ghost image presence and extension on the observer inclination angle, we construct numerically the function for the radial coordinate of the radiating point of the Keplerian disc $u_e(l,\theta_o;g)$ corresponding to the photon with the impact parameter $l$ reaching the distant observer. This function is crucial for the analysis of the ghost and direct images and their relation and is implicitly given by the equation
\begin{equation}
	\pi = \psi_0 + \int_{u_o}^{u_t}\frac{l\diff u}{\sqrt{1-f(1/u;g)l^2 u^2}}+\int_{u_e(l;g)}^{u_t}\frac{l\diff u}{\sqrt{1-f(1/u;g)l^2 u^2}}
\end{equation} 
with $\psi_0=\pi/2-\theta_o$. This function gives the radial position of the emitter, $r_e = 1/u_e$, for a given spacetime parameter $g$, photon impact parameter $l$, and position of the observer $(r_o = 1/u_o,\theta_o)$. The resulting functions $u_e(l,\theta_o;g)$, and the corresponding images of the innermost parts of the Keplerian discs are presented in Figures 10-12. The corresponding frequency shift ranges and values of the impact parameters $l_{min}$ and $l_{max}$ related to the local extrema of the function $u_{e}(l;g,\theta_{o})$ and the values $l_0, l_1, l_2$ of the solutions of the equation $u_{\Omega} = u_{e}(l;g,\theta_{o})$, where $u_{\Omega} = 1/r_{\Omega}$ represent the inner edge of the Keplerian disc, are presented in Tables 1-2. 

Clearly, from Figures 10-12 we see that the position function $u_{e}(l,\theta_o;g)$ can have one local maximum and one local minimum, and allows then, for the properly chosen values of the starting position $r_e = 1/u_e$, even three values of the impact parameter corresponding to photons reaching the distant observer, if $r_e > r_{in}$.  For decreasing value of the observer inclination angle $\theta_o$ both  extrema, $u_{e max}$ and $u_{e min}$, of the position function $u_e(l,\theta_o;g)$ are increasing. Denoting by $u_{in}=1/r_{in}=1/r_{\Omega}$ the inner edge of the disc, we can see in Figure 9 that for $\theta_o=80^\circ$ there is $u_{in} < Min(u_{e min},u_{e max})$ and there is only one intersection of the line $u_{in}$ with the function $u_e(l,\theta=80^o;g)$. Then only the standard direct image is generated. For the inclination angles $\theta_o=85^\circ$ and $\theta_o=89^\circ$ there is $u_{in} > u_{e min}(\theta_o;g)$ and there are at least two intersections of the line $u_{in}$ with the function $u_e(l,\theta_o;g)$ implying arising of the ghost images (in Figures 9-11 represented by black regions).

We can see that for $u = u_{e min}(l_{min},\theta_o;g)$ one photon with $l=l_0=l_1=l_{min}$ creates the ghost image of the point, and the photon with impact parameter being the other solution of the equation $u = u_{e min}(l_{min},\theta_o;g) = u_{e}(l,\theta_o;g)$, $l=l_2$, creates the direct image of this point. For a point with a radial coordinate $u>u_{in}$, there are three solutions of the equation $u = u_{e}(l,\theta_o;g)$, denoted as $l_0 < l_1 < l_2$, and the photons with $l=l_0$ and $l=l_1$ create two ghost images of the source point, while the photon with $l=l_2$ creates one point of the direct image of the source. This procedure works up to the radius for which there is $u = u_{e max}(l_{max},\theta_o;g)$ when the photon with $l=l_1=l_2=l_{max}$ creates a common point of the ghost image and the direct image, while the photon with $l=l_0$, where $l_0$ is the second, inner solution of the equation $u = u_{e max}(l_{max},\theta_o;g) = u_{e}(l,\theta_o;g)$, creates the second ghost image of the source point. For $u>u_{e max}$ there si only one solution of the equation $u = u_{e}(l,\theta_o;g)$ for the impact parameter, $l=l_0$, representing the coalescing direct and ghost image. Clearly, the occurrence of the ghost image, and its coalescence with the direct image is determined by the relation of the inner edge of the Keplerian disc $u_{in}$ and the local extrema of the function $u_{e}(l,\theta_o;g)$. For $u_{in} > u_{e min}$ the ghost images occur, and for $u_{in} > u_{e max}$ the ghost image coalesces with the direct image. We can convince ourselves in Figure 12 that the situation when the ghost image is merged with the standard direct image occurs when the condition $u_{in} \ge  u_{e max}$ is fulfilled.  

Behavior of the frequency shift range of the direct and ghost images is relatively complex for the range of the large inclination angles allowing for existence of the ghost images in the Bardeen no-horizon spacetimes as demonstrated in Table 1. Concerning the maximal frequency shift $(1+z)_{max}$ of the direct image, its dependence on the inclination angle is the same for all the considered spacetimes, having a maximum for the inclination angle $\theta_{o} \sim 85^{o}$, while the minimal frequency shift $(1+z)_{min}$ of the direct image decreases with increasing inclination angle for $g/m = 1.5$, but it has a minimum at $\theta_{o} \sim 85^{o}$ for large charge parameters $g/m = 2$ and $g/m = 2.5$. For the ghost images, we observe a substantial reduction of the frequency shift range at both edges: $(1+z)_{maxG} < (1+z)_{max}$ and $(1+z)_{minG} > (1+z)_{min}$. The range of the frequency shift of the ghost images always increases with increasing inclination angle of the observer in a given spacetime -- the minima (maxima) are decreasing (increasing) with increasing inclination angle $\theta_{o}$. 


\begin{figure}[H]
	\begin{center}
	\begin{tabular}{cc}
		\includegraphics[scale=0.6]{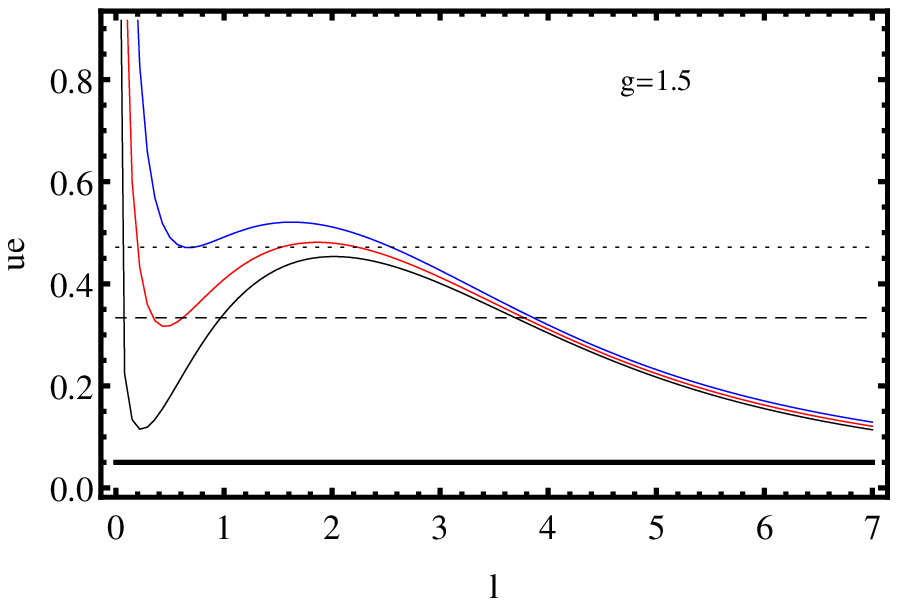} & \includegraphics[scale=0.5]{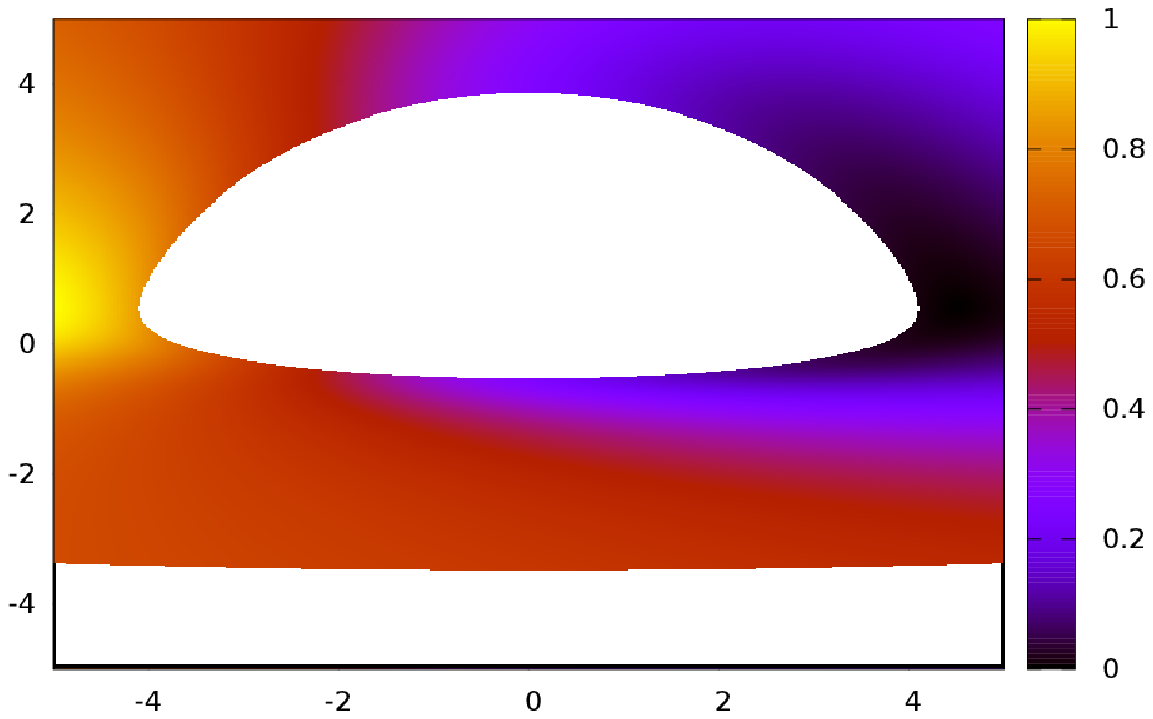}\\
		\includegraphics[scale=0.5]{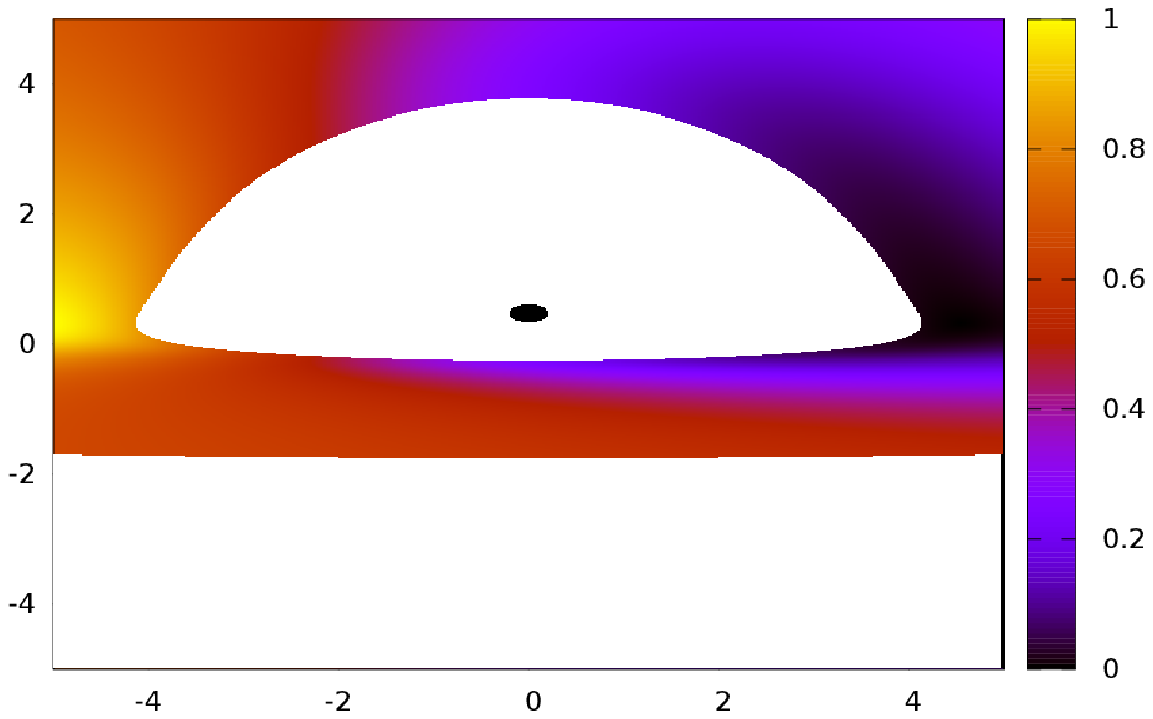} & \includegraphics[scale=0.5]{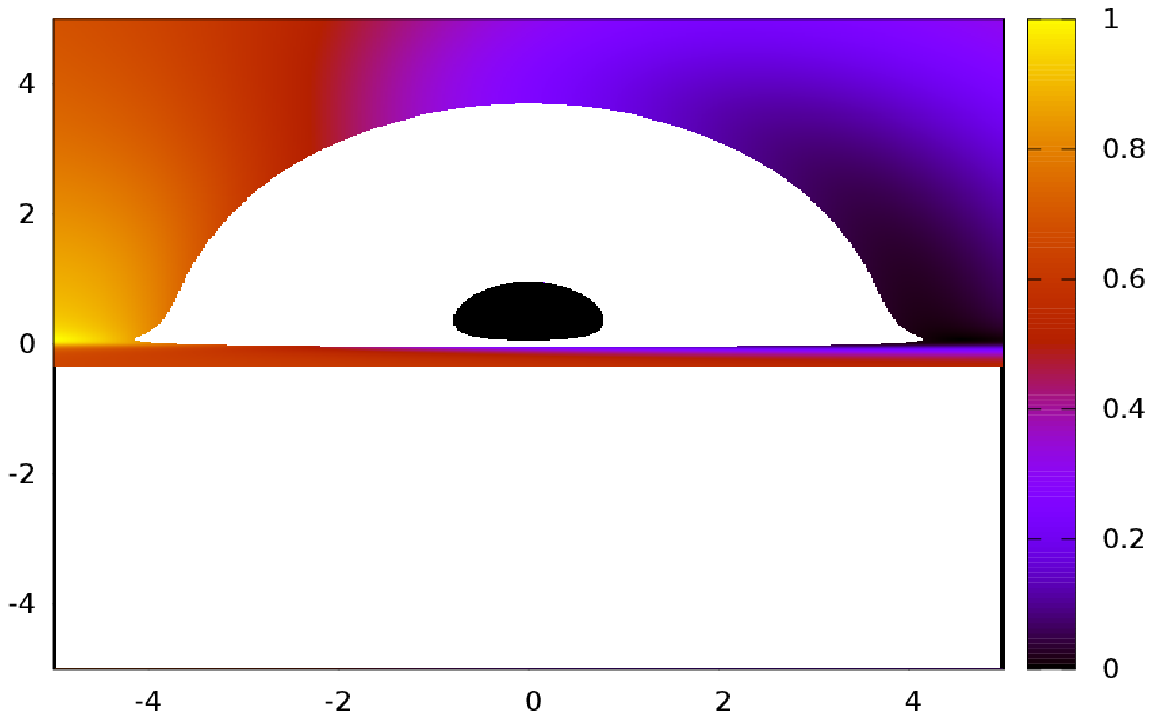}
	\end{tabular} 
	\caption{The position function $u_e=u_e(l,\theta_o;g)$ (left top) and the direct images of the innermost regions of the Keplerian discs are given for three representative observer inclination angles $\theta_o=80^\circ$ (blue), $85^\circ$ (red), and $89^\circ$ (black). For these three values of the inclination angle the direct images are constructed, including the ghost image in the cases when the image exists ( $85^\circ$ and $89^\circ$). The charge parameter is chosen to be $g/m=1.5$. The black region represents the ghost image part (for $\alpha=0$ the minimal and maximal values of $\beta$ correasponding to ghost image from particular $r_e$ are $\beta_{min}=l_0$ and $\beta_{max}=l_2$). There are three characteristic radii in the figure: $u_e=u_{stat}=1/r_{stat}$ (dotted), $u_e=u_{\Omega}=1/r_{\Omega}$ (dashed), and $u_e=u_{out}=1/r_{out}=1/20$ (black, thick). }
	\end{center}
\end{figure}

\begin{figure}[H]
	\begin{center}
	\begin{tabular}{cc}
		\includegraphics[scale=0.5]{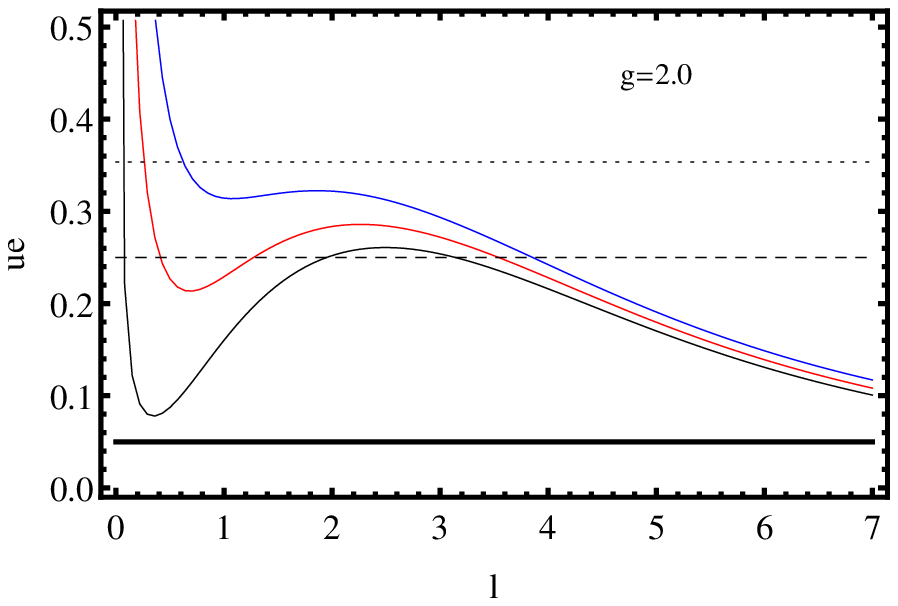} & \includegraphics[scale=0.3]{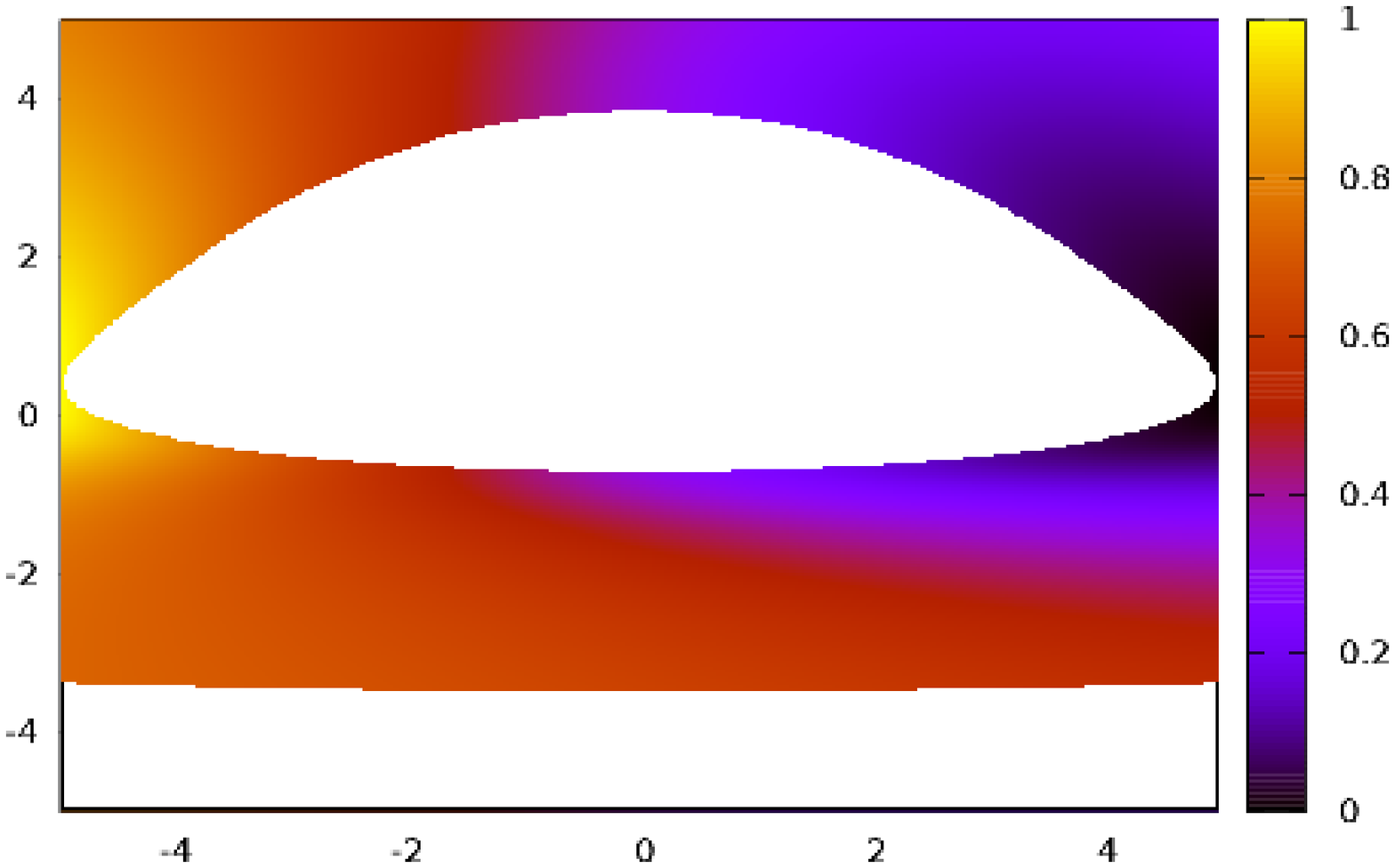}\\
		\includegraphics[scale=0.3]{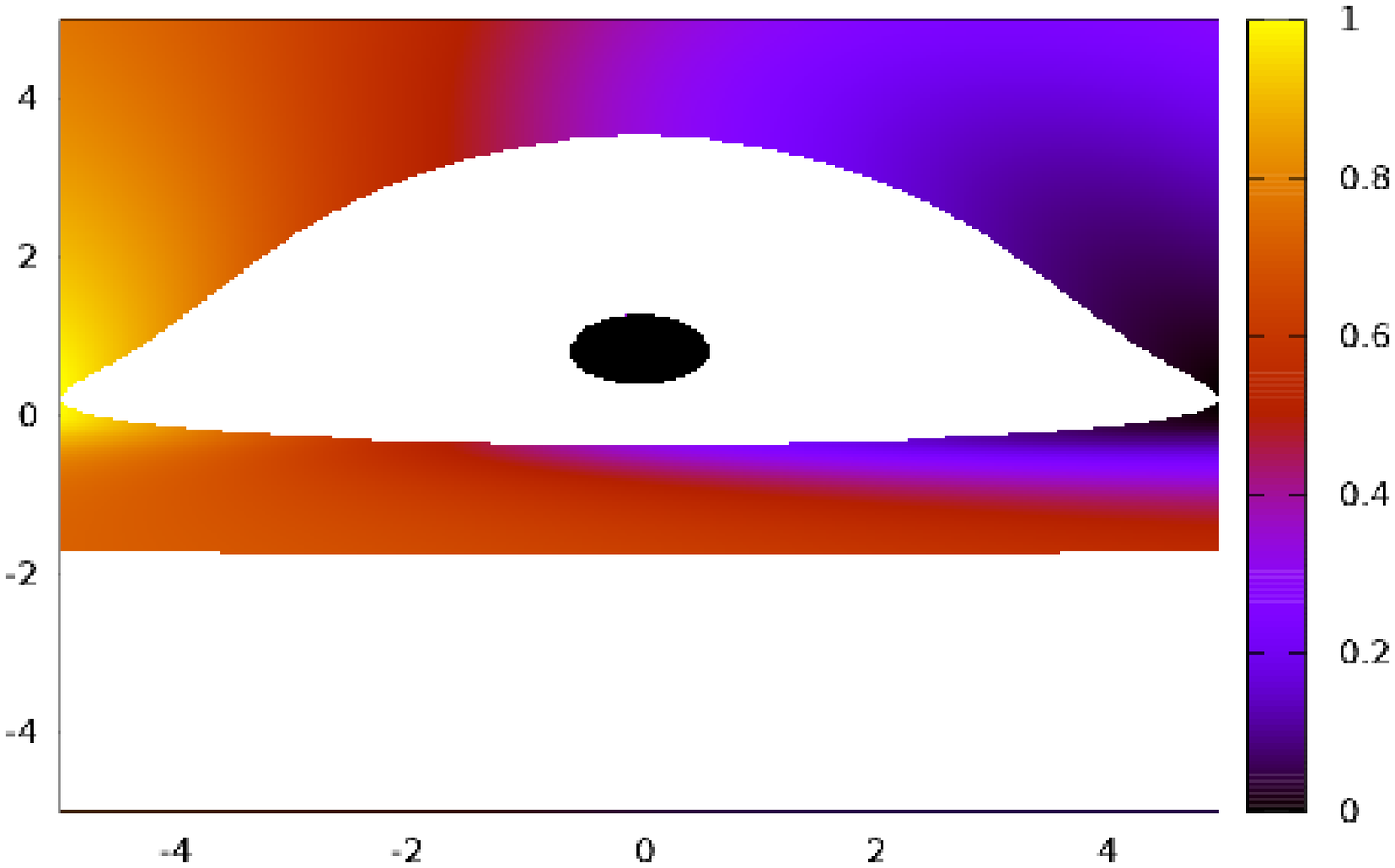} & \includegraphics[scale=0.5]{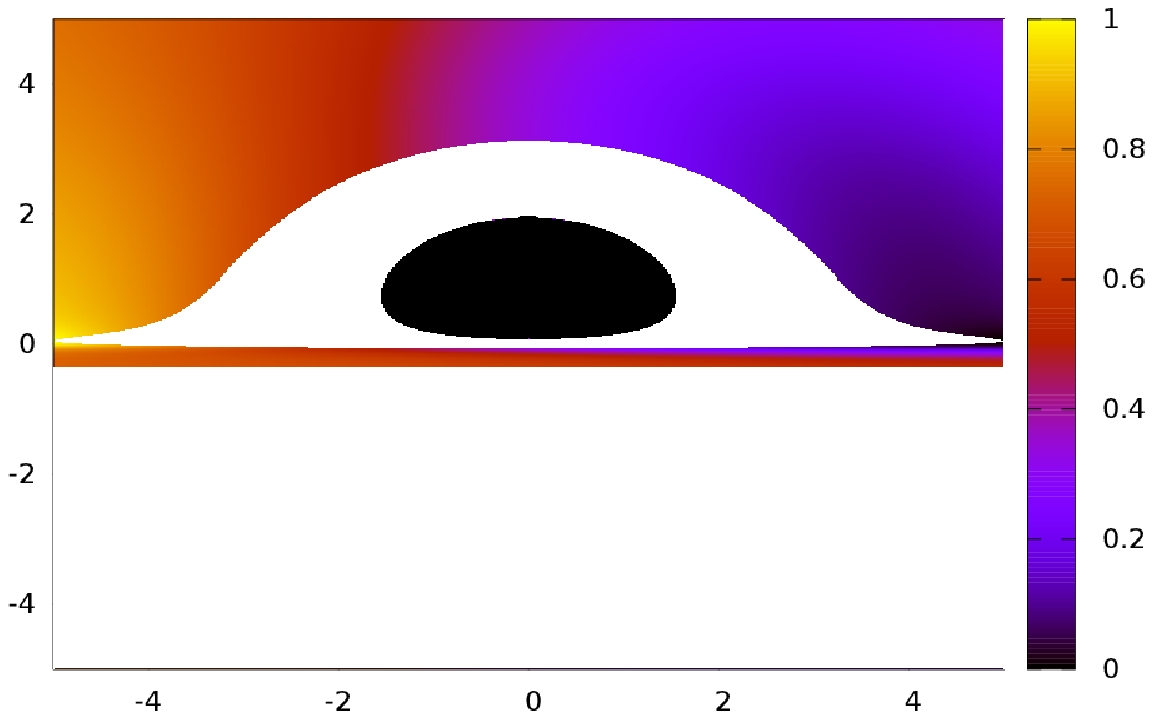}
	\end{tabular}
	\caption{The position function $u_e=u_e(l,\theta_o;g)$ (left top) and the direct images of the innermost regions of the Keplerian discs are given for three representative observer inclination angles $\theta_o=80^\circ$ (blue), $85^\circ$ (red), and $89^\circ$ (black). The charge parameter is chosen to be $g/m=2.0$. The black region represents the ghost image part (for $\alpha=0$ the minimal and maximal values of $\beta$ correasponding to ghost image from particular $r_e$ are $\beta_{min}=l_0$ and $\beta_{max}=l_2$). There are three characteristic radii in the figure: $u_e=u_{stat}=1/r_{stat}$ (dotted), $u_e=u_{\Omega max}=1/r_{\Omega max}$ (dashed), and $u_e=u_{out}=1/r_{out}=1/20$ (black, thick).}
	\end{center}
\end{figure}
		
\begin{figure}[H]
	\begin{center}
	\begin{tabular}{cc}
		\includegraphics[scale=0.5]{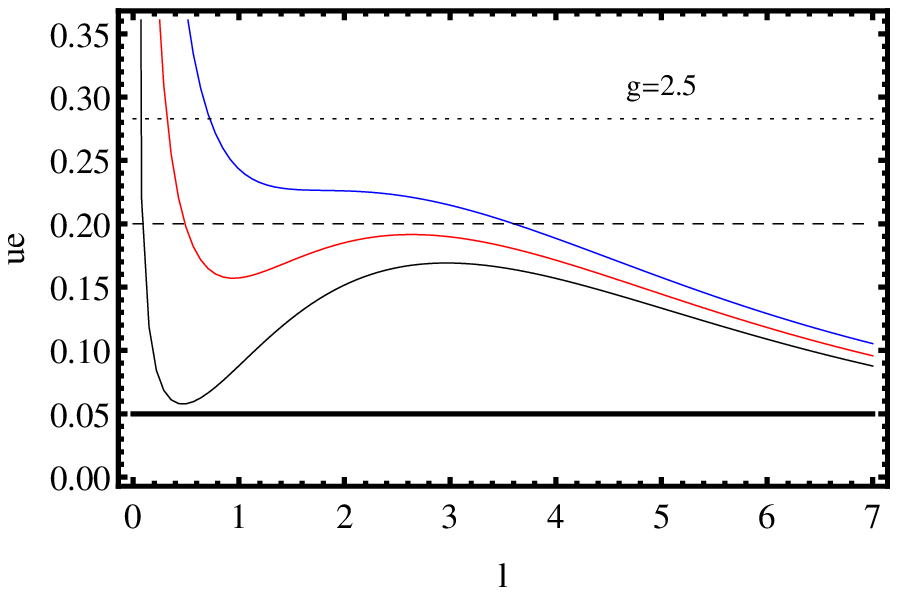} & \includegraphics[scale=0.3]{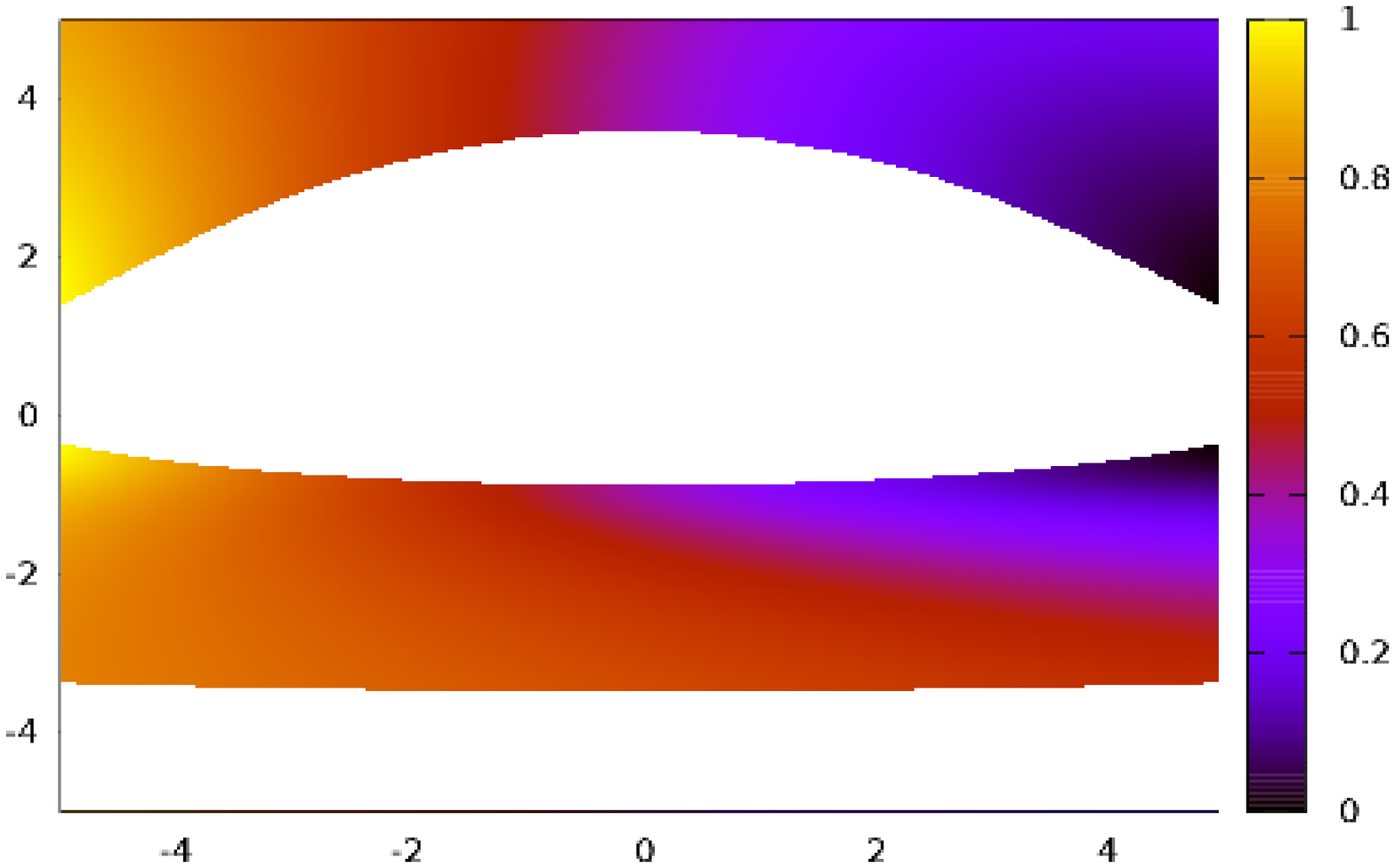}\\
		\includegraphics[scale=0.25]{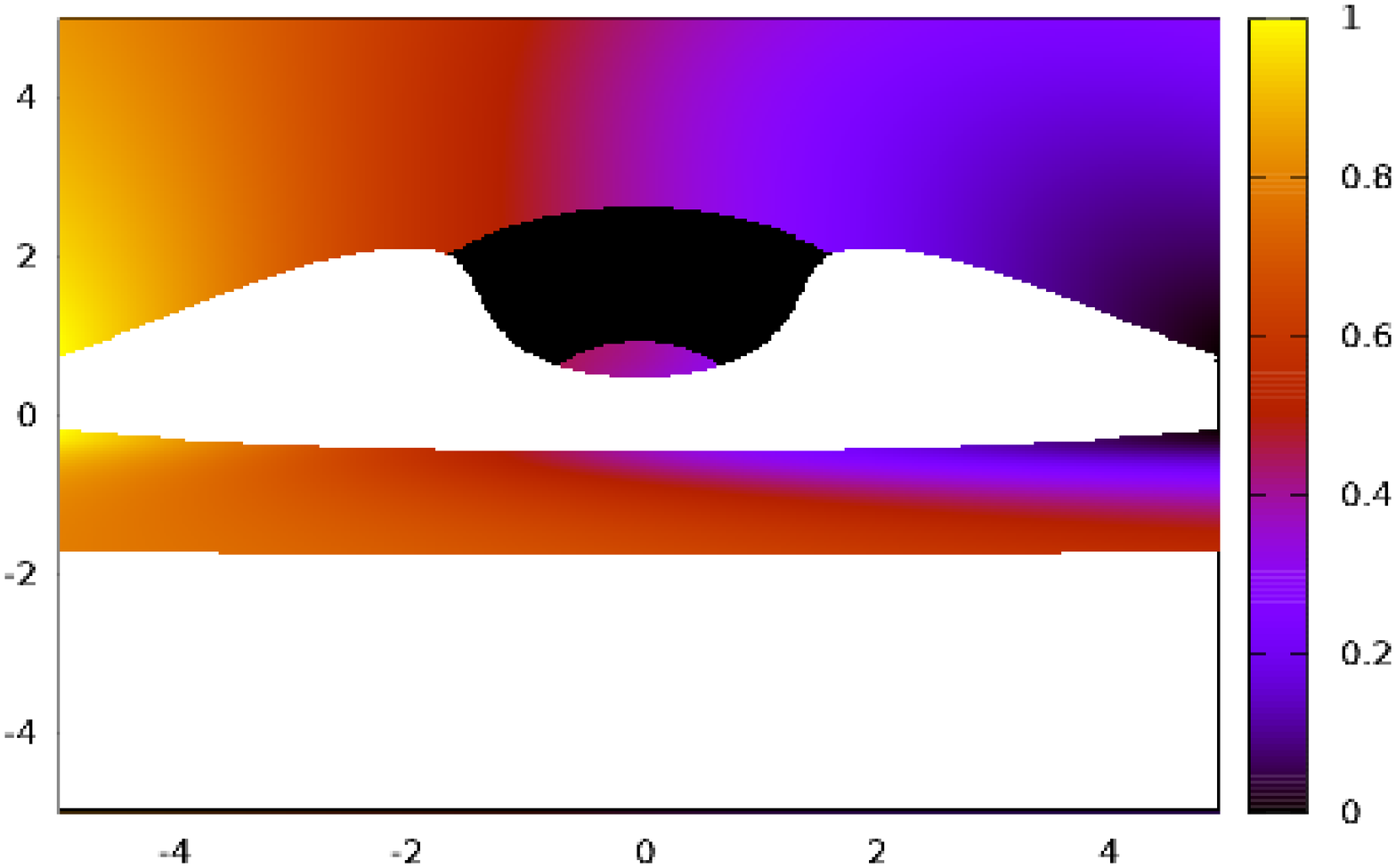} & \includegraphics[scale=0.3]{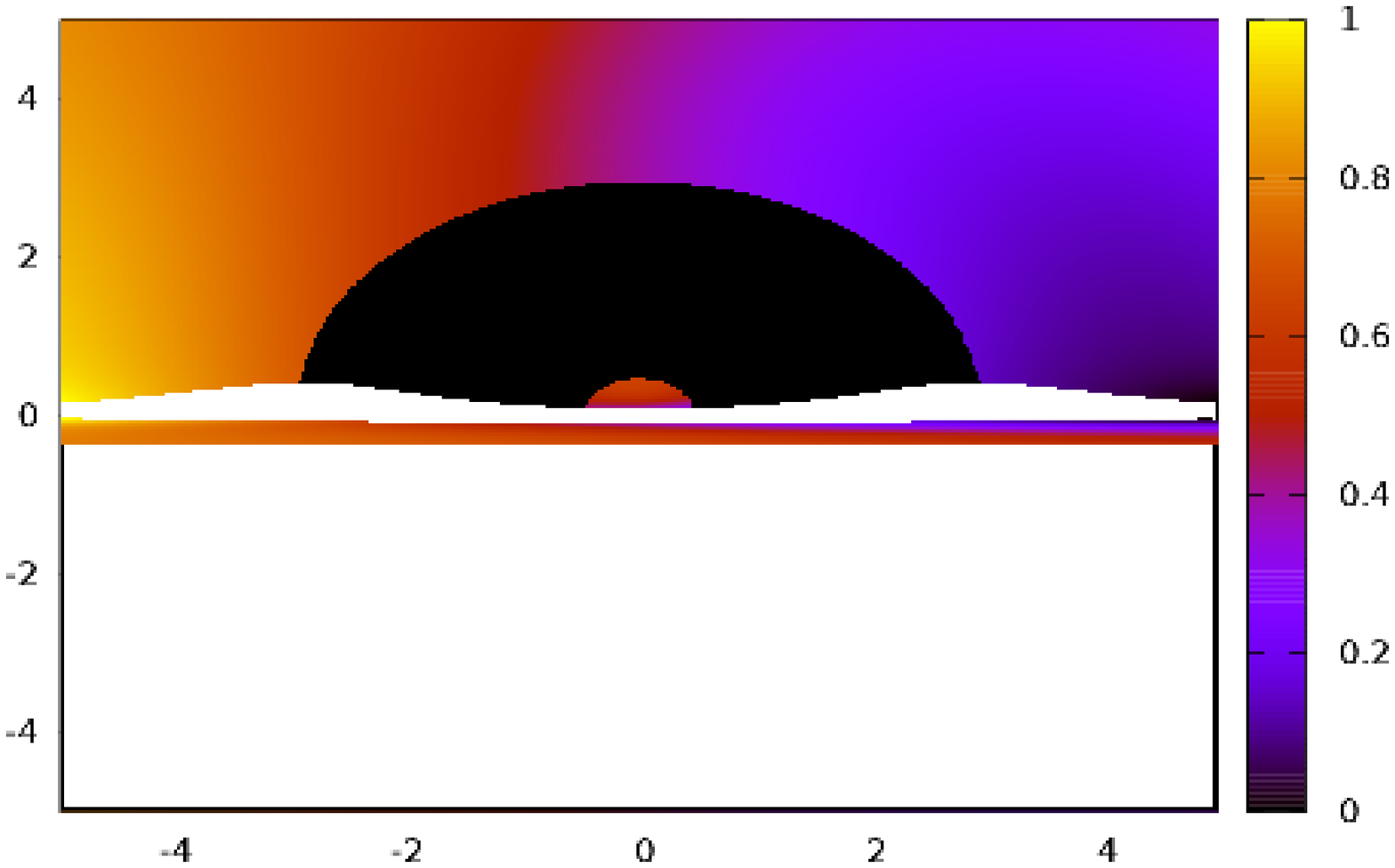}
	\end{tabular}
	\caption{The position function $u_e=u_e(l,\theta_o;g)$ (left top) and the direct images of the innermost regions of the Keplerian discs are constructed for three representative observer inclination angles $\theta_o=80^\circ$ (blue), $85^\circ$ (red), and $89^\circ$ (black). The charge parameter is chosen to be $g/m=2.5$. The black region represents the ghost image part joined to the standard direct image. There are three characteristic radii in the figure: $u_e=u_{stat}=1/r_{stat}$ (dotted), $u_e=u_{\Omega max}=1/r_{\Omega max}$ (dashed), and $u_e=u_{out}=1/r_{out}=1/20$ (black, thick). From the plot of the position function $u_e(l,\theta_{o};g)$ one can see that the coloured area inside of the black region corresponds to the primary image constructed for $l<l_{max}$ (see Table 2) and for $r_e < r(l_{max})$.}
	\end{center}
\end{figure}

\begin{table}[H]
	\begin{center}
	\caption{Frequency shift range of the whole primary image and of the ghost primary image alone.}
	\begin{tabular}{|c|c|cc|cc|}
	\hline
	$g/m$ & $\theta_o$ & $(1+z)_{min}$ & $(1+z)_{max}$ & $(1+z)_{minG}$ & $(1+z)_{maxG}$\\
	\hline 
	$1.5$ & $80^\circ$ & $0.4556$ & $1.2949$ &-&-\\
	      & $85^\circ$ & $0.4531$ & $1.3084$ &-&-\\
	      & $89^\circ$ & $0.4528$ & $1.3009$ &$0.6119$& $0.8140$\\
	\hline
	$2.0$ & $80^\circ$ & $0.5599$ & $1.2534$ & - & - \\
		  & $85^\circ$ & $0.5579$ & $1.2616$ & $0.7467$ & $0.7682$\\
		  & $89^\circ$ & $0.5614$ & $1.2499$ & $0.6881$ & $0.8812$\\
	\hline
	$2.5$ & $80^\circ$ & $0.6228$ & $1.2281$ & - & - \\
		  & $85^\circ$ & $0.6210$ & $1.2346$ & $0.7961$ & $0.8308$\\
		  & $89^\circ$ & $0.6250$ & $1.2281$ & $0.7318$ & $0.9143$\\
	\hline
	\end{tabular}
	\end{center}
\end{table}

\begin{table}[H]
\begin{center}
	\caption{The impact parameters $l_{min}$ and $l_{max}$, of the local minimum and maximum of the position function $u_e=u_e(l,\theta_{o})$ and the intersections, $l_0$, $l_1$, and $l_2$, of the line $u_e=u_{\Omega}$ with the position function $u_e=u_e(l;\theta_{o})$.}
	\begin{tabular}{|c|c|cc|ccc|}
	\hline
	$g/m$ & $\theta_o$ & $l_{min}$ & $l_{max}$ & $l_0$ & $l_1$ & $l_2$\\
	\hline
	$1.5$ & $80^\circ$ & $0.6754$ & $1.6247$ & - & - & $3.8626$\\
	      & $85^\circ$ & $0.4558$ & $1.8624$ & $0.3473$ & $0.6182$ & $3.7760$\\
	      & $89^\circ$ & $0.2434$ & $2.0175$ & $0.0681$ & $0.9688$ & $3.6953$\\
	\hline
	$2.0$ & $80^\circ$ & $1.0782$ & $1.8592$ & - & - & $3.8559$\\
		  & $85^\circ$ & $0.6841$ & $2.2630$ & $0.4065$ & $1.2727$ & $3.5413$\\
		  & $89^\circ$ & $0.3650$ & $2.4931$ & $0.0762$ & $1.9488$ & $3.1405$\\
	\hline
	$2.5$ & $80^\circ$ & - & - & - & - & $3.6038$\\
		  & $85^\circ$ & $0.9464$ & $2.6345$ & $0.4903$ & - & - \\
		  & $89^\circ$ & $0.4741$ & $2.9653$ & $0.08256$ & - & -\\
	\hline
	\end{tabular}
\end{center}
\end{table}

\subsection{Region of the disc corresponding to the direct ghost images} 

Finally, we determine the regions of the Keplerian discs that could create the ghost images for large inclination angle of the distant observers. Now we have to use the complex situation where both the impact parameters $l,q$ are taken into account, and we have to limit the radiating part of the Keplerian discs in both the radial $r$ and axial $\phi$ coordinates. Therefore, we have to use the trajectory functions in full complexity. We also give for completeness the distribution map of the frequency shift across the Keplerian disc region creating the ghost image. 

The results of the numerical calculations are presented in Figure 13. We can see that the ghostly imaged region increases with increasing inclination angle of the observer and increasing charge parameter of the spacetime. For the Bardeen no-horizon spacetimes admitting for the existence of the circular photon geodesics no ghost images were obtained for the Keplerian discs, since in such spacetimes the ISCO is located above the region where the ghost imaging can work. Nevertheless, ghost images could occur for radiating region located at $r < r_{ISCO}$, e.g., in the region of unstable circular geodesics, or in the region of the stable circular geodesics located close to the static radius. 

\begin{figure}[H]
	\begin{center}
	\begin{tabular}{c}
			\includegraphics[scale=0.7]{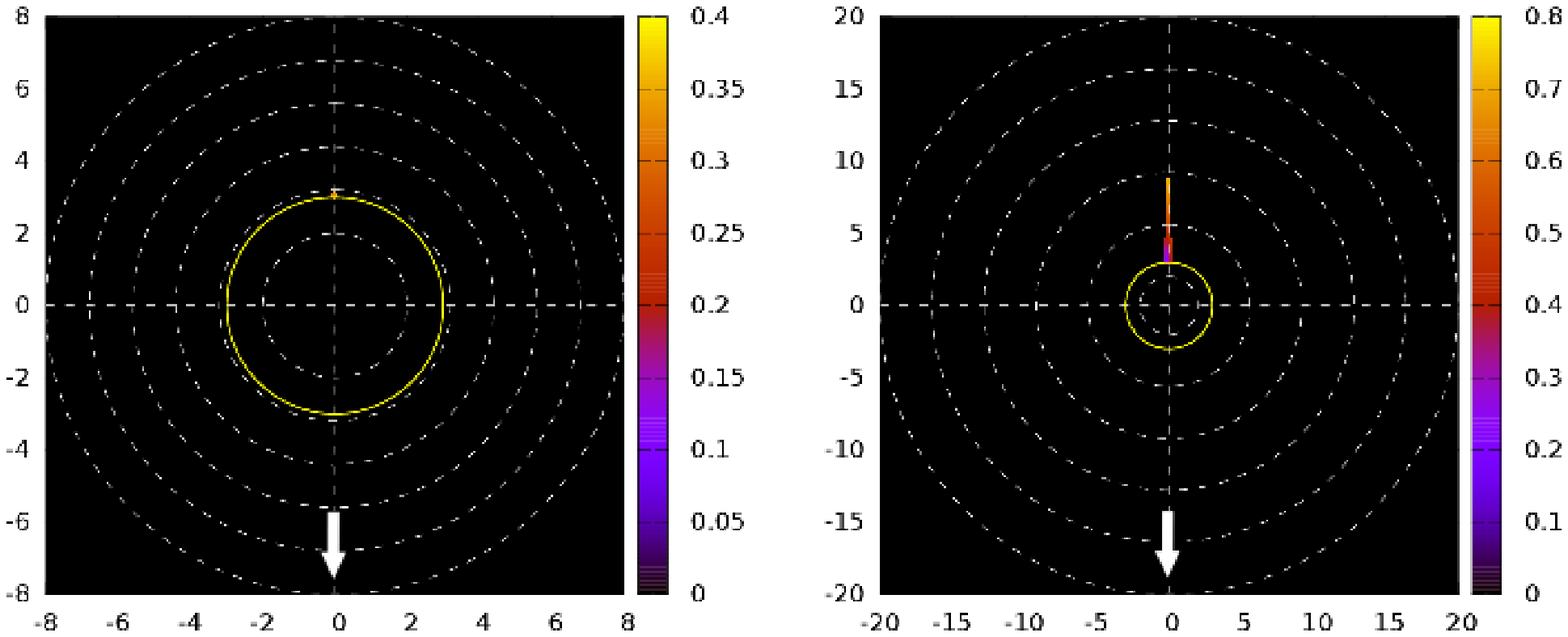}\\
			\includegraphics[scale=0.7]{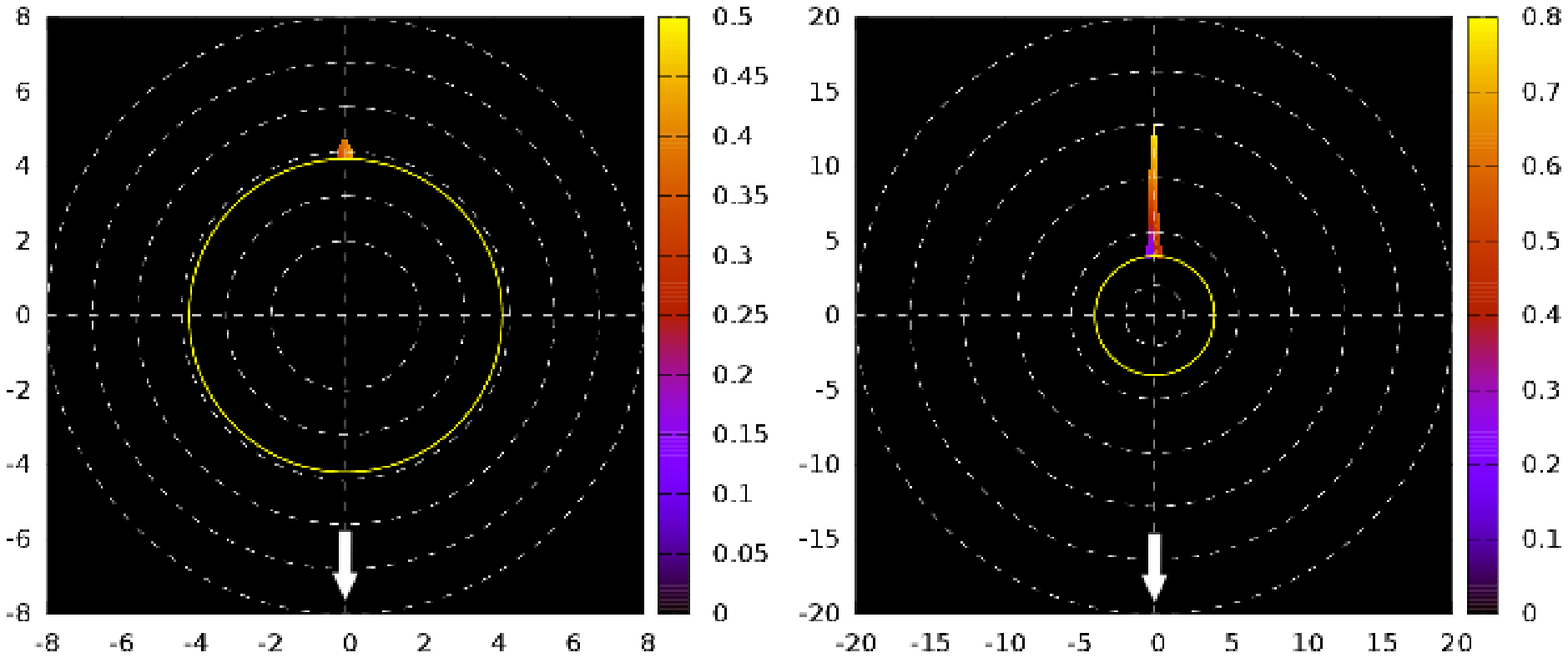}\\
			\includegraphics[scale=0.7]{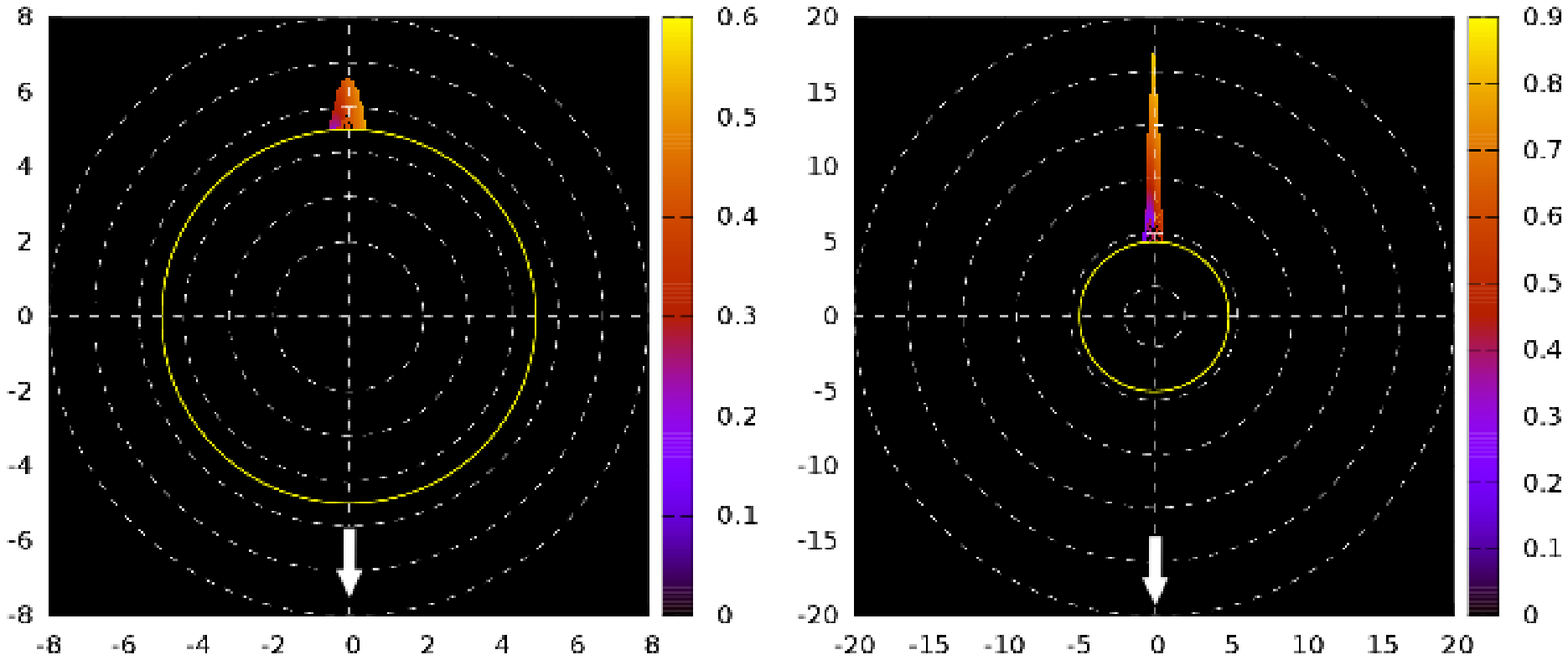}
	\end{tabular}
		\caption{Loci of the disk segment where the ghost image originates are constructed for the three representative values of the charge parameter $g/m=1.5$, $2.0$, and $2.5$. The yellow ring represents the inner edge of the disk at $r=r_{\Omega}$. The observer inclination is set to $\theta_o=85^\circ$ (left) and $89^\circ$ (right). White arrow shows the direction towards the distant observer.}
	\end{center}
\end{figure}

\begin{table}[H]
\begin{center}
	\caption{The values of azimuthal, $\Delta\phi$, and radial, $\Delta r$, width of the disk segment generating ghost image for three representative values of parameter $g/m=1.5$, $2.0$, and $2.5$ and two representative values of the observer inclination angle $\theta_o=85^\circ$, and $89^\circ$. }
	\begin{tabular}{|c|c|c|c|}
	\hline
	$\theta_o$ & $g/m$ & $\Delta\phi$ & $\Delta r$\\
	\hline
	$85^\circ$ & $1.5$ & $0.078$ & $0.175$\\
	& $2.0$ & $0.151$ & $0.702$\\
	& $2.5$ & $0.206$ & $1.393$\\
	\hline
	$89^\circ$ & $1.5$ & $0.178$ & $5.8$\\
			   & $2.0$ & $0.226$ & $8.97$\\
			   & $2.5$ & $0.268$ & $12.61$\\
	\hline
	\end{tabular}
\end{center}
\end{table}

\section{Indirect images in the Bardeen no-horizon spacetimes}

Now we study creation and properties of the secondary, indirect images and related indirect ghost images that are created by photons crossing once the equatorial plane. 

\subsection{Construction of the indirect images}

The indirect (2nd order) images are formed by photons satisfying the following conditions
\begin{eqnarray}
	2\pi&=&\psi_0+\int^{u_t}_{u_e}\frac{l \diff u}{\sqrt{U}} +\int^{u_t}_{u_o}\frac{l \diff u}{\sqrt{U}}\textrm{ for }l>0,\\
	-\pi&=&\psi_0-\left|\int^{u_t}_{u_e}\frac{l \diff u}{\sqrt{U}} +\int^{u_t}_{u_o}\frac{l \diff u}{\sqrt{U}}\right|\textrm{ for }l<0
\end{eqnarray} 
where the initial angle is $\psi_0=\pi/2-\theta_o$ with $\theta_o$ being the observer inclination.
Construction of the indirect images given by photons with the impact parameter $l>0$ and $l<0$ is illustrated in Figure 14. 
\begin{figure}[H]
	\begin{center}
		\includegraphics[scale=0.8]{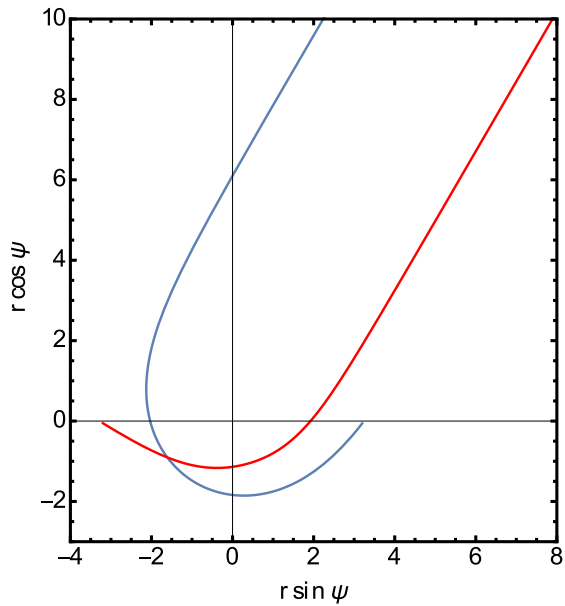}
		\caption{The light rays creating the secondary images. Here they are constructed for the observer inclination $\theta_o=30^\circ$.}
	\end{center}
\end{figure}

During the construction of the Keplerian disc images we have encountered the moment when, for particular values of $g$ and $\theta_o$, the secondary image "vanishes". The reason is clear - for a given value of the charge parameter $g$, the $\phi$ coordinate of photon trajectory does not reach neither $2\pi$ (condition 1 for $l>0$), or $-\pi$ (condition 2 for $l<0$). To describe this phenomenon for both the conditions, we have constructed in Figure 15 a series of curves $\phi=\phi(l;\theta_o,g)$ to see the moment when the deflection angle $\phi$ fails to satisfy the above conditions. 

 \begin{figure}[H]
 	\begin{center}
 		\begin{tabular}{cc}
 			\includegraphics[scale=0.6]{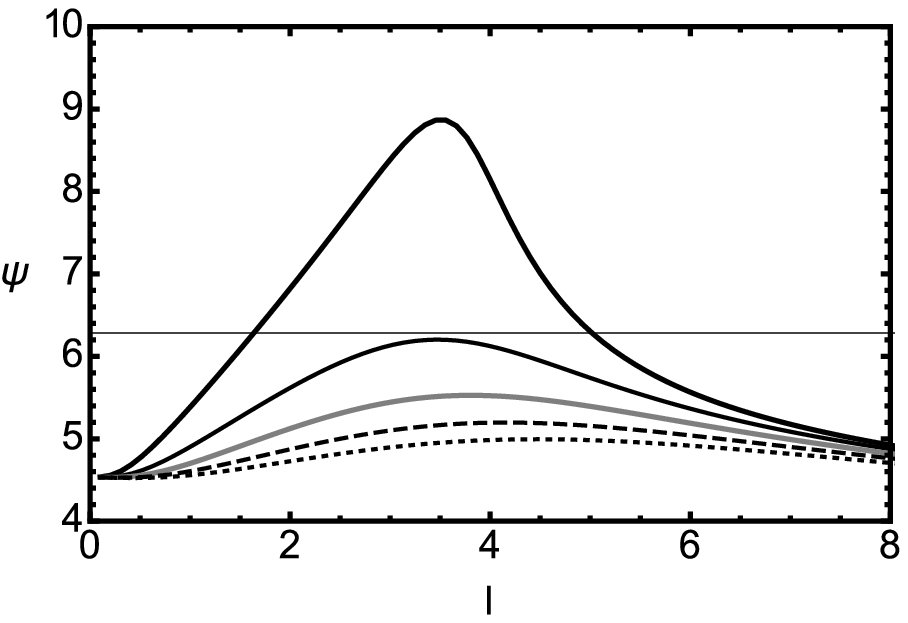}&\includegraphics[scale=0.1]{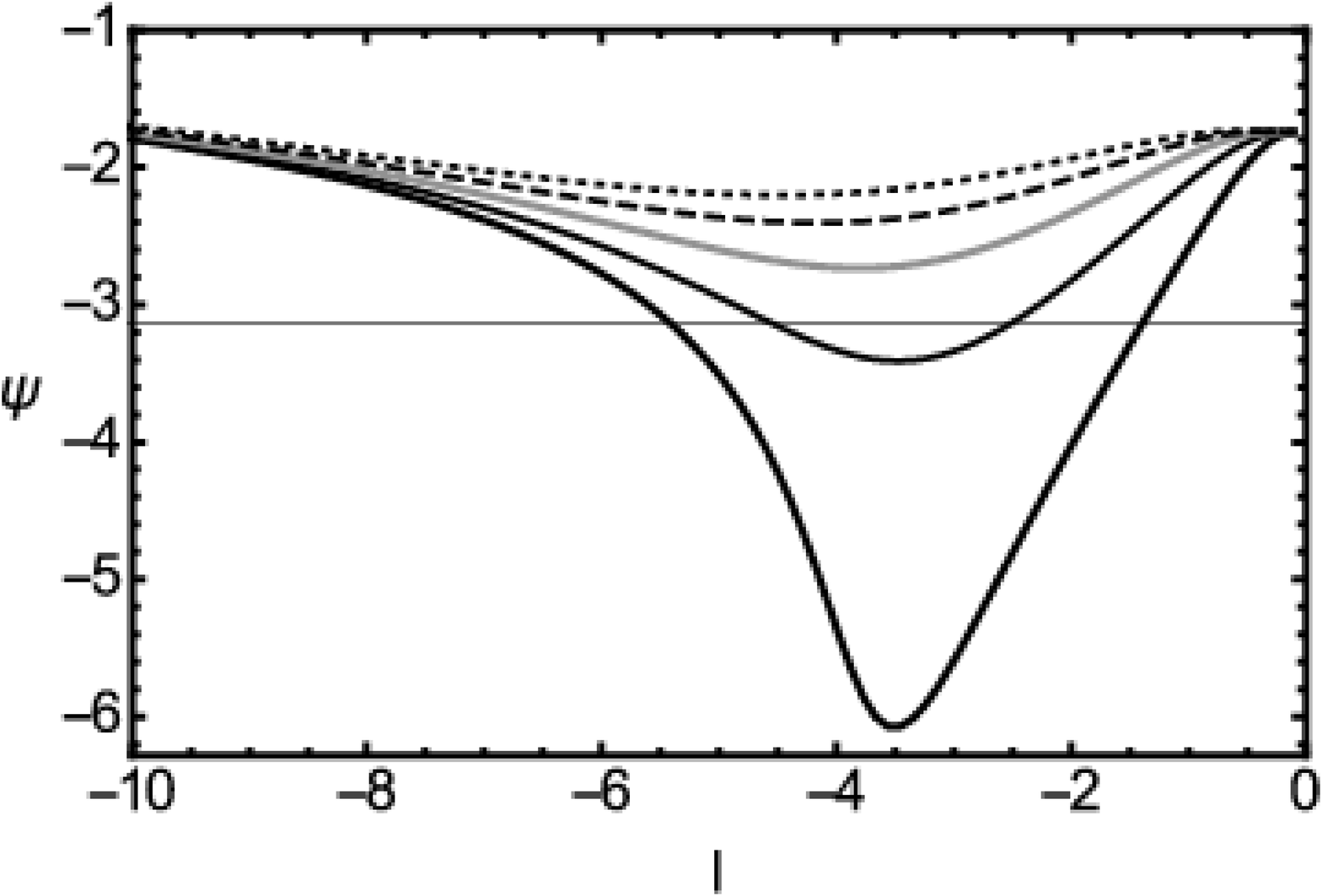}
 		\end{tabular}
 		\caption{Plots of the deflection function $\theta(l;u_e,g,\theta_o)$ constructed for five representative values of $g/m=1.0$ (black, thick), $1.375$ (black), $1.75$ (gray,thick), $2.125$ (dashed), $2.5$ (dotted). The emitter is fixed at radial coordinate $r_e=20$.}
 	\end{center}
 \end{figure}

The secondary image photons fail to satisfy the condition 1 first; then the image starts to be deformed and its "upper" part vanishes. Therefore, the secondary image vanishes totally when the condition 2 (related to photons with $l<0$) is not satisfied. By evaluating minima of the deflection function $\phi(l;g,u_e,\theta_o)$ we obtain function $\phi_{min}(g;u_e,\theta_o)$. The critical value of the spacetime parameter for vanishing of the indirect images, $g=g_c(\theta_o)$, is given implicitly by the equation 
\begin{equation}
	\psi_{min}(g_c;u_e,\theta_o)=-\pi.
\end{equation}

For a given radial coordinate of emitter $r_{e}=20m$ we calculated the corresponding values $(g_c,\theta_o)$ that are presented in the Table 4.

\begin{table}[H]
	\begin{center}
	\caption{Critical values of the charge parameter $g_{c}(\theta_{o})$ for vanishing of the indirect images. They are determined by the minima od the deflection function $\psi(l;u_e,g,\theta_o)$. The emitter is fixed at radial coordinate $r_e=20m$.}
	\begin{tabular}{|c|cccccccc|}
		\hline
		$\theta_o$ & $10^\circ$ & $20^\circ$ &$30^\circ$ & $40^\circ$ & $50^\circ$
		 & $60^\circ$ & $70^\circ$ & $80^\circ$\\
		\hline
		$g_c/m$ & $1.469$ & $1.577$ & $1.708$ & $1.859$ & $2.063$ & $2.358$ & $2.776$ & $3.419$\\
		\hline
	\end{tabular}
	\end{center}
\end{table}

We can see that the critical charge parameter for the Bardeen spacetimes, when the secondary images start to be forbidden due to small ray-deflection effect, strongly depends on the inclination angle of the observer. The critical charge increases with increasing inclination angle. Therefore, the deflection of light rays weakens with increasing Bardeen charge $g$, indicating suppression of the gravitational effects with increasing charge parameter. 

\subsection{Indirect images of Keplerian discs}

We have demonstrated that the direct images of Keplerian discs in the field of no-horizon objects contain a ghost image. Its structure is different from the one generated in naked singularity spacetimes (like those occuring in the RN naked singularity spacetime -- see next section) and can be therefore considered as fingerprint of the regular no-horizon spacetimes. We are now looking for a fingerprint of the spacetime regularity in the secondary disc images. For the purpose of illustration of the character of secondary images we have constructed the three series of the indirect images corresponding to alredy generated primary disk images; for comparison we include also the secondary images in the Bardeen black hole spacetimes when the ghost images do not occur. The indirect (secondary) images are presented in Figure 16 for $\theta_{o}=30^\circ$, in Figure 17 for $\theta_{o}=60^\circ$, and in Figure 18 for $\theta_{o}=85^\circ$.

\begin{figure}[H]
	\begin{tabular}{cc}
		\includegraphics[scale=0.1]{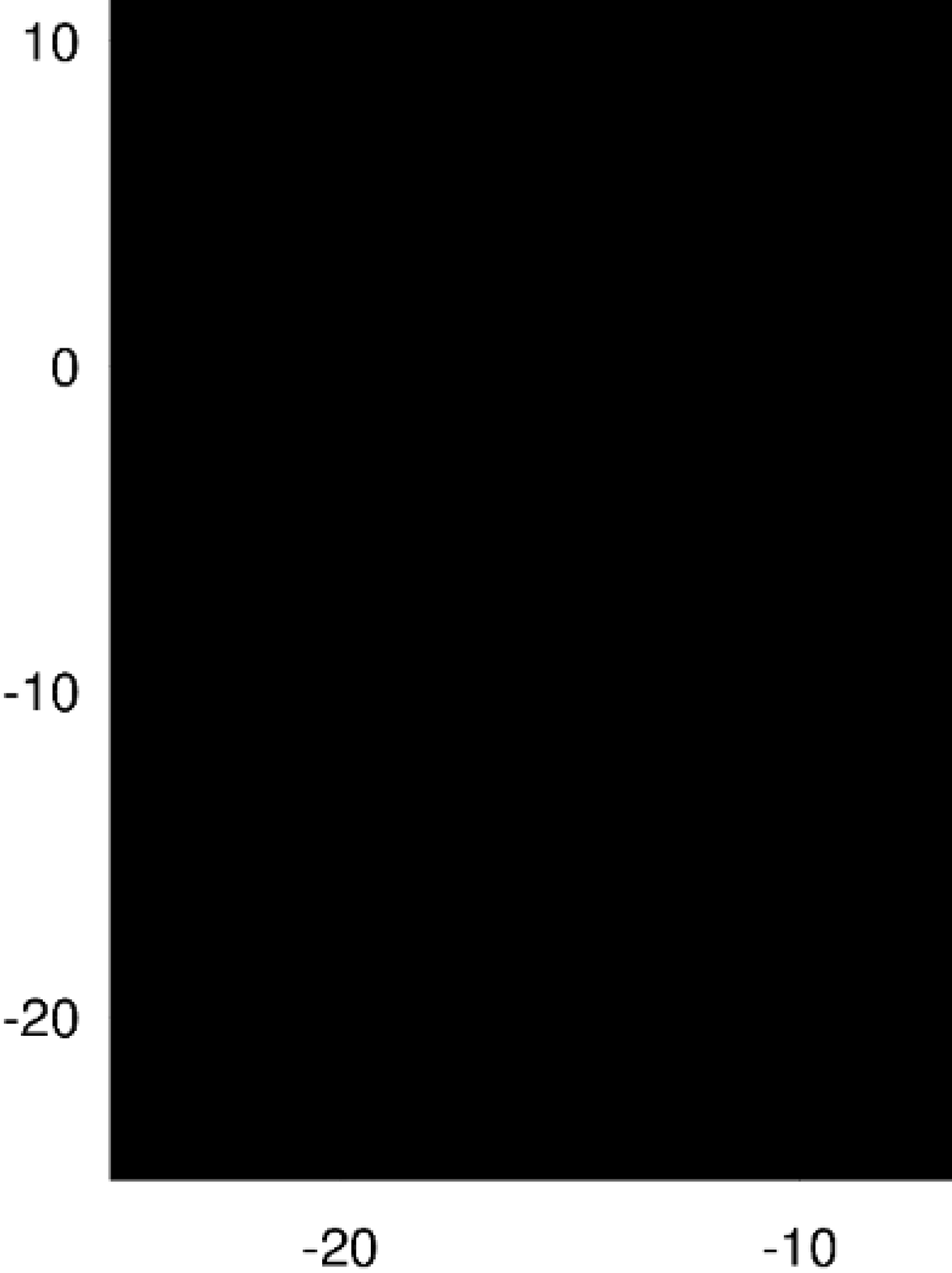}&\includegraphics[scale=0.1]{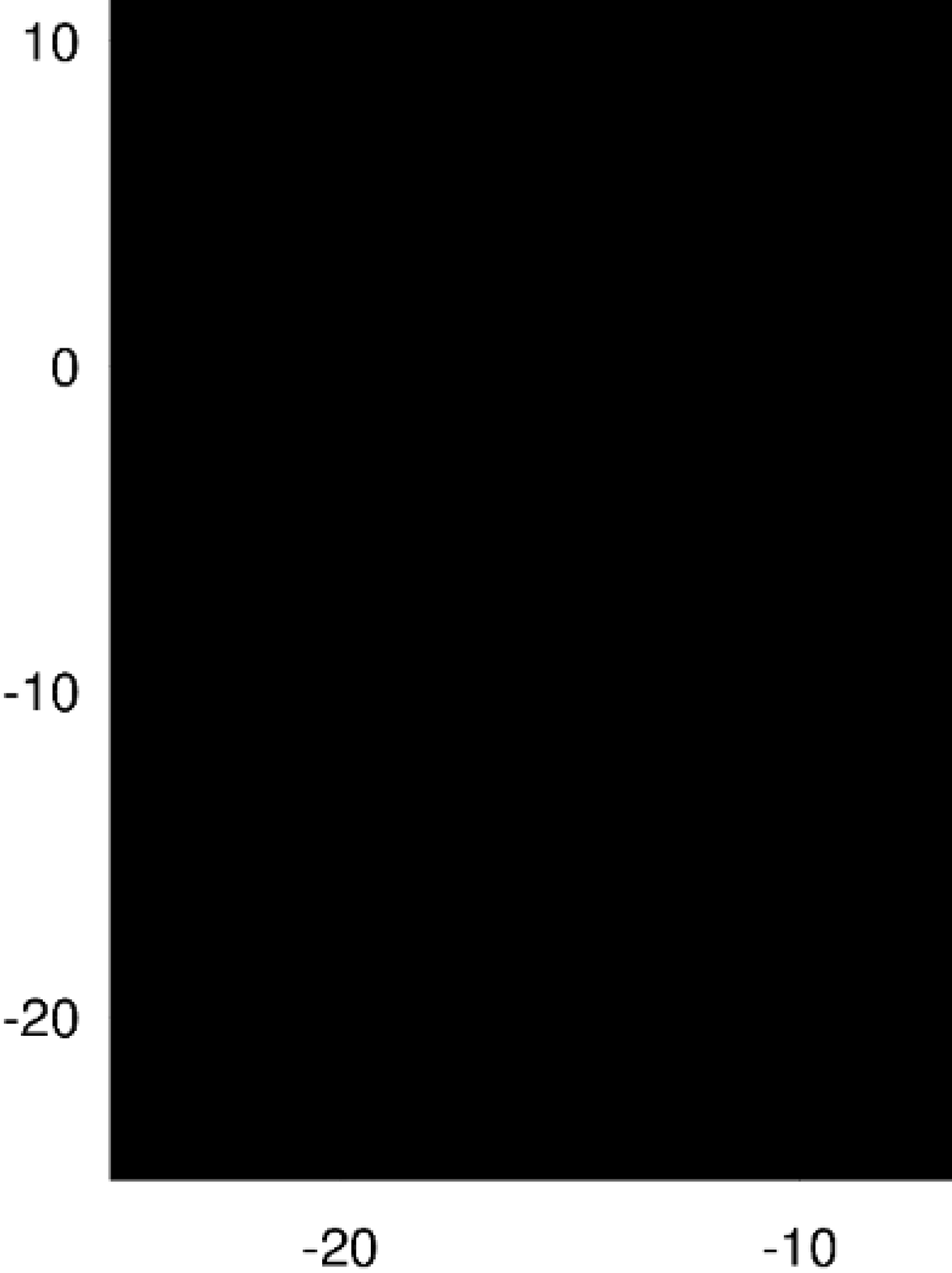}\\
		\includegraphics[scale=0.15]{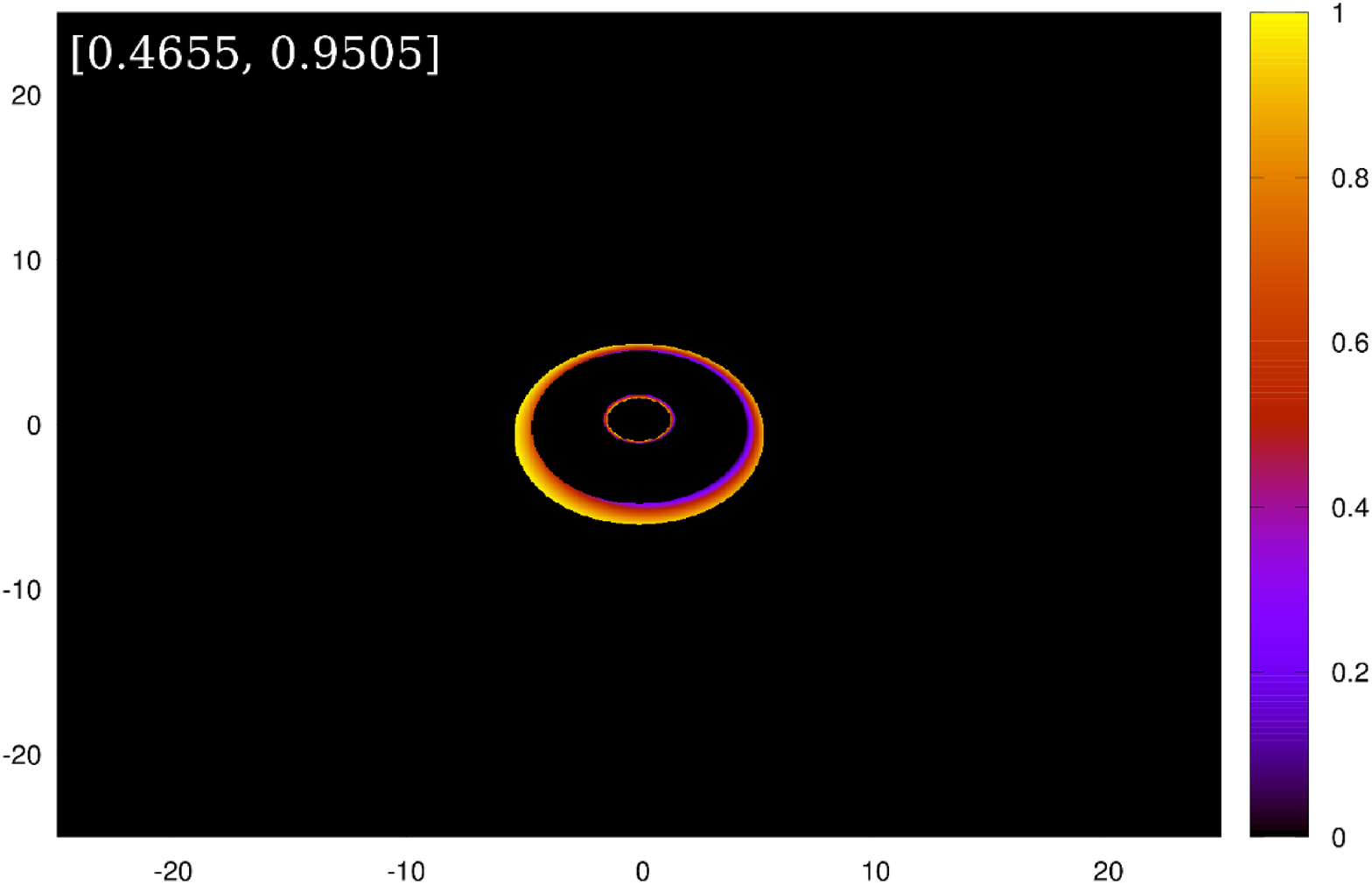}&\includegraphics[scale=0.1]{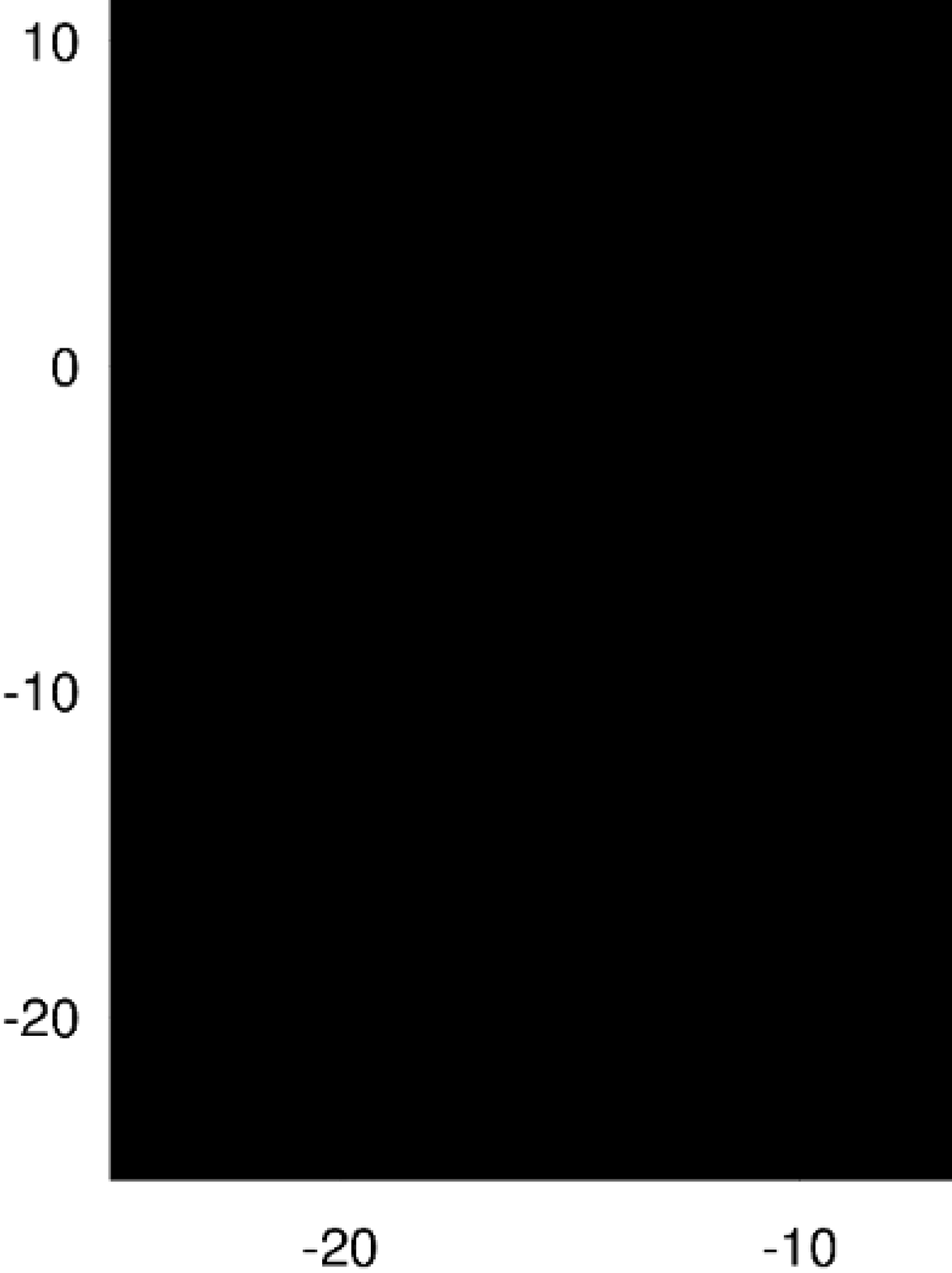}\\
		\includegraphics[scale=0.1]{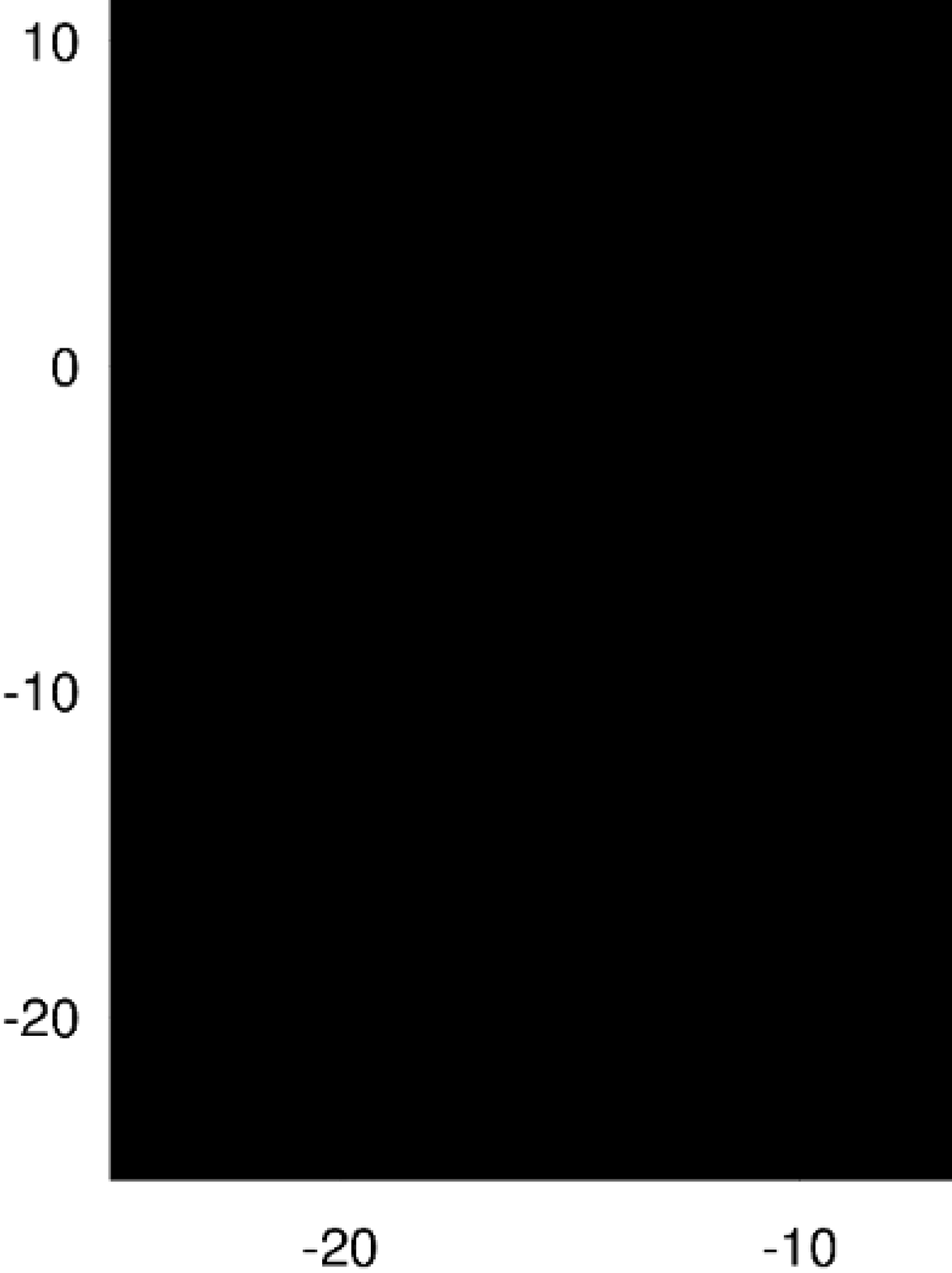}&\includegraphics[scale=0.1]{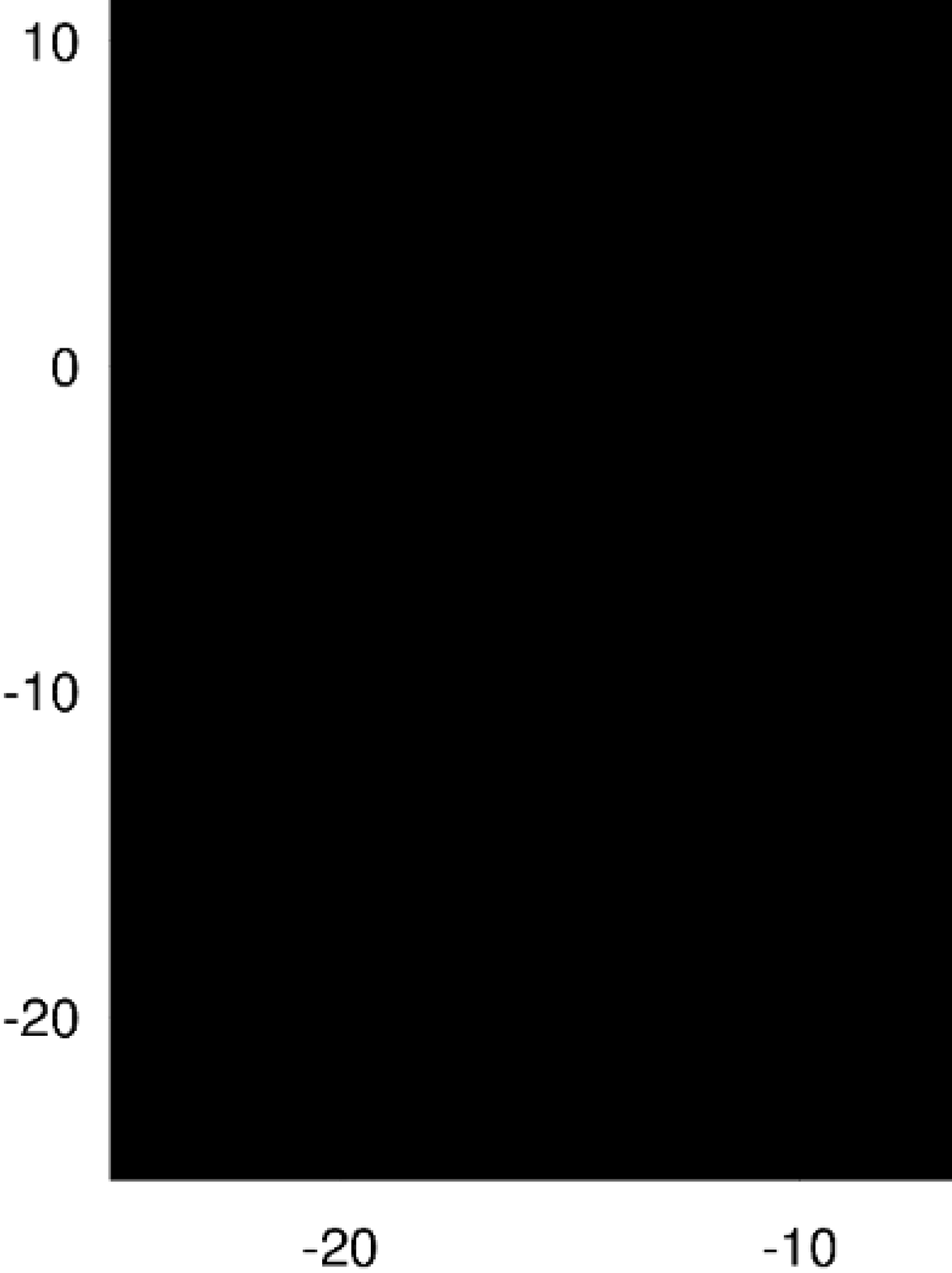}
	\end{tabular}
	\caption{The secondary disc images of Keplerian discs are constructed for the Bardeen spacetimes including the black hole case. The particular choice of the charge parameter reads $g/m=0.5$, $(g_{NoH}/m + g_p/m)/2$, $(g_p/m + g_s/m)/2$, $1.0$, $1.5$, $2.0$. The observer inclination angle is chosen to be $\theta_o=30^\circ$. No secondary image occurs for $g/m=2.0$ for this inclination. Notice that the secondary images in the Bardeen spacetimes are similar to those generated in the RN spacetimes.}
\end{figure} 


\begin{figure}[H]
	\begin{tabular}{cc}
		\includegraphics[scale=0.1]{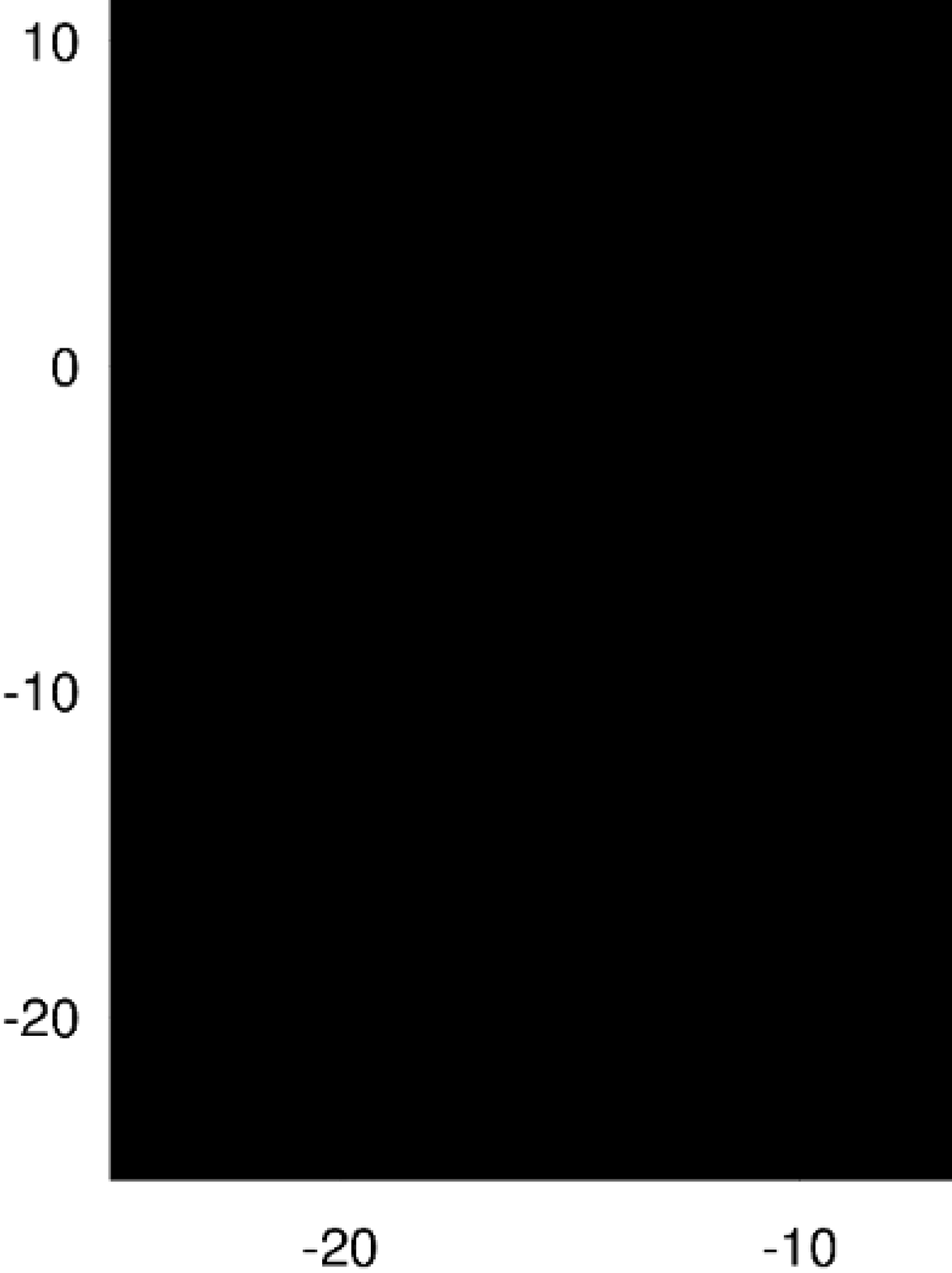}&\includegraphics[scale=0.1]{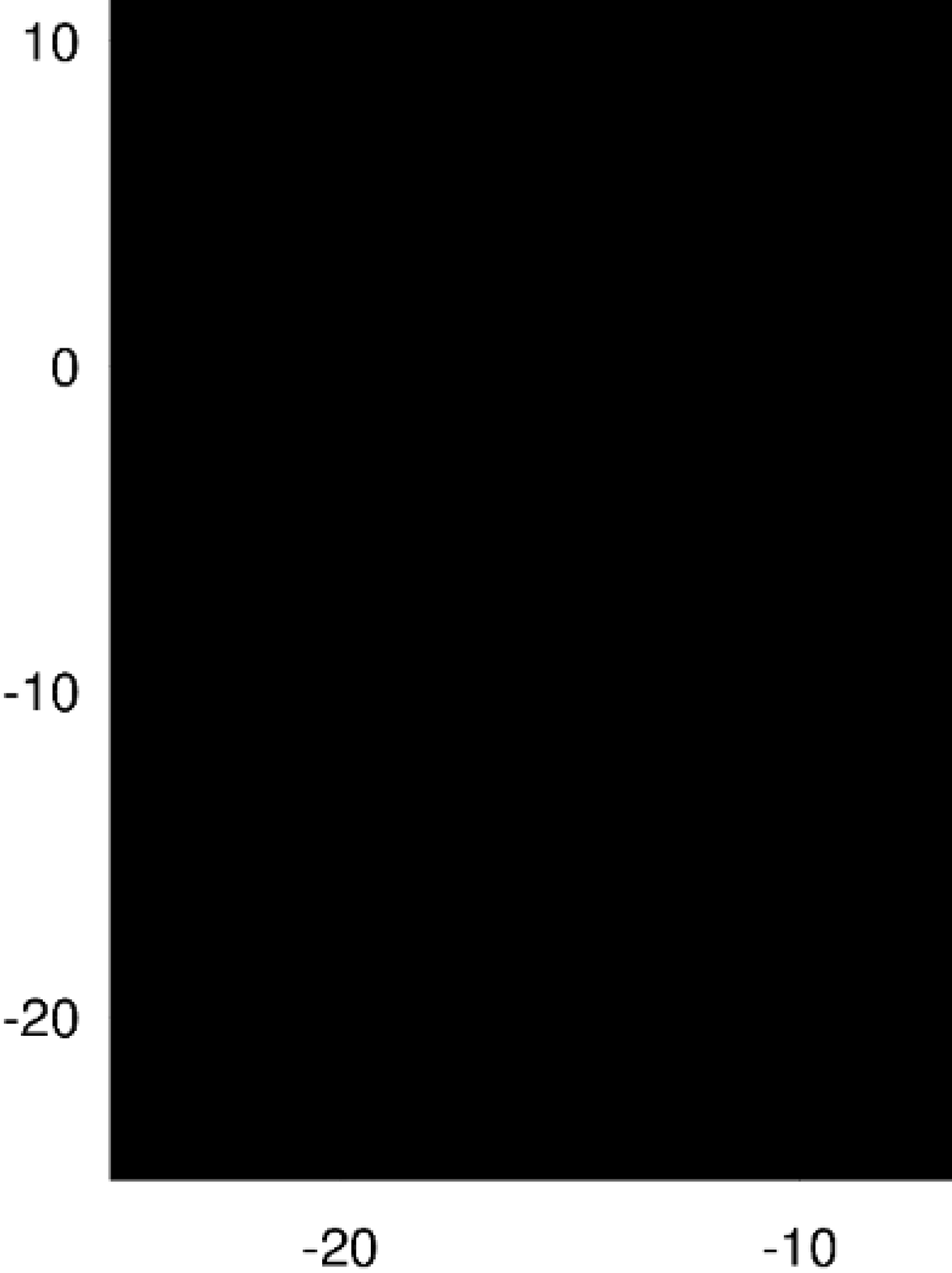}\\
		\includegraphics[scale=0.1]{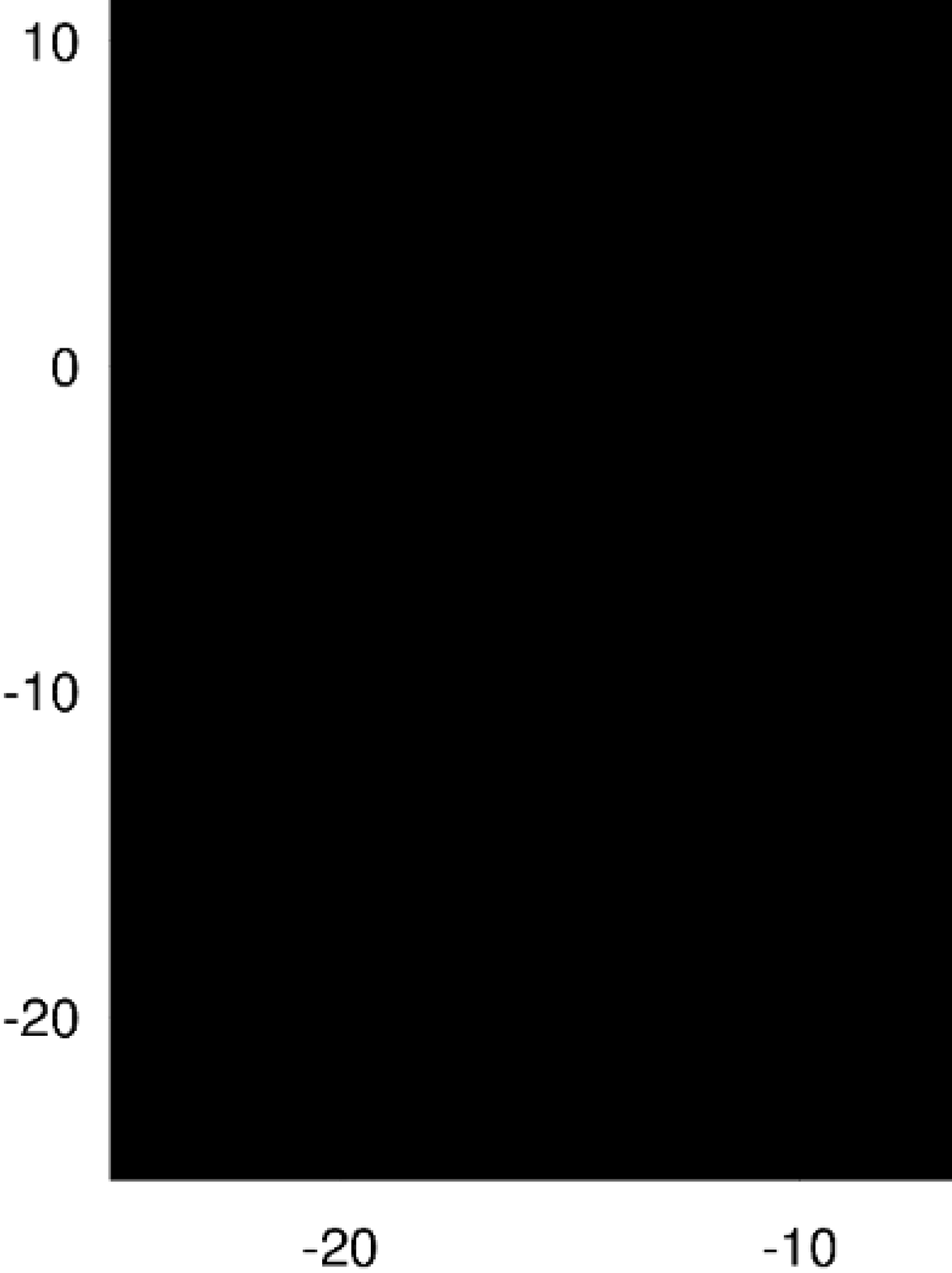}&\includegraphics[scale=0.1]{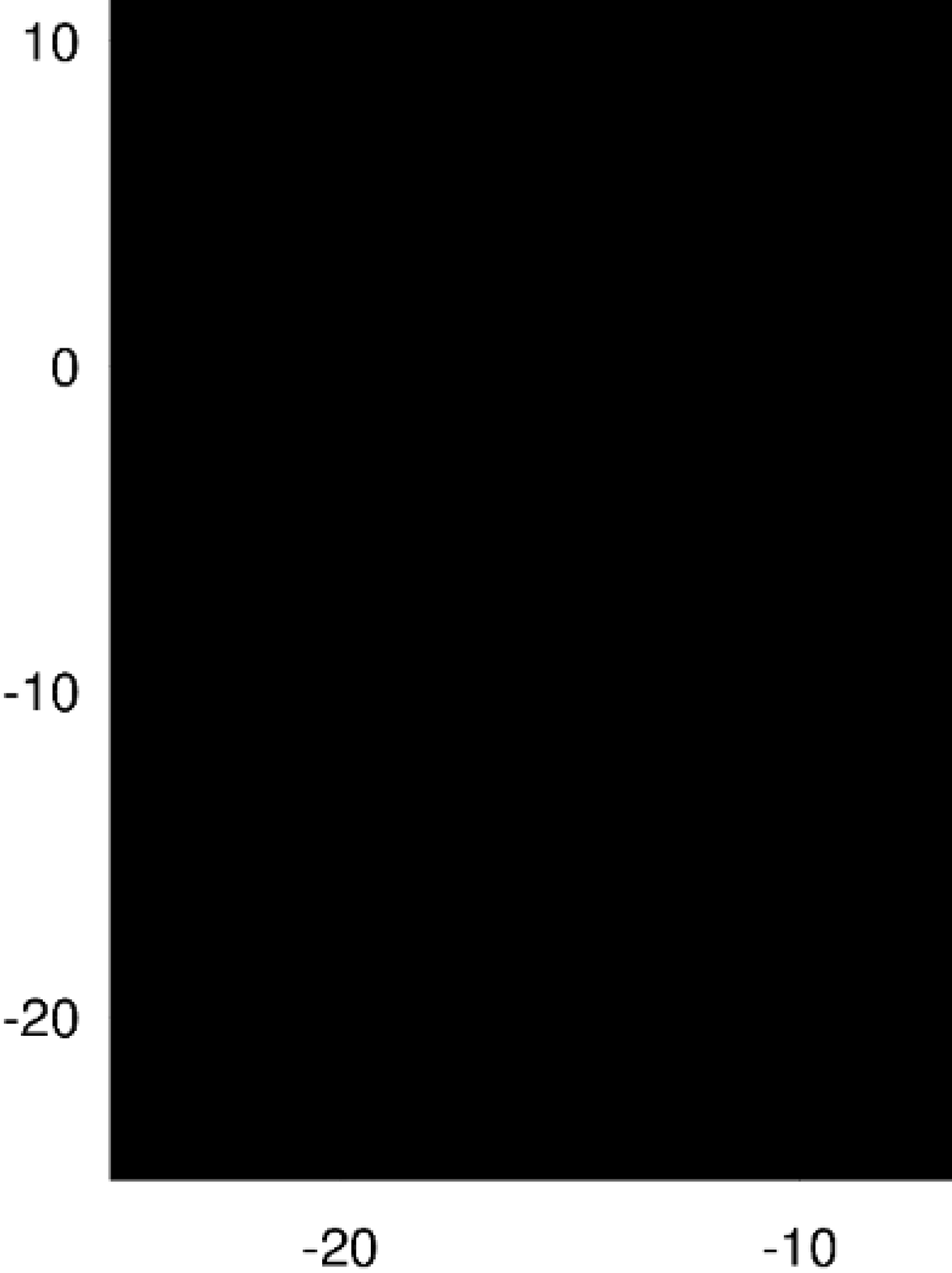}\\
		\includegraphics[scale=0.1]{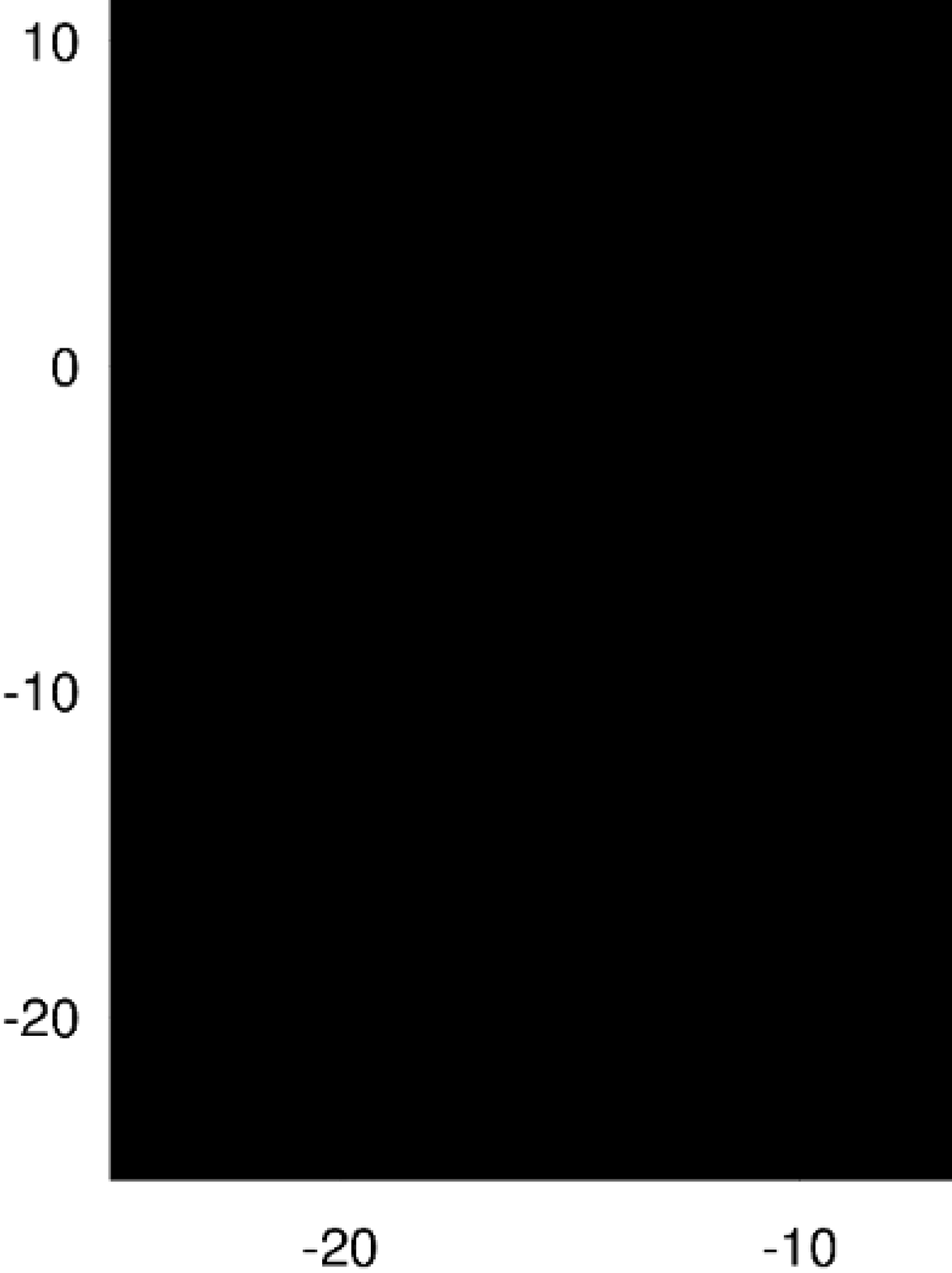}&\includegraphics[scale=0.1]{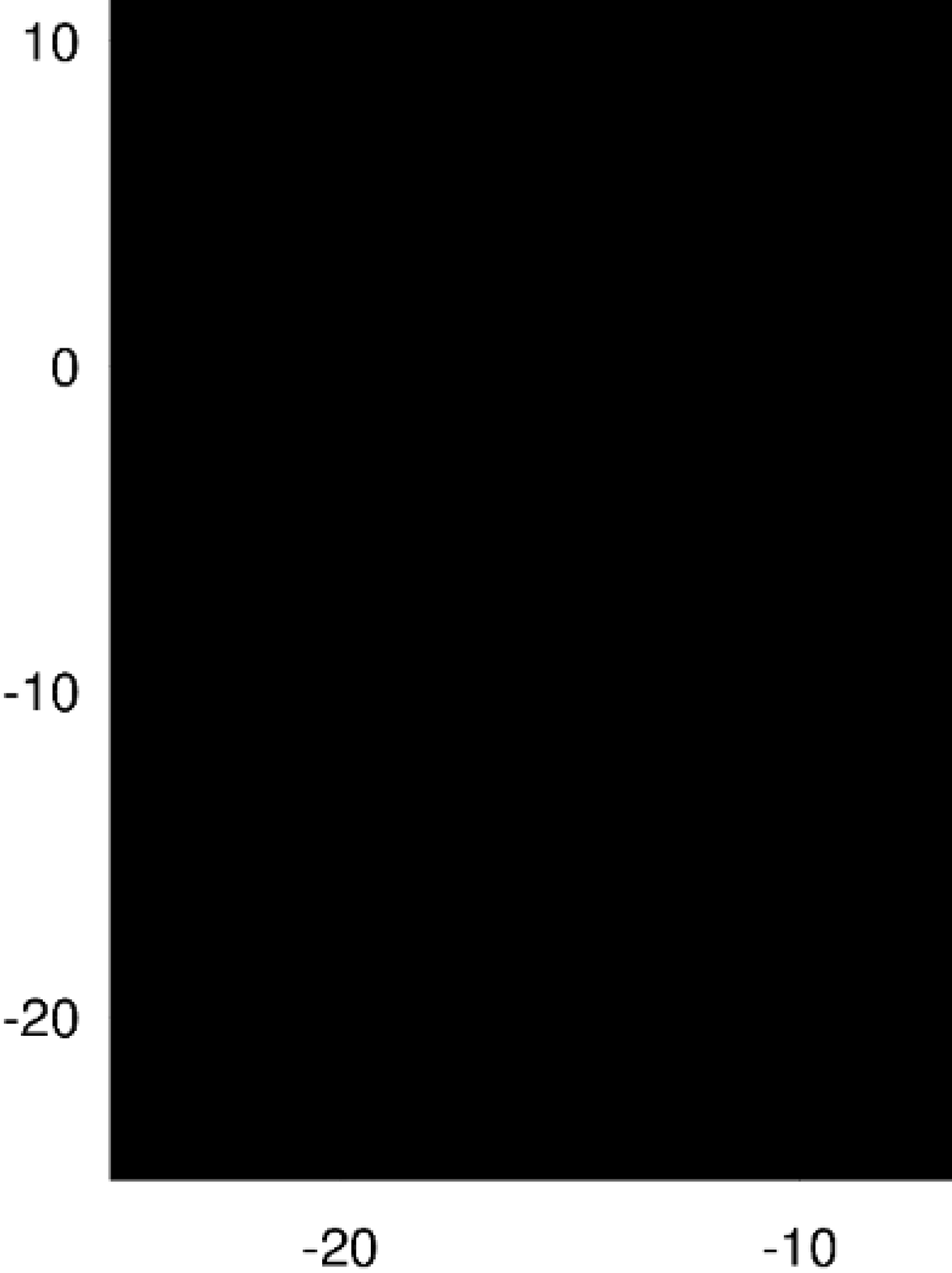}
	\end{tabular}
	\caption{The secondary disk images constructed for the Bardeen spacetimes with the particular choice of the charge parameter $g/m=0.5$, $(g_{NoH}/m + g_p/m)/2$, $(g_p/m + g_s/m)/2$, $1.0$, $1.5$, $2.0$. The observer inclination angle is chosen to be $\theta_o=60^\circ$. Now the secondary images occur in all the considered spacetimes, being restricted for $g/m=1.5,2.0$ when the ghost image vanishes.}
\end{figure} 


\begin{figure}[H]
	\begin{tabular}{cc}
		\includegraphics[scale=0.1]{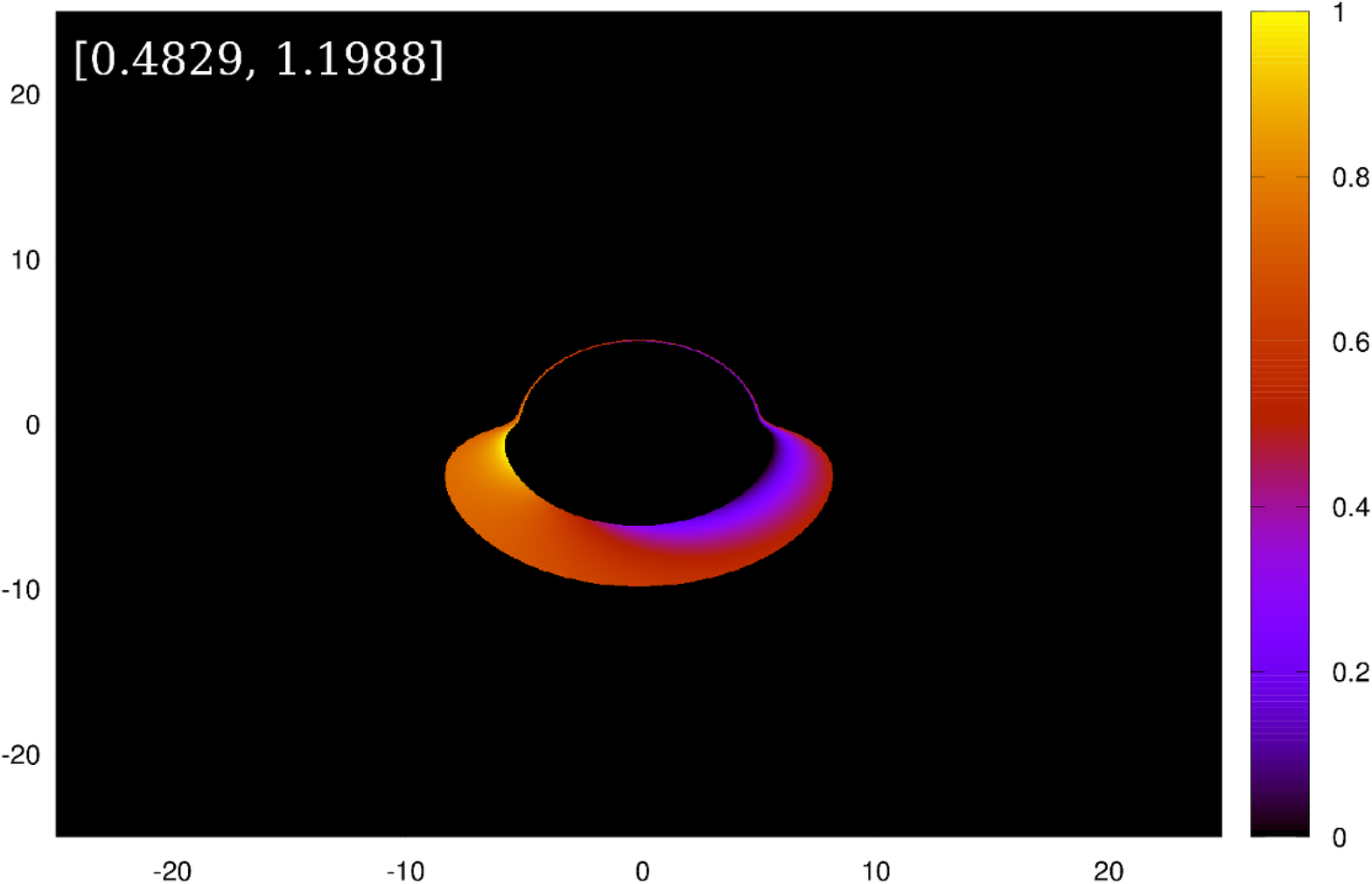}&\includegraphics[scale=0.1]{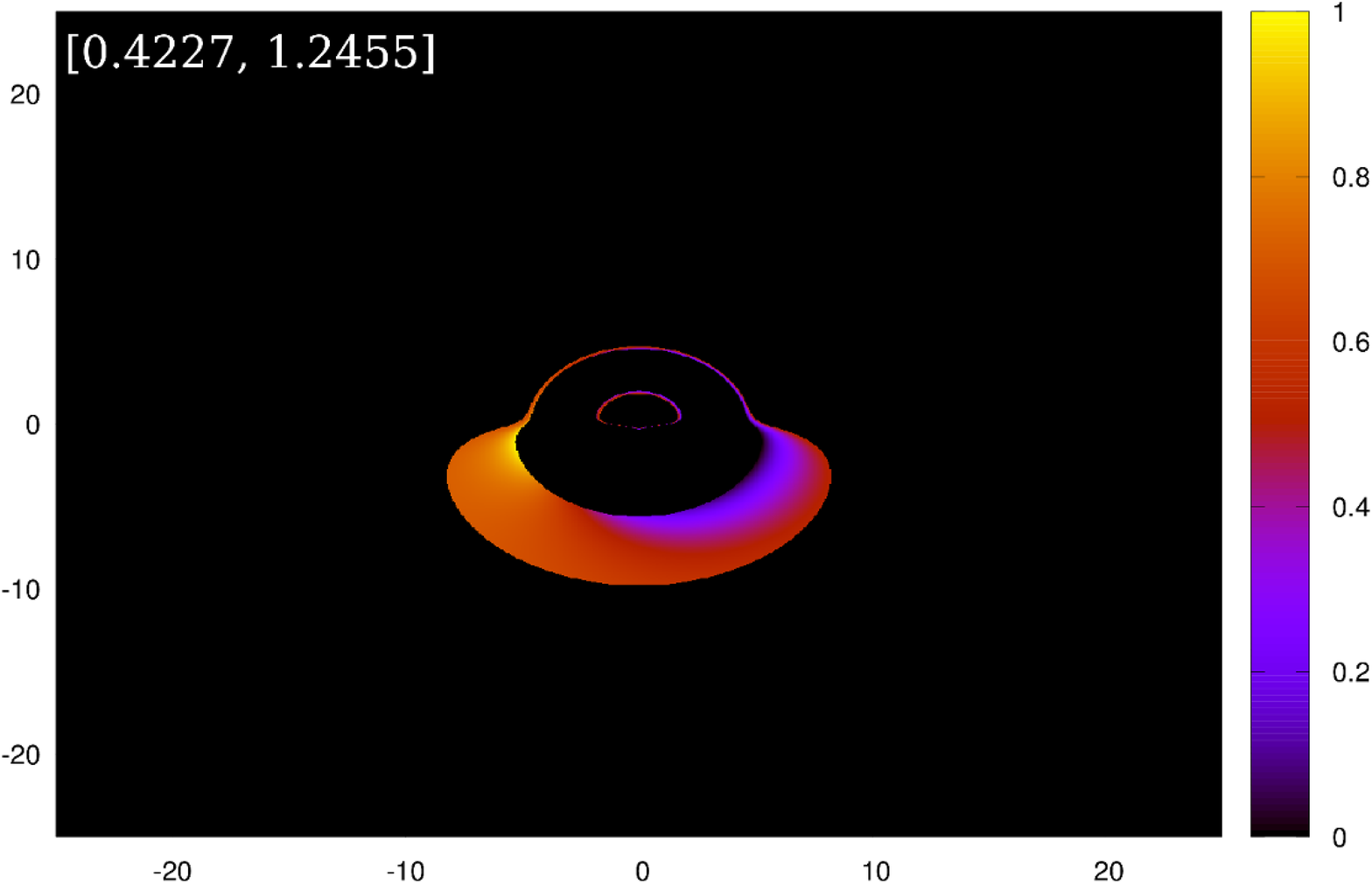}\\
		\includegraphics[scale=0.3]{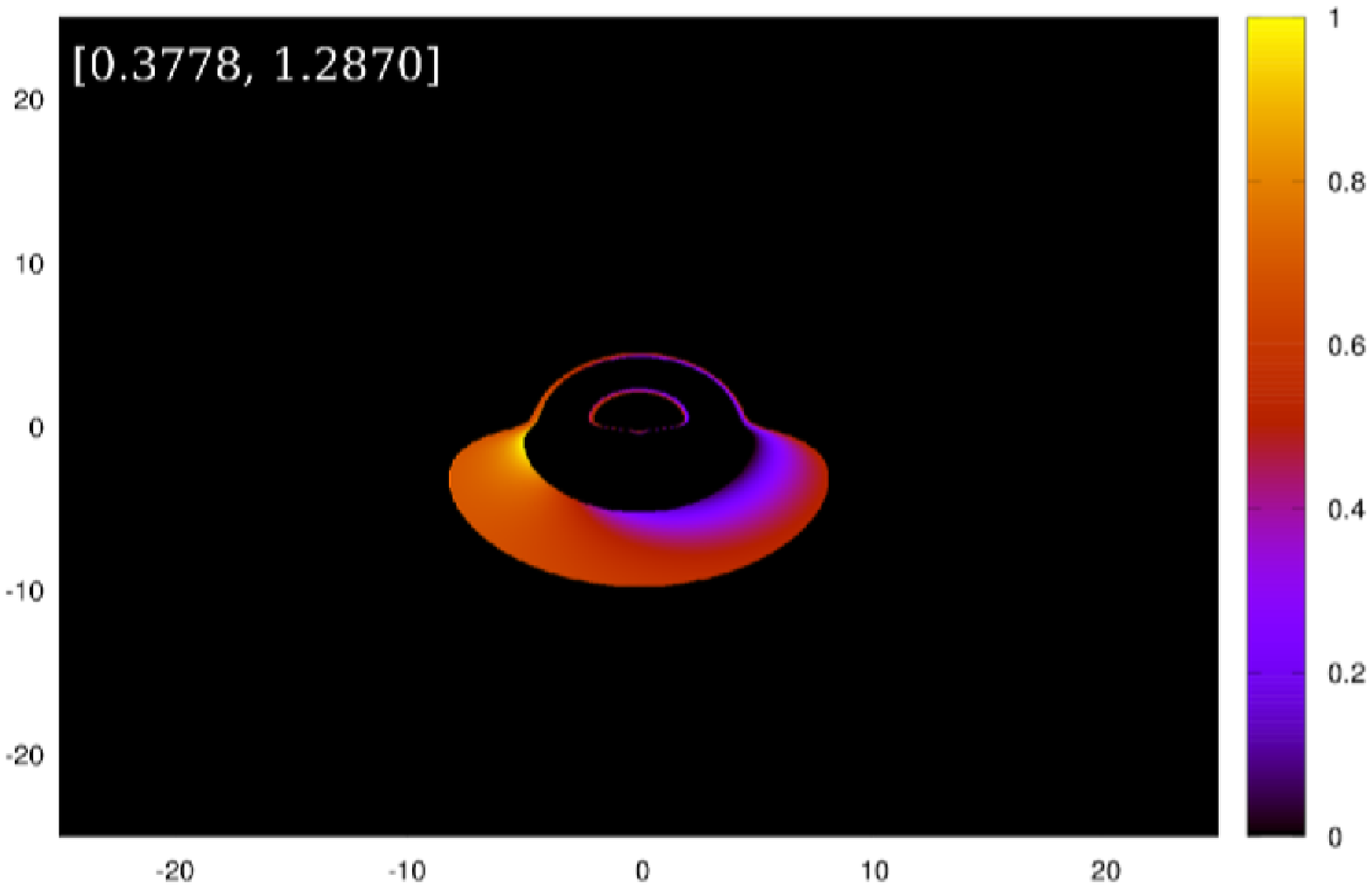}&\includegraphics[scale=0.1]{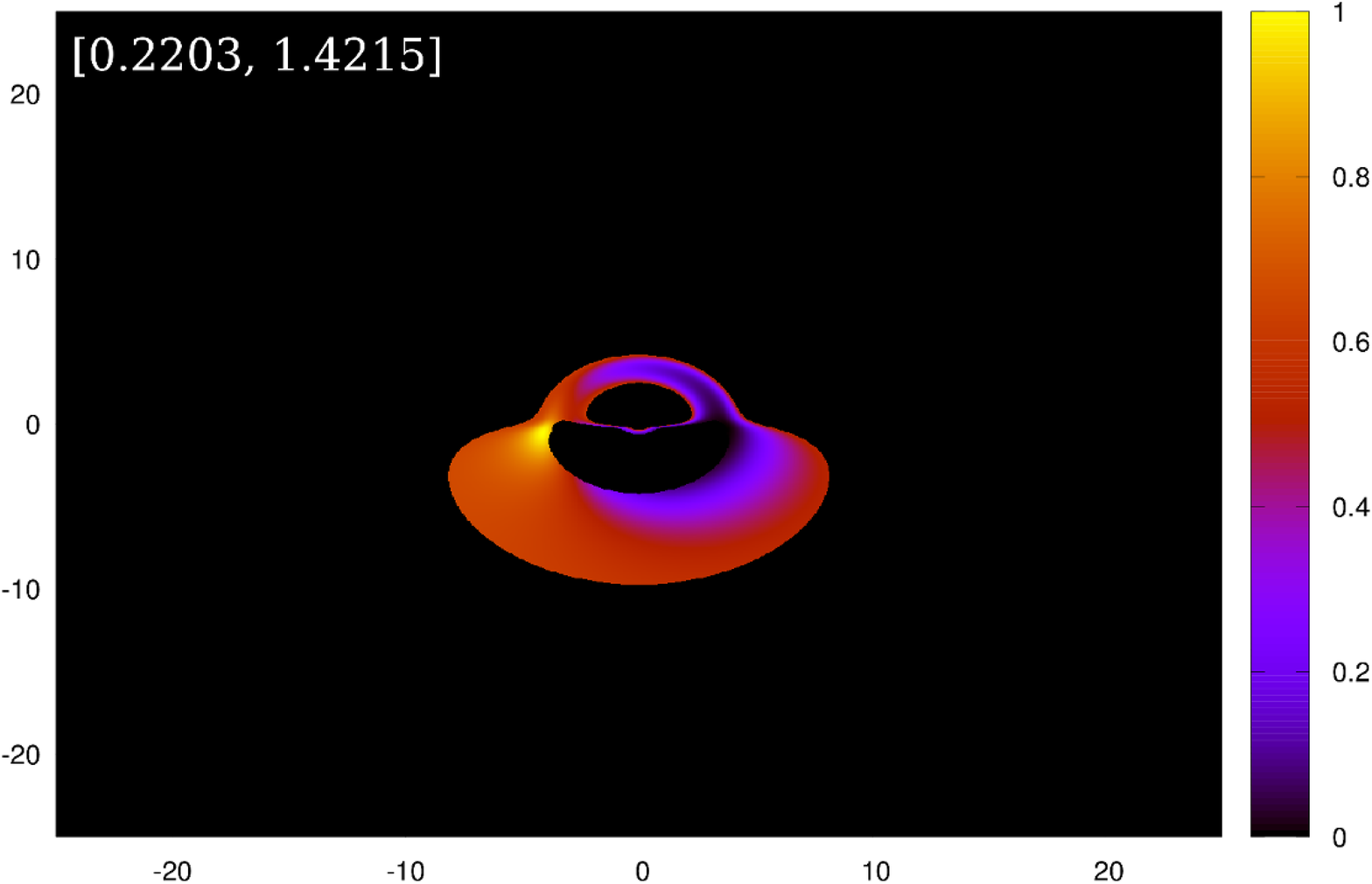}\\
		\includegraphics[scale=0.1]{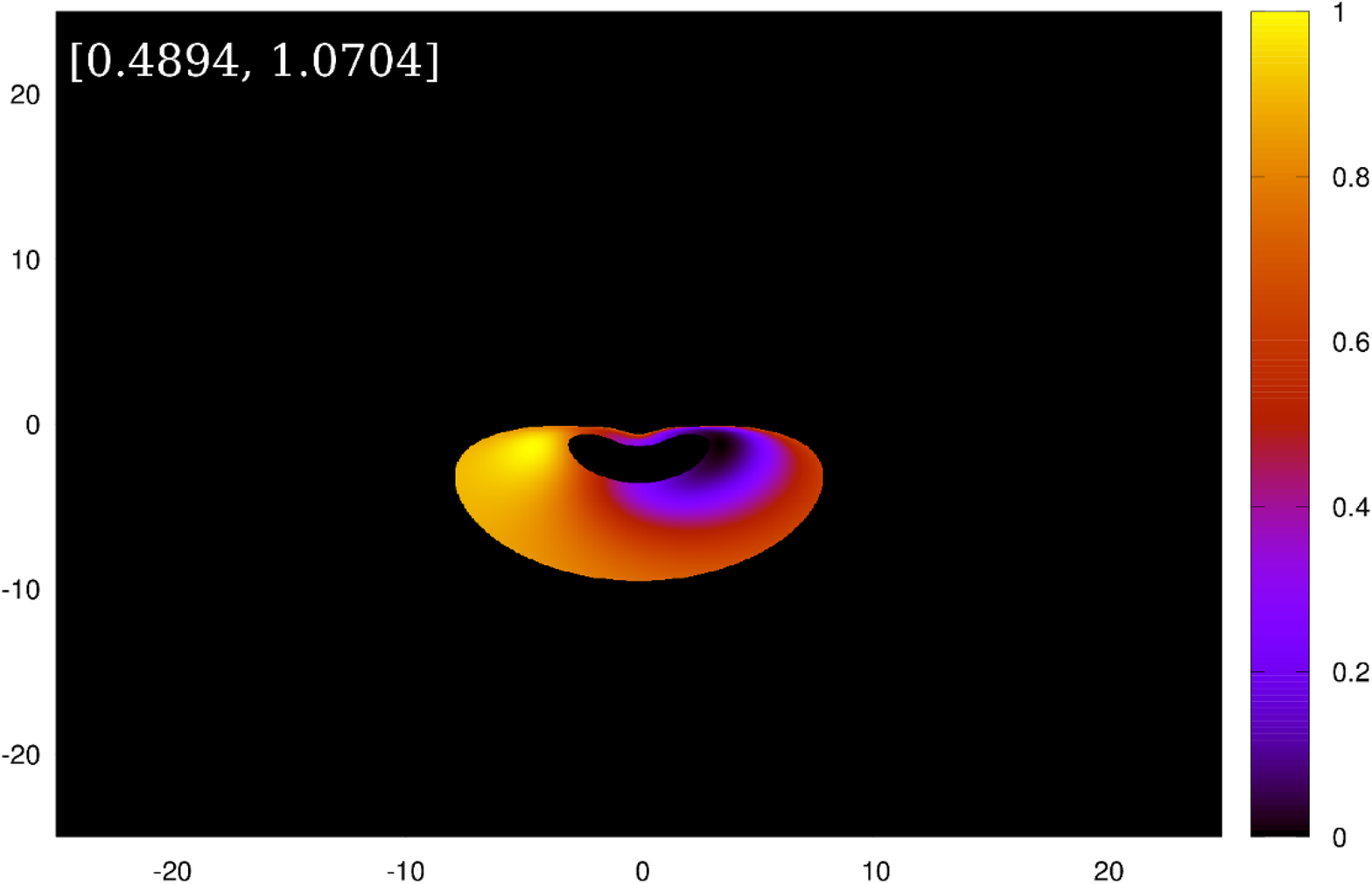}&\includegraphics[scale=0.3]{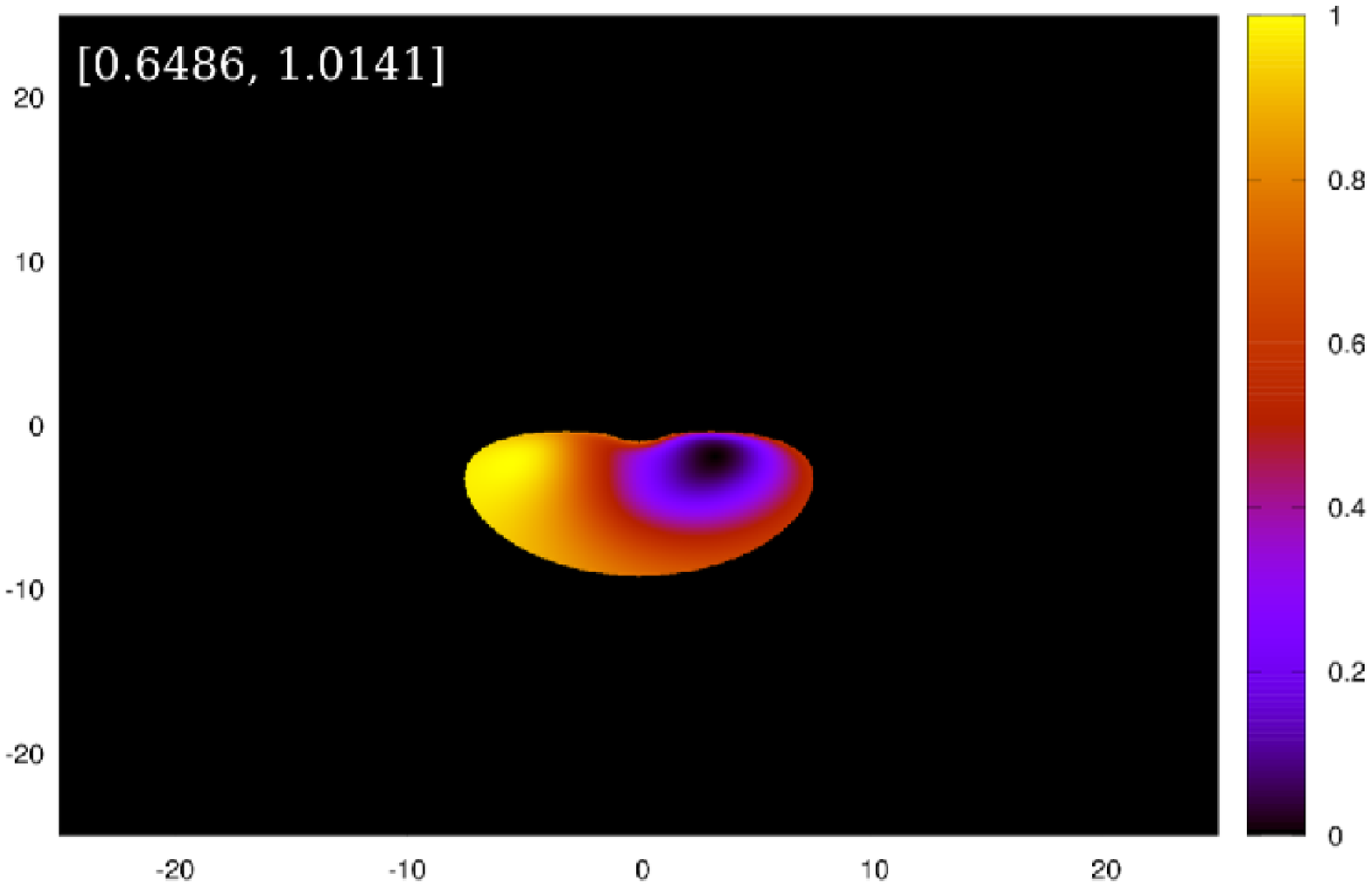}
	\end{tabular}
	\caption{The secondary disk images constructed for the Bardeen spacetimes with the particular choice of the charge parameter $g/m=0.5$, $(g_{NoH}/m + g_p/m)/2$, $(g_p/m + g_s/m)/2$, $1.0$, $1.5$, $2.0$. The observer inclination angle is chosen to be $\theta_o=85^\circ$. The ghost image vanishes again in the spacetimes with $g/m=1.5,2.0$ when the secondary image undergoes the deformation related to the insufficiency of the ray deflection.}
\end{figure} 


The indirect ghost images occur even for small inclination angles, contrary to the direct ghost images that can be created only for very large inclination angles; moreover, the secondary ghost images can cover whole the radiating part of the Keplerian disc. For charges appropriately smaller than the critical charge related to a given inclination angle of the observer, the secondary images are restricted to the "bottom" parts due to the insufficient deflection of the photons coming from the "upper" parts of the discs. (Such a deformation of the secondary image can occur only in the bardeen spacetimes with the specific charge $g/m>1$.) No secondary images occur in the Bardeen spacetime with $g>g_{c}(\theta_{o})$. 

The extrema of the frequency shift in the secondary images of Keplerian discs demonstrate monotonous behavior in dependence on the inclination angle. The extrema $z_{max}$ ($z_{min}$) increase (decrease) with increasing inclination angle in a given spacetime. On the other hand, for fixed inclination angle, the dependence of $z_{max}$ and $z_{min}$ on the charge parameter $g$ is complex and demonstrates a break as shown above. 

\subsection{Complete images of Keplerian discs in the Bardeen no-horizon spacetimes}

For the Keplerian discs orbiting in the Bardeen no-horizon spacetimes we finally give the complete image composed from the direct and indirect images including the ghost images and higher-order images, if they exist. To construct the images, we shall use the method presented in \cite{Stu-Sche:2010:CLAQG:} that is convenient for ray-tracing in the curved spacetimes having no event horizons. 

\subsubsection{Complete images in the no-horion spacetimes admitting circular photon orbits}

We demonstrate first appearance of the Keplerian discs in the regular no-horizon spacetimes allowing for existence of the photon circular orbits. Note that for construction of the Keplerian disc complete image only the unstable outer photon circular orbit is relevant, while the stable inner circular photon orbit is crucial for the trapped photons, being irrelevant for direct observational phenomena \cite{Stu-Sche:2010:CLAQG:,Stu-Sche:2014:CLAQG:}. In constructing the complete image we use the ray-tracing method taking into account also the higher order ($n \geq 3$) images. To visualize the complete Keplerian disc we follow the approach of Luminet where the colors represent the bolometric flux of radiation coming from the disk \cite{Lum:1979:AAP}. Here we use arbitrary units being interested in relative luminosity of the higher-order images; for physical definition see \cite{Lum:1979:AAP}. The results are illustrated in Figure 19. 

\begin{figure}[H]
	\begin{center}
		\includegraphics[scale=0.6]{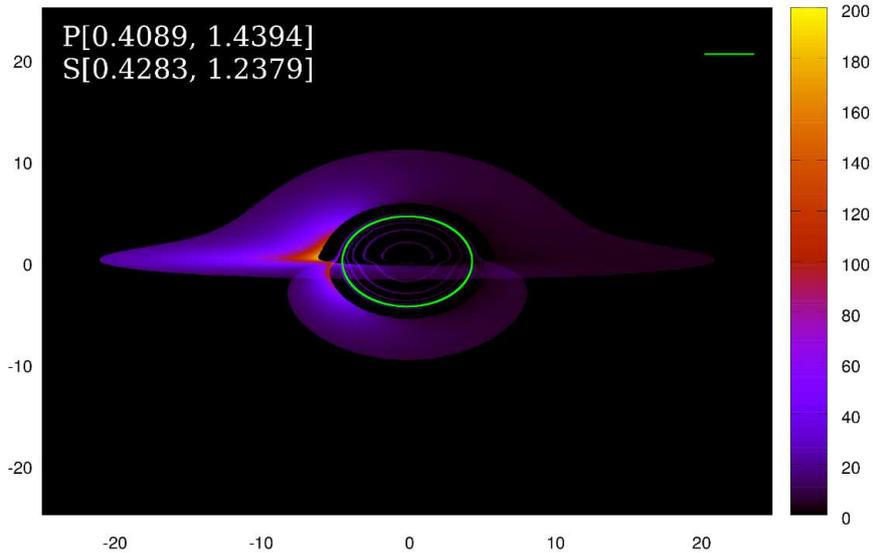}
		\caption{Complete image of Keplerian disc formed in the Bardeen spacetime with the charge parameter $g/m=0.8$ allowing for existence of the circular photon orbits. The disc spans from $r_{in}=r_{ISCO}$ to $r_{out}=20$. The observer inclination angle reads $\theta_o=85^\circ$. The red circle determines the unstable photon circular orbit. The frequency shift range is given for both the primary (P) and secondary (S) images. The colors represent the bolometric flux.}
	\end{center}
\end{figure}

In Figure 20 we present enlarged region of the complete image located close to the unstable photon circular orbit with impact parameter $l=l_{ph(u)}(g)$, representing a boundary between the images created by photons with $l>l_{ph(u)}(g)$ and $l<l_{ph(u)}(g)$. In fact, an infinite number of images have to be located in the vicinity of $l_{ph}$, however, with increasing number of the order of the image, its detectability decreases (for details see \cite{Mis-Tho-Whe:1973:Gravitation:,Vir-Elli:2002:PHYSR4:,Vir-Kee:2008:PHYSR4:}). We can see that the distribution of the higher order images inside and outside the photon circular orbit differs being more dense outside the photon orbit (images with $l>l_{ph(u)}(g)$). 

\begin{figure}[H]
	\begin{center}
		\begin{tabular}{c}
		\includegraphics[scale=0.42]{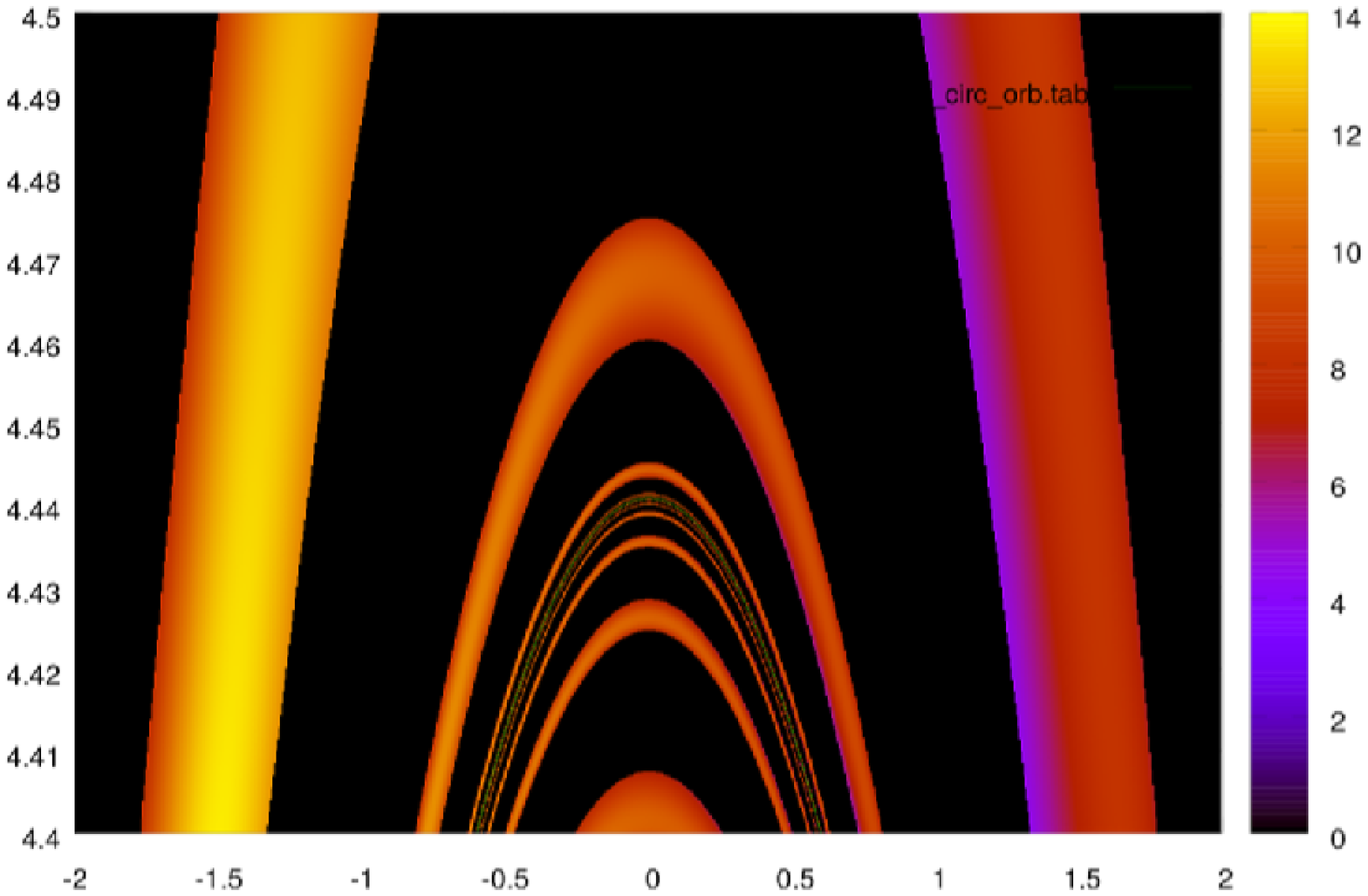}\\
		\includegraphics[scale=0.42]{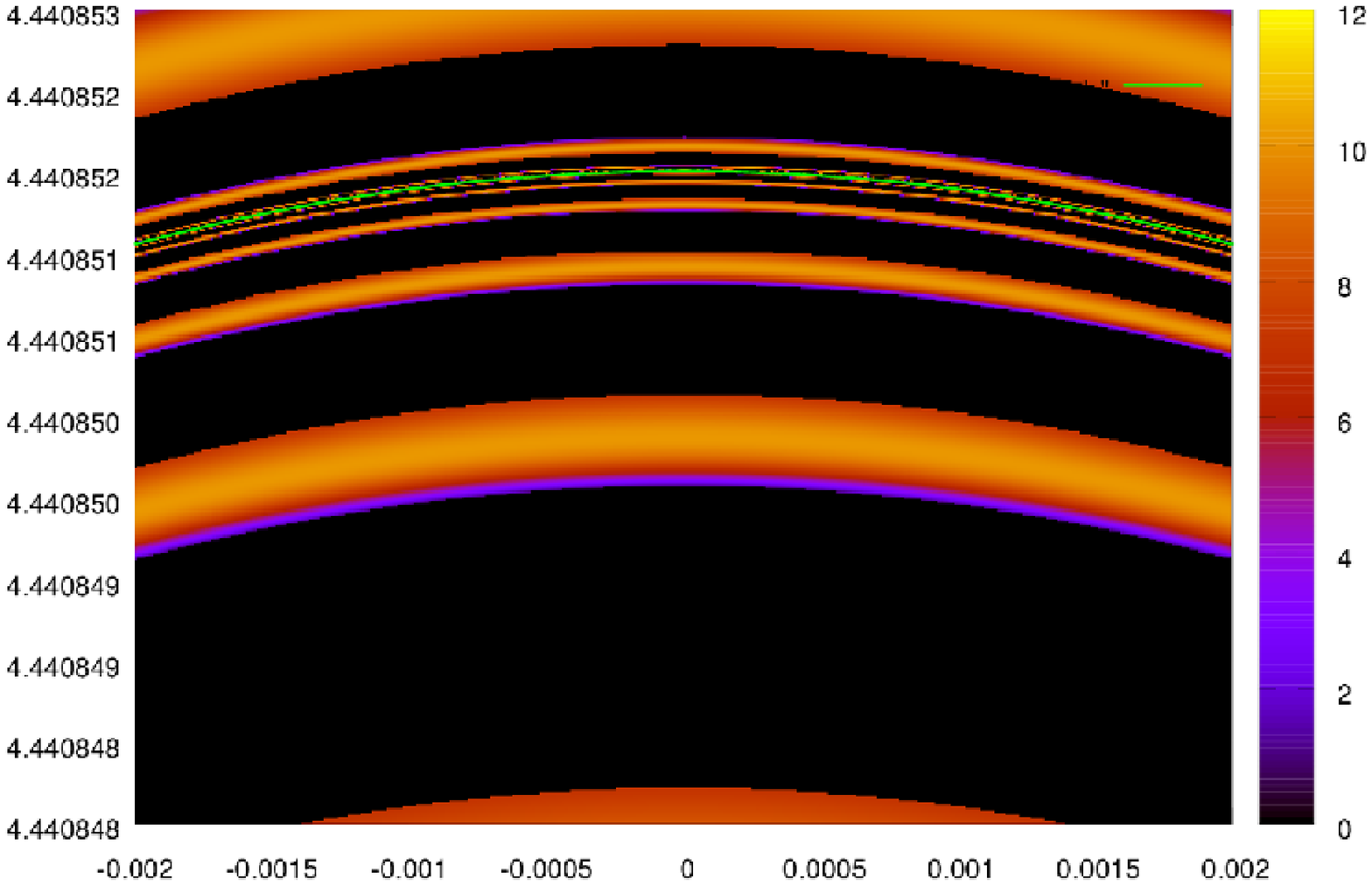}
		\end{tabular}
		\caption{The enlarged regions of the Keplerian disc complete image in close vicinity of unstable  circular photon orbit (green line) in the Bardeen spacetime with the charge parameter $g/m=0.8$. The disc extension is the same as in the previous figure. The observer inclination angle is again $85^\circ$. The colors represent the bolometric flux.}
	\end{center}
\end{figure}  

\subsubsection{Complete image in the no-horizon spacetimes without the circular photon orbits}

In such Bardeen spacetimes the Keplerian disc complete image consists only from the direct (primary) and indirect (secondary) images containing the corresponding ghost images. Higher order images do not exist because of the lack of the circular photon orbit. The typical complete image is constructed for inclination angle allowing for existence of the direct ghost image and is presented in Figure 21. In the central region the direct ghost image occurs being a clear signature of the regular no-horizon spacetime. 

\begin{figure}[H]
	\begin{center}
		\includegraphics[scale=0.4]{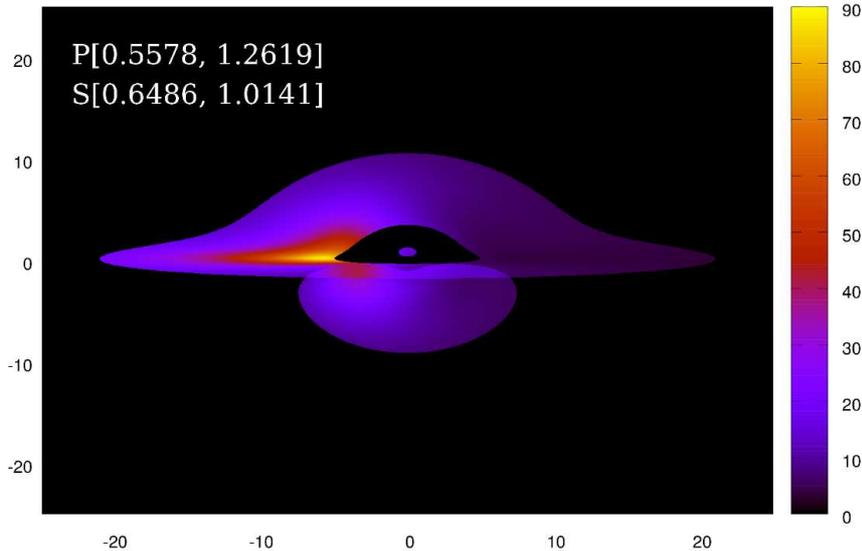}
		\caption{Complete image of the Keplerian disc orbiting in the Bardeen spacetime with the charge parameter $g/m=2.0$ not allowing existence of photon circular orbits. The disc spans from $r_{in}=r_{\Omega max}$ to $r_{out}=20$. The observer inclination angle is chosen $\theta_o=85^\circ$. In this case no higher order images can occur. The direct ghost image is the clear qualitative signature of the presence of the Bardeen no-horizon spacetime. The frequency shift range is given for both the primary (P) and secondary (S) images. The colors represent the bolometric flux.}
	\end{center}
\end{figure}

\section{Keplerian disc images in the Reissner-Nordstrom naked singularity spacetimes}

We compare the situation related to the creation of ghost images in the regular no-horizon Bardeen spacetimes to the complementary case of the Reissner-Nordstrom (RN) naked singularity spacetimes. Due to the existence of the naked singularity, in the RN spacetimes the ghost images are generated in a different way than in the regular Bardeen no-horizon spacetimes. The non-existence of the horizon implies existence of ghost images that are generated due to the repulsive effect of the effective potential of the photon motion in the vicinity of the singularity at $r=0$. Existence of the ghost images has been discovered and studied in the case of the Kerr naked singularity spacetimes \cite{Stu-Sche:2010:CLAQG:}, where some additional effects related to the circular geodesic properties input some additional effects interesting from the observational point of view \cite{Stu:1980:BAC:,Stu-Hle:2000:CLAQG:,Stu-Sche:2012a:CLAQG:,Stu-Sche:2013:CLAQG:}. 

We first shortly summarize properties of the geodesic circular motion in the RN naked singularity spacetimes. Then we discuss the photon motion, concentrating on the creation of the ghost images related to both the direct and indirect appearance of the Keplerian discs, and the vanishing of the indirect images due to insufficient deflection angle. 

\subsection{Reissner-N\"{o}rdstr\"{o}m spacetime} 

The solution of the Einstein-Maxwell equations for spherically symmetric and static spacetime is governed by the lapse function 
\begin{equation}
f(r;Q)=1-\frac{2}{r}+\frac{Q^2}{r^2},
\end{equation}
where $Q = \tilde{Q} / M$ is the  specific charge of the RN spacetime; $\tilde{Q}$ is the electric charge, and $M$ is the gravitational mass of the spacetime. We do not consider here the "Coulomb" electromagnetic field complementary to the RN spacetime -- for details see \cite{Mis-Tho-Whe:1973:Gravitation:}. We use the dimensionless radial coordinate and dimensionless specific charge $Q$, by putting $M=1$. Contrary to the regular spacetimes where the geometry is nearly flat at $r \sim 0$, in the RN spacetimes the lapse function diverges (with its derivatives) at $r=0$, generating thus a repulsive barrier. \footnote{Notice that in the Kehagias-Sfetsos spacetimes of the modified Horava quantum gravity the lapse function is finite at $r=0$, but its derivative diverges indicating a physical singularity \cite{Stu-Sche-Abd:2014:PHYSR4:}.}

The horizons, satisfying $f(r;Q)=0$, are located at radii 
\begin{equation}
r_\pm=1 \pm \sqrt{1-Q^2}.
\end{equation}
In the following we focus attention to the RN naked singularity spacetimes, i.e., we assume (considering positively charged RN backgrounds) 
\begin{equation}
  Q>Q_{NS}=1. 
\end{equation}

\subsection{Keplerian orbits of neutral test particles} 

The circular Keplerian orbit at a given radius $r$ has the specific angular momentum $l_c$ and the specific covariant energy $E_c$ given by the relations \cite{Stu-Hle:2002:ActaPhysSlov:,Pug-Que-Ruf:2011:PHYSR4:} 
\begin{equation}
 l_c^2=\frac{r^2(r-Q^2)}{2Q^2+r(r-3)} 
\end{equation}
and 
\begin{equation}
E^2_c=V_{eff}=\frac{r^2-2r+Q^2}{r^2(2Q^2+r(r-3))}.
\end{equation} 

The static radius of the RN naked singularity spacetimes (where $l_{c}=0$) is located at 
\begin{equation}
r_{stat}=Q^2.
\end{equation}
The photon circular orbits are located at 
\begin{equation}
r_{ph\pm}=\frac{1}{2}(3 \pm \sqrt{9-8Q^2}).
\end{equation}
The critical value of the charge parameter of the RN spacetimes allowing for existence of circular photon orbits reads 
\begin{equation}
  Q_{ph}=\frac{3}{2\sqrt{2}}. 
\end{equation}
The marginally stable circular orbits are located at radii given by the condition $dV_{eff}/dr=0$ and $d^2V_{eff}/dr^2=0$ which leads to the equation \cite{Stu-Hle:2002:ActaPhysSlov:,Pug-Que-Ruf:2011:PHYSR4:,Pug-Que-Ruf:2011b:PHYSR4:}   
\begin{equation}
	r^3-6r^2+9Q^2 r-4 Q^4=0.
\end{equation}
The corresponding inner and outer marginally stable orbits radii then read
\begin{equation}
r_{ms-out}= 2\left[1+\sqrt{4-3Q^2}\cos\left(\frac{\gamma}{3}\right)\right] 
\end{equation}
and
\begin{equation}
r_{ms-in}= 2\left[1+\sqrt{4-3Q^2}\cos\left(\frac{\gamma+4\pi}{3}\right)\right] 
\end{equation}
where 
\begin{equation}
	\gamma=\cos^{-1}\left[\frac{8-9Q^2+2Q^4}{(4-3Q^2)^3}\right].
\end{equation}
The critical value of the charge parameter of the RN spacetimes allowing for existence of unstable circular orbits reads 
\begin{equation}
  Q_{s}=\frac{\sqrt{5}}{2} 
\end{equation}

The angular frequency of the Keplerian orbits related to distant observers reads 
\begin{equation}
\Omega=\frac{U^\phi}{U^t}=-\frac{f}{r^2}\frac{l_c}{E_c}=\sqrt{\frac{r-Q^2}{r^4}}.
\end{equation}
There is a maximum of the angular frequency $\Omega$, located at
\begin{equation}
 r_{\Omega}=\frac{4}{3}Q^2.
\end{equation}
The structure of the circular geodesics represented in the $r-Q$ space is given in Figure 22 where regions corresponding to stable, unstable and no circular geodesic orbits are presented. 
\begin{figure}[H]
	\begin{center}
		\includegraphics[scale=0.9]{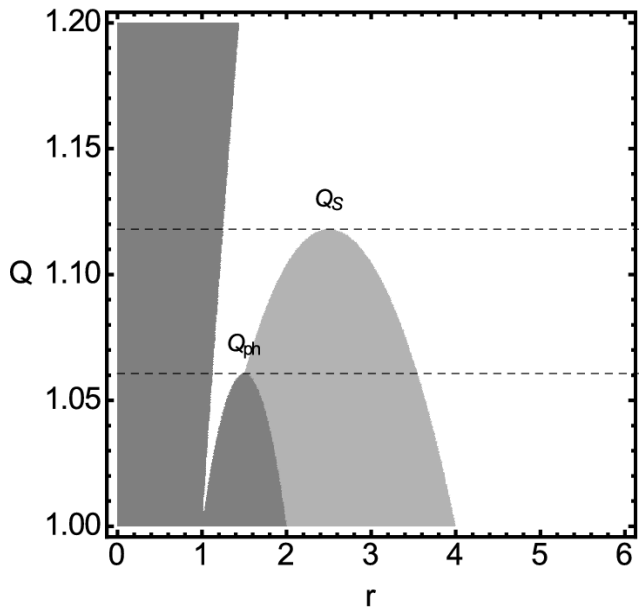}
		\caption{Structure of the circular geodesic orbits in the RN naked singularity spacetimes reflected in the $r-Q$ parameter space. There are three regions corresponding to no-circular orbits (dark gray), unstable circular orbits (gray), and stable circular orbits (white). The two characteristic points of the RN naked singularity spacetimes are given by $Q_{ph}=3/(2\sqrt{2})\doteq 1.061$ and $Q_{s}=\sqrt{5}/2=1.118$. The black line finishing at the point $Q_{ph}$ governs the radius where gradient of the angular velocity of the Keplerian orbits vanishes}
	\end{center}
\end{figure}
The structure is close similar to the KH naked singularity spacetimes of the modified quantum Horava gravity \cite{Vie-etal:2014:PHYSR4:,Stu-Sche-Abd:2014:PHYSR4:}, or to the regular Bardeen of ABG no-horizon spacetimes \cite{Stu-Sche:2015:IJMPD:}. 

\subsection{Keplerian disc images}

We construct Keplerian disc images formed in the vicinity of RN naked singularities. We use the same technique (methods and formulae) as for construction of images in the regular Bardeen spacetimes; we shall not repeat the details, presenting only the results of the calculations. \footnote{In the RN black hole spacetimes the photon motion and the gravitational lensing were studied in \cite{Stu-Cal:1991:GenRelGrav:,Eir-Rom-Tor:2002:PHYSR4:,Ser:2009:PHYSRL:,Eir-Sen:2014:EJPC:}. In similar braneworld black hole spacetimes the optical effects were studied in \cite{Sche-Stu:2008:IJMPD:,Sche-Stu:2008:GenRelGrav:,Nun:2010:PHYSR4:,Ama-Eir:2012:PHYSR4:}.} Up to the value of charge parameter $Q=Q_{s}\simeq 1.118$ the assumed Keplerian disc spans between $r_{in}=r_{ISCO}$ and $r_{out}=20$. For $Q>Q_{s}$ we put the inner edge of the disk to $r_{\Omega max}$. Note that in the RN spacetimes the photon motion can be expressed in terms of the standard elliptical integrals -- details can be found in \cite{Eir-Rom-Tor:2002:PHYSR4:}. Here we shall not repeat those details. 

\subsubsection{Primary images} 

Results of the construction of the standard and ghost primary (direct) images in the typical RN naked singularity spacetimes by the ray-tracing method are presented in Figures 23-25. 

\begin{figure}[H]
	\begin{center}
		\begin{tabular}{cc}
			\includegraphics[scale=0.25]{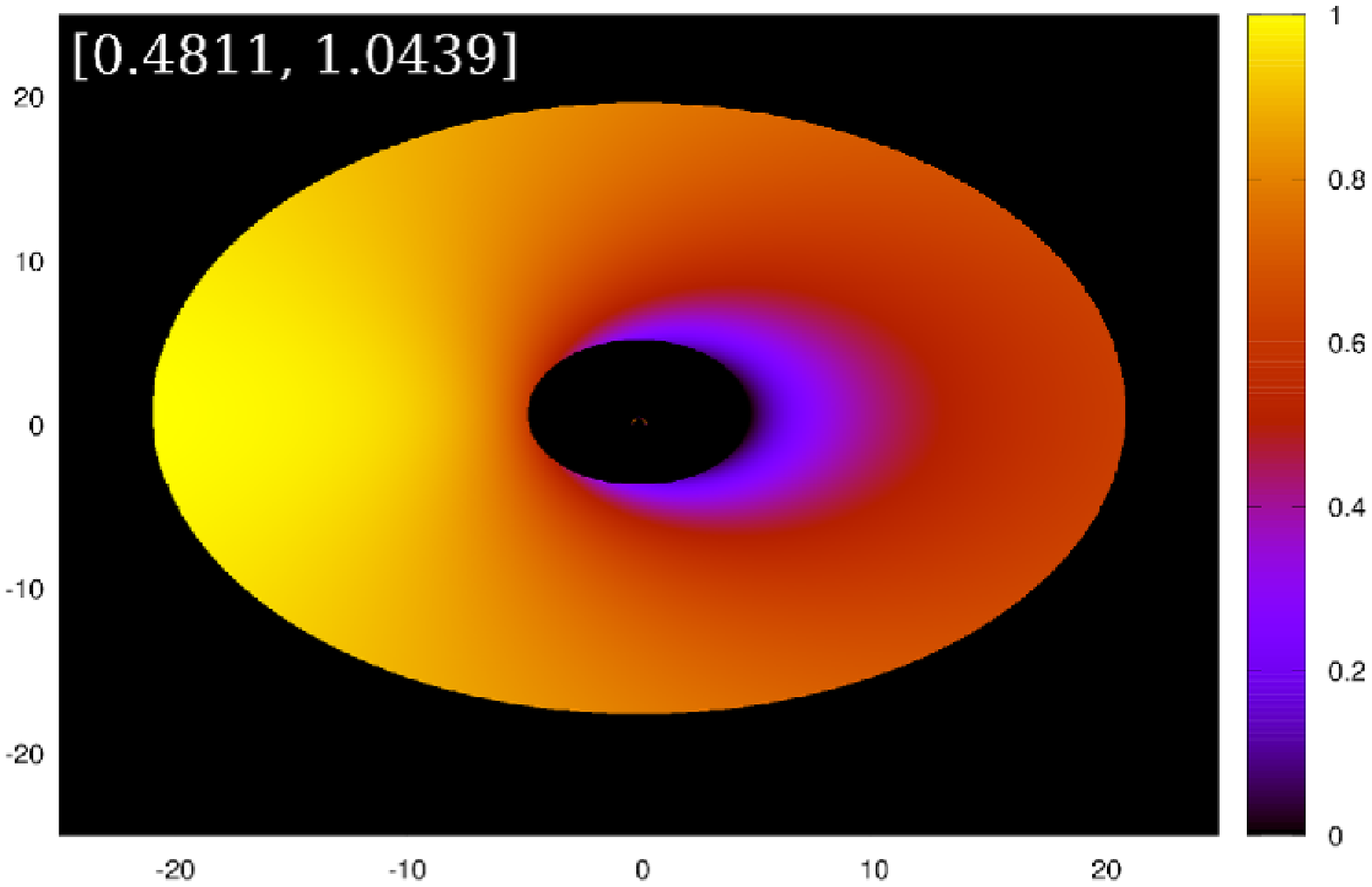}&\includegraphics[scale=0.25]{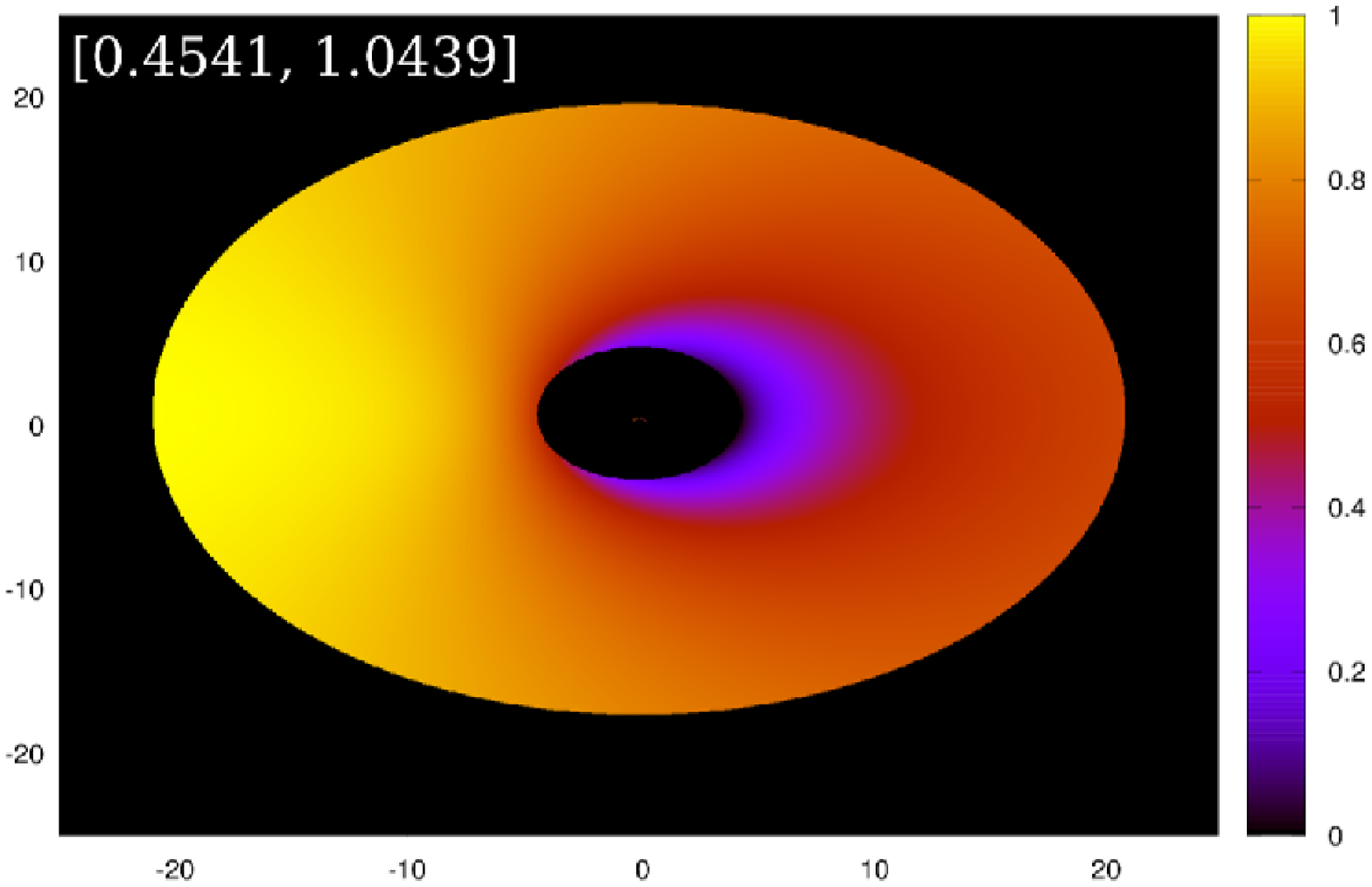}\\
			\includegraphics[scale=0.25]{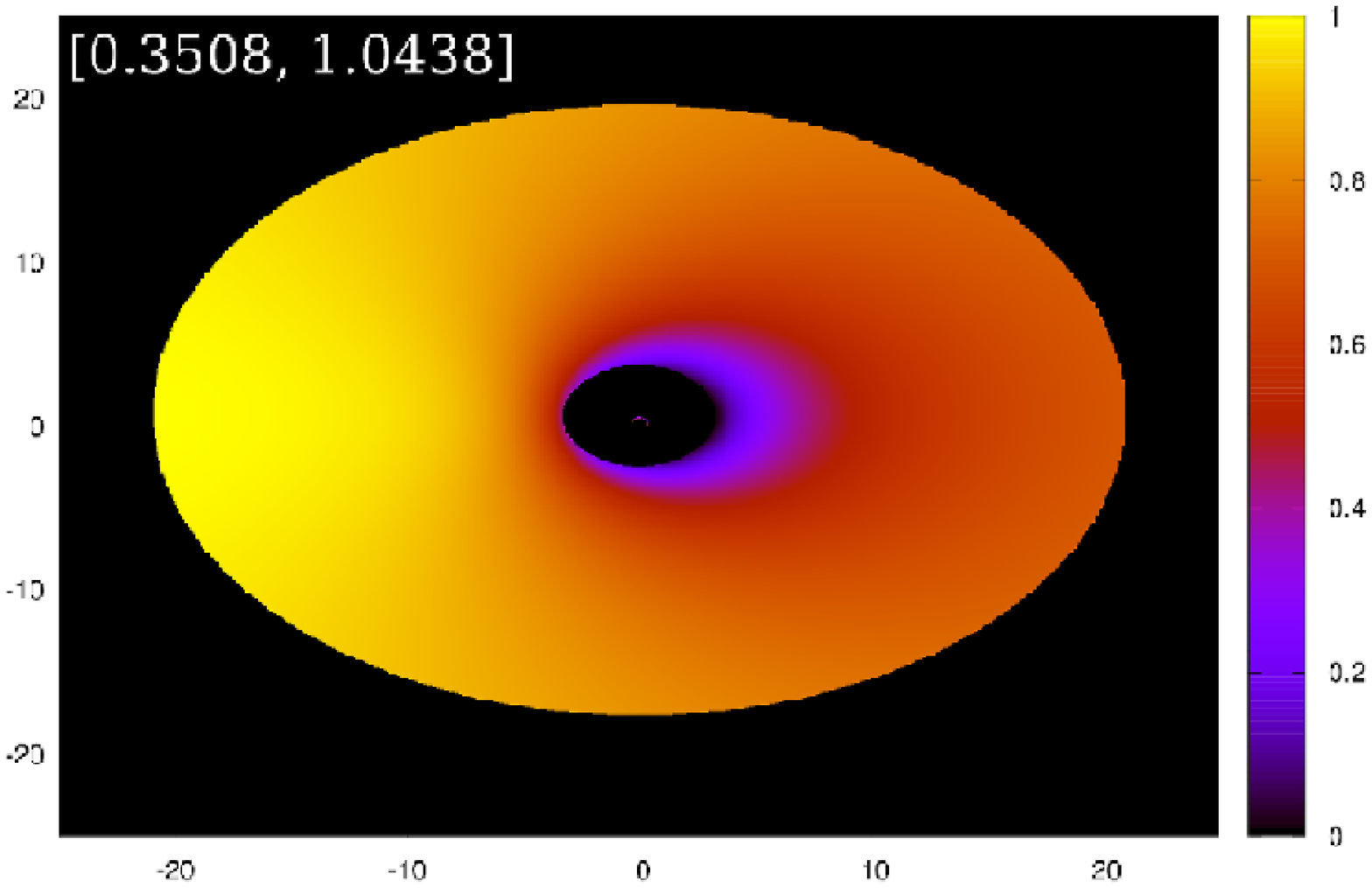}&\includegraphics[scale=0.25]{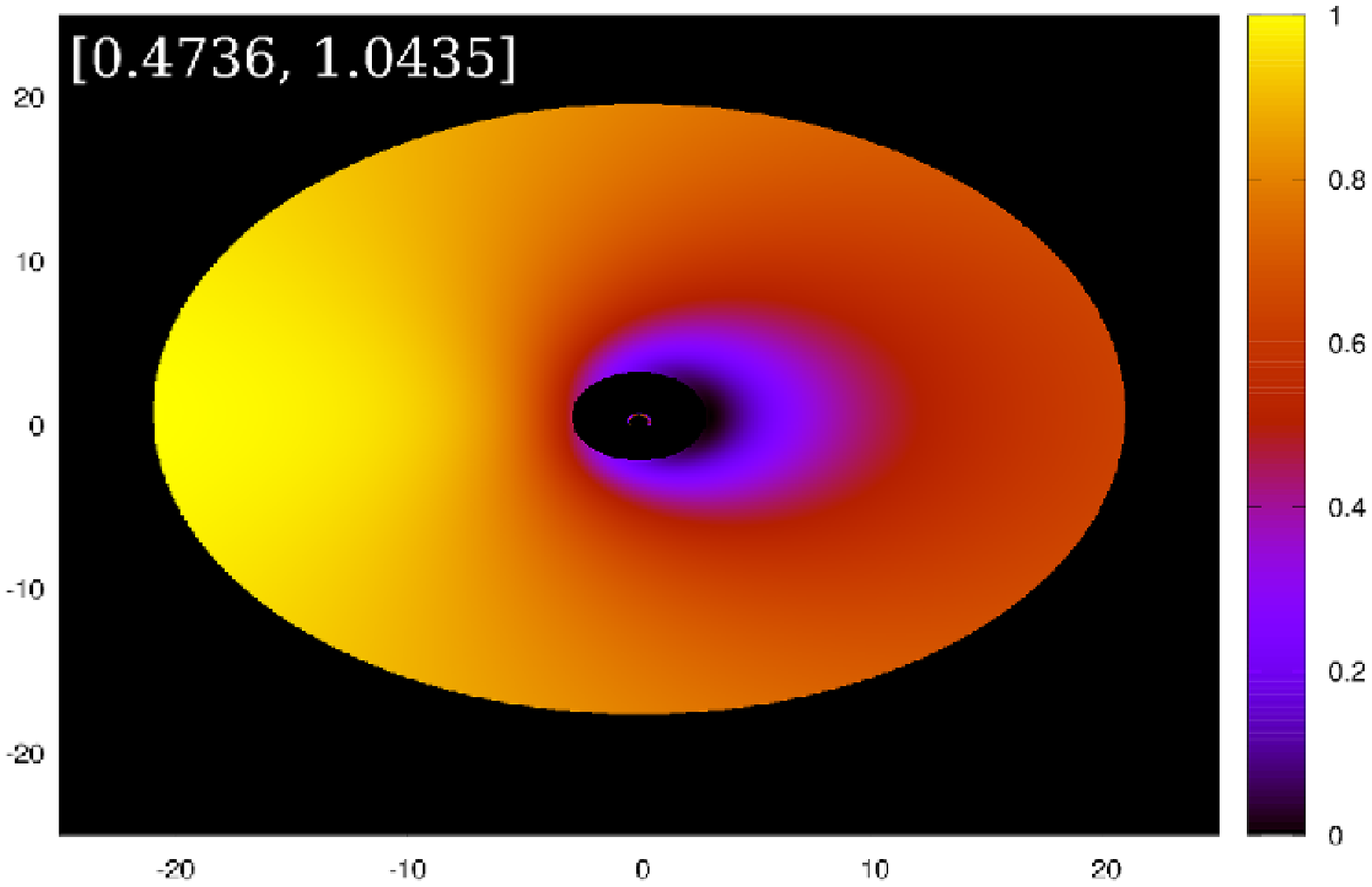}	
		\end{tabular}
		\caption{Frequency shift map of the Keplerian disc \emph{primary} image in the vicinity of RN naked singularity determined by charge parameter $Q=1.01$, $3/(2\sqrt{2})$, $1.118$, and $1.3$. The observer inclination angle is $30^\circ$.}
	\end{center}	
\end{figure}

\begin{figure}[H]
	\begin{center}
		\begin{tabular}{cc}
			\includegraphics[scale=0.1]{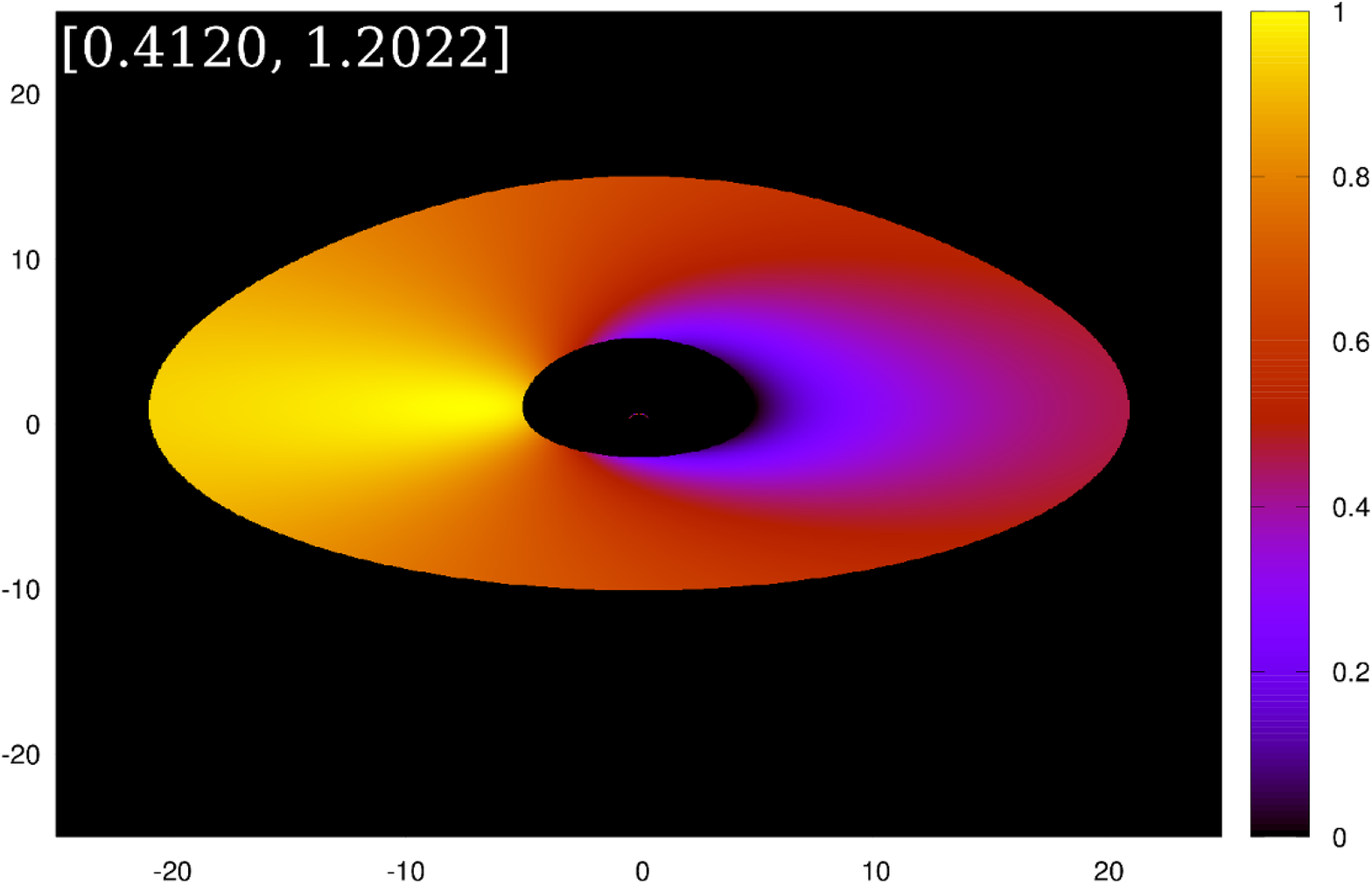}&\includegraphics[scale=0.1]{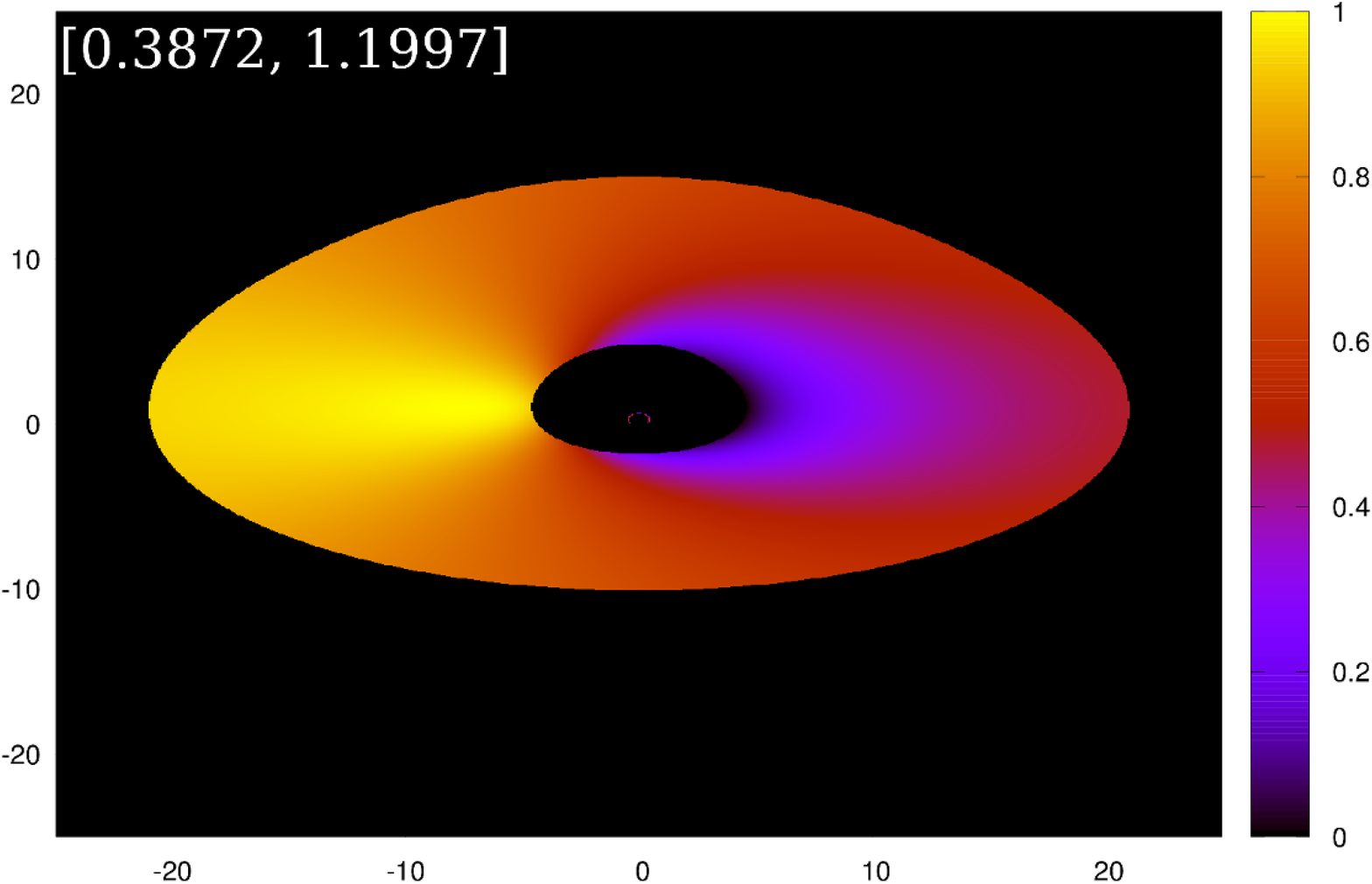}\\
			\includegraphics[scale=0.15]{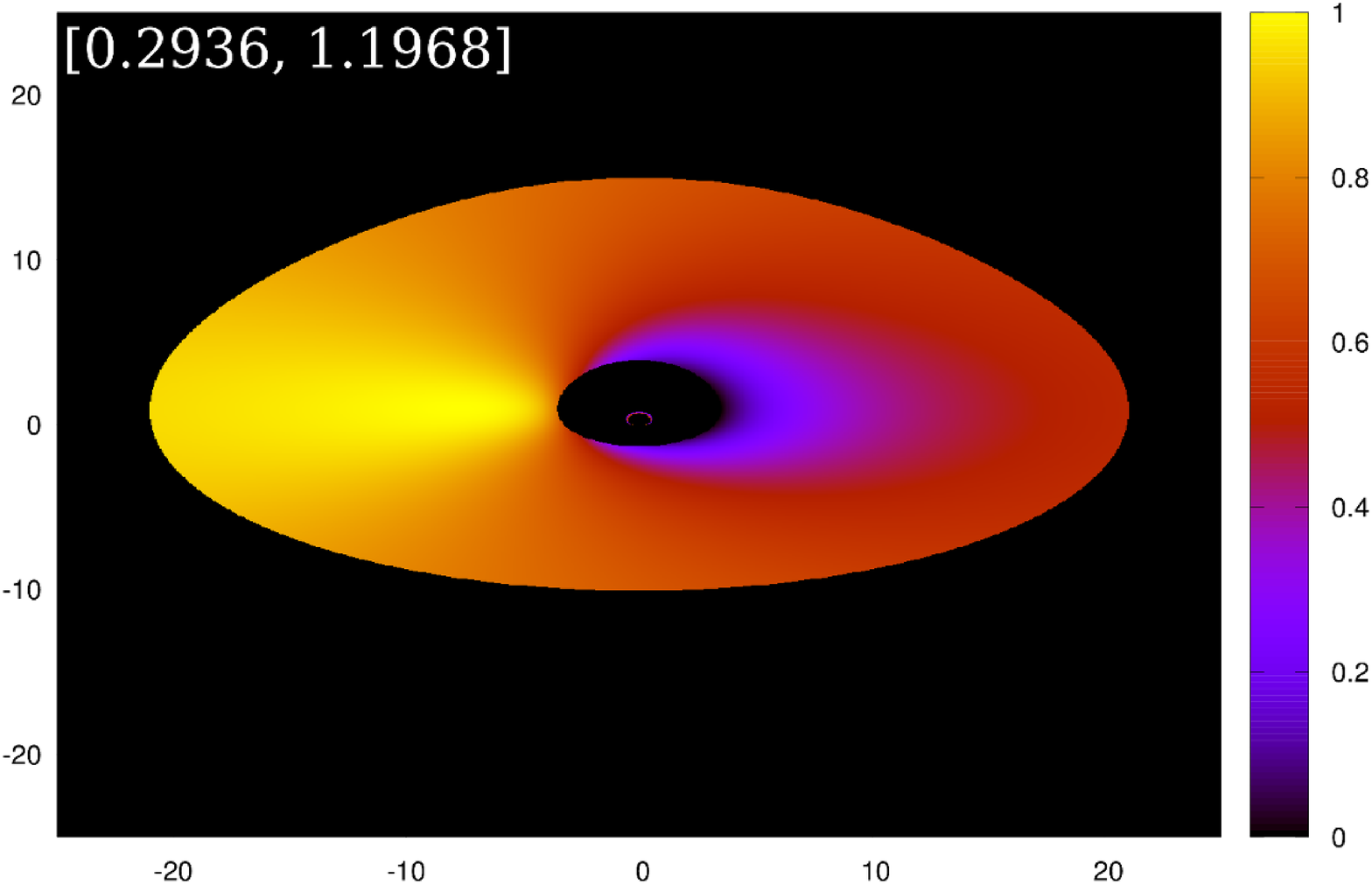}&\includegraphics[scale=0.3]{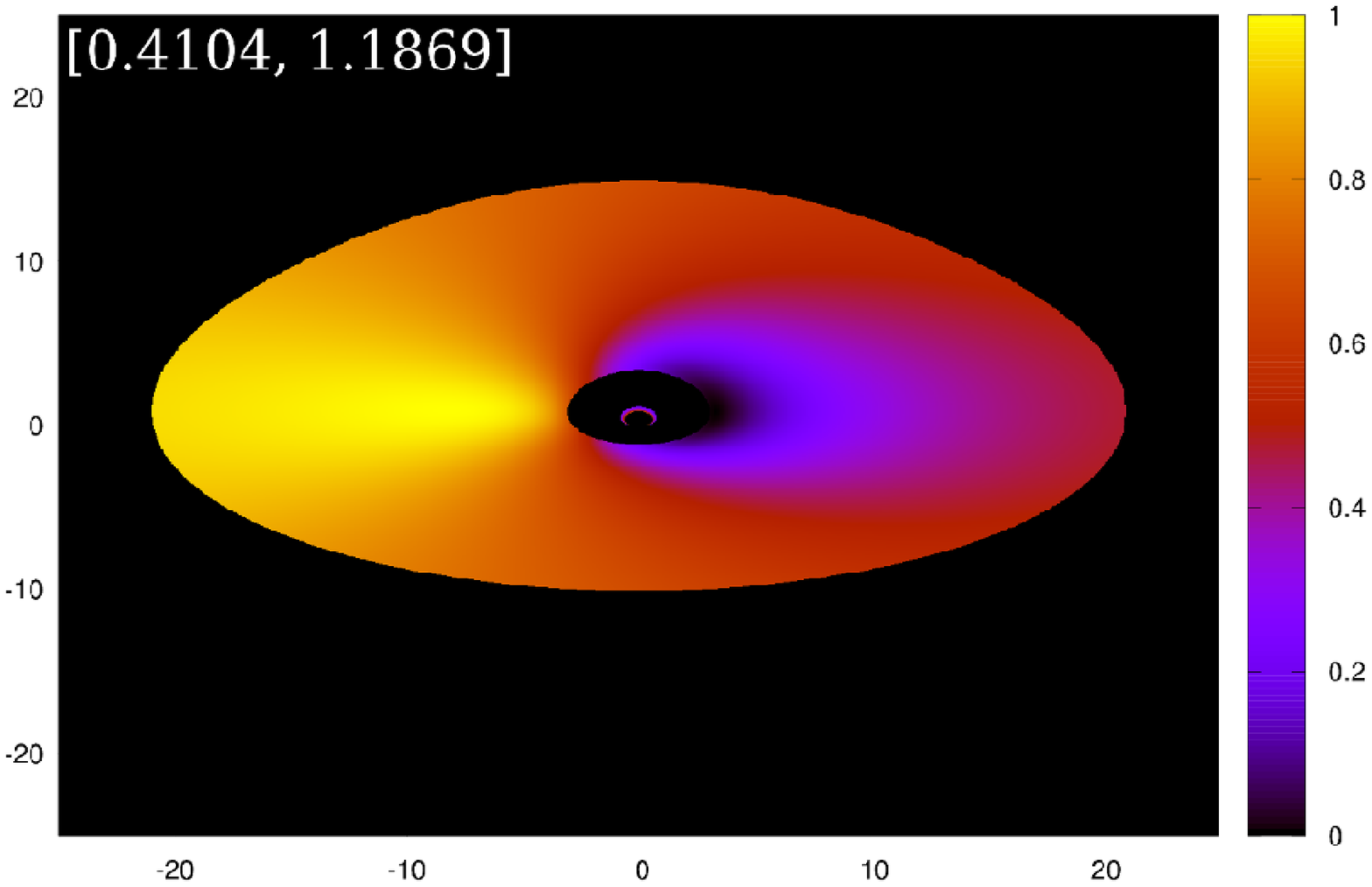}	
		\end{tabular}
		\caption{Frequency shift map of the Keplerian disc \emph{primary} image in the vicinity of RN naked singularity determined by charge parameter $Q=1.01$, $3/(2\sqrt{2})$, $1.118$, and $1.3$. The observer inclination angle is $60^\circ$.}
	\end{center}	
\end{figure}

\begin{figure}[H]
	\begin{center}
		\begin{tabular}{cc}
			\includegraphics[scale=0.15]{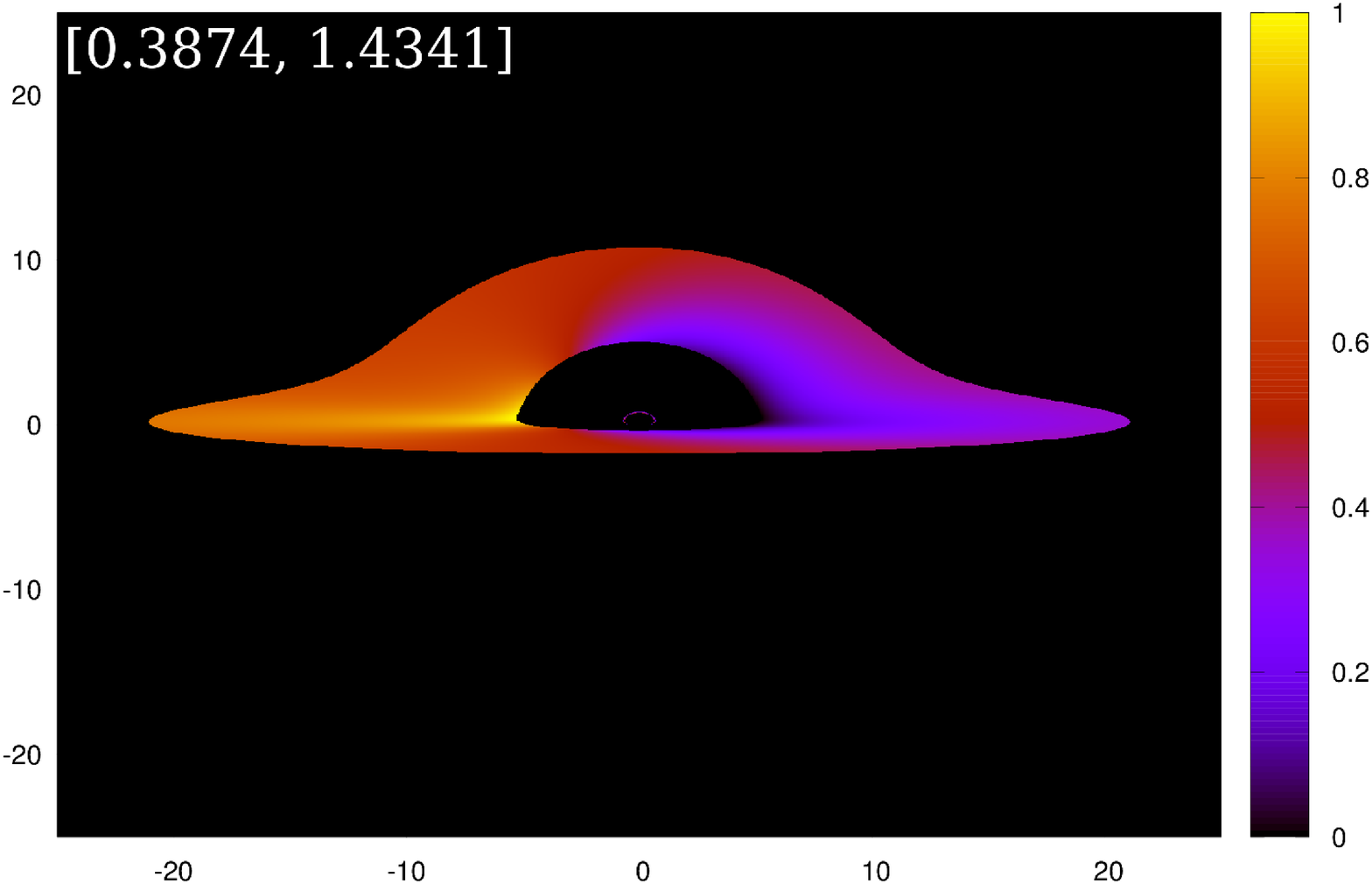}&\includegraphics[scale=0.1]{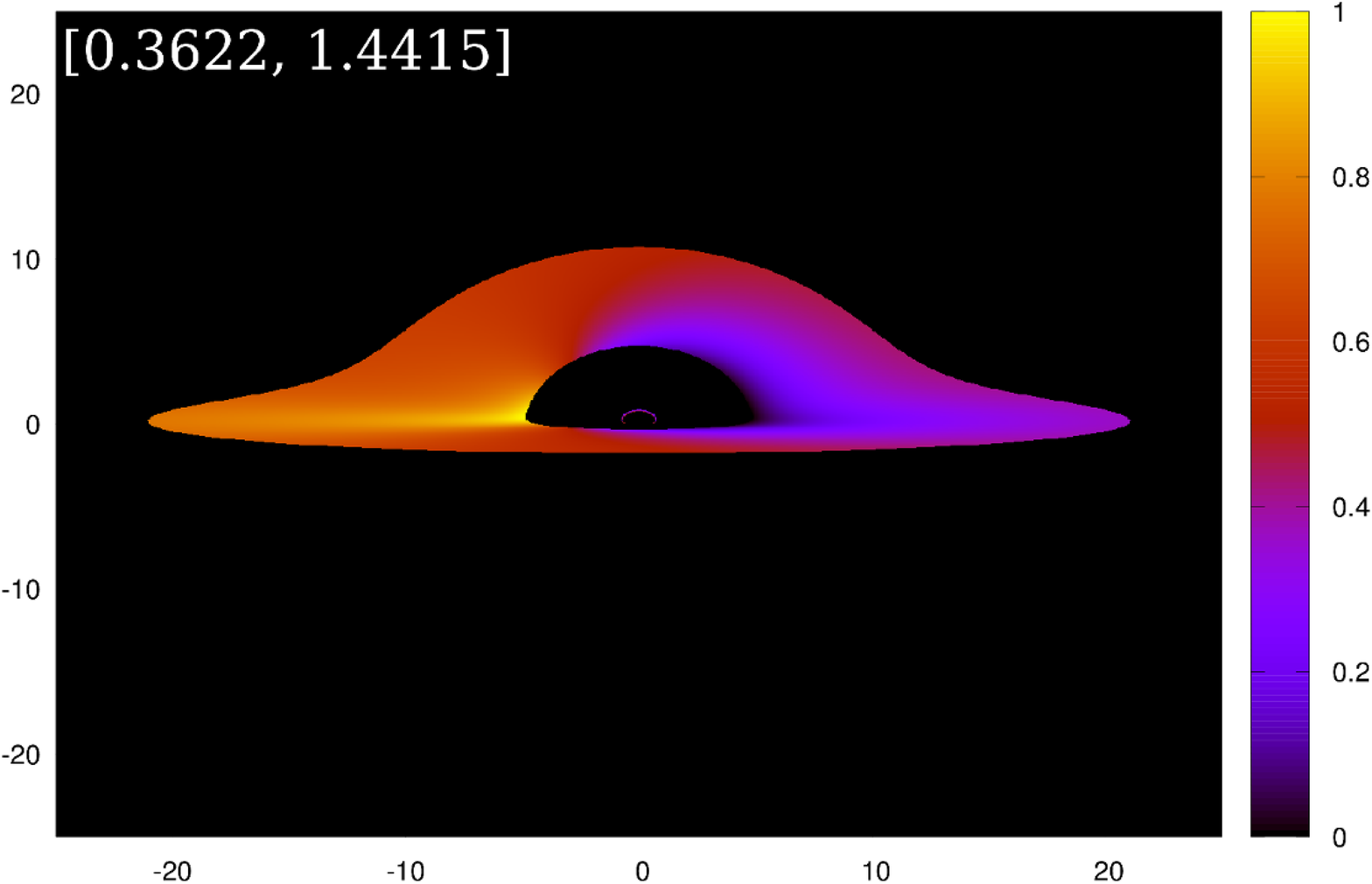}\\
			\includegraphics[scale=0.15]{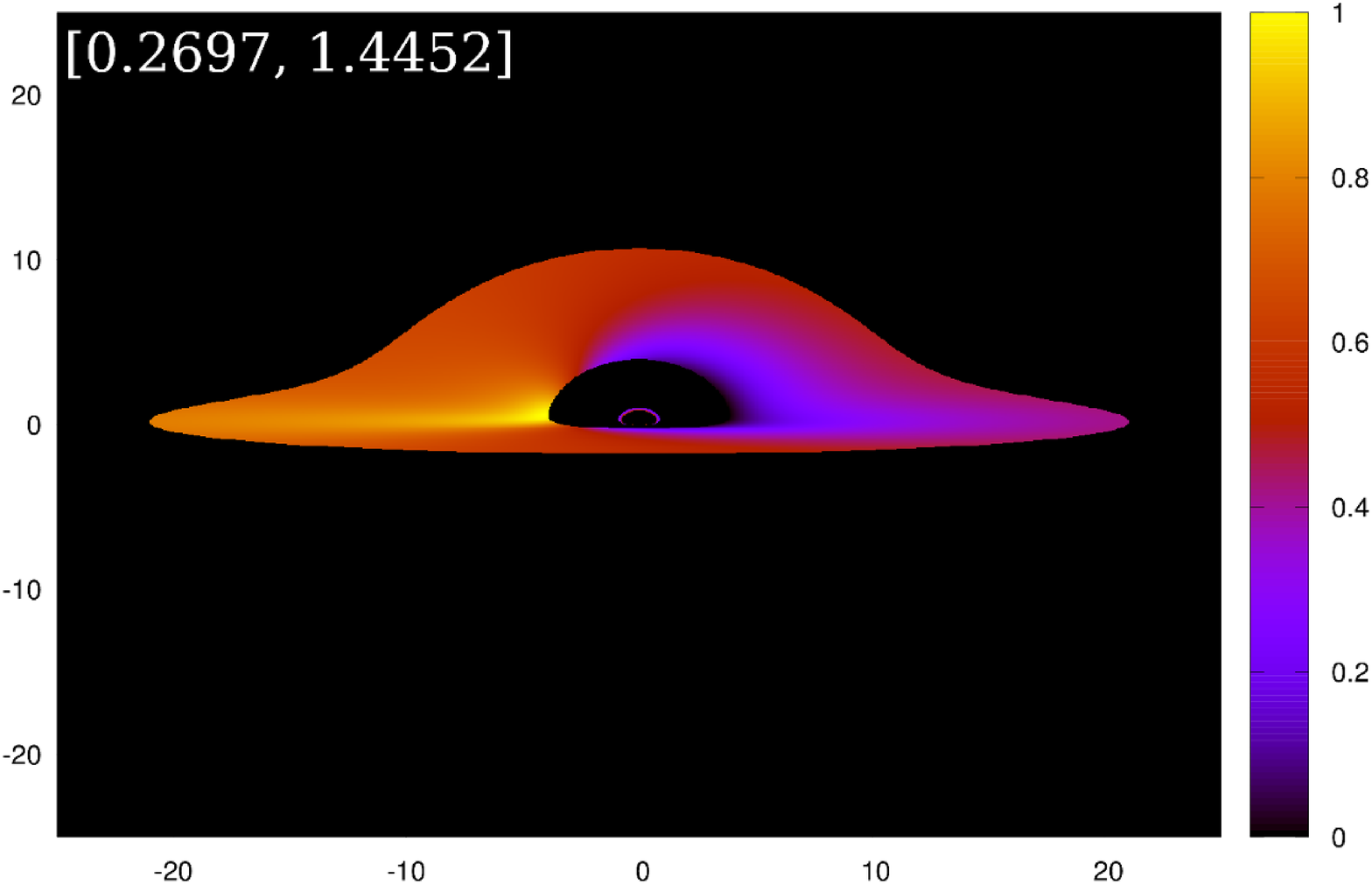}&\includegraphics[scale=0.1]{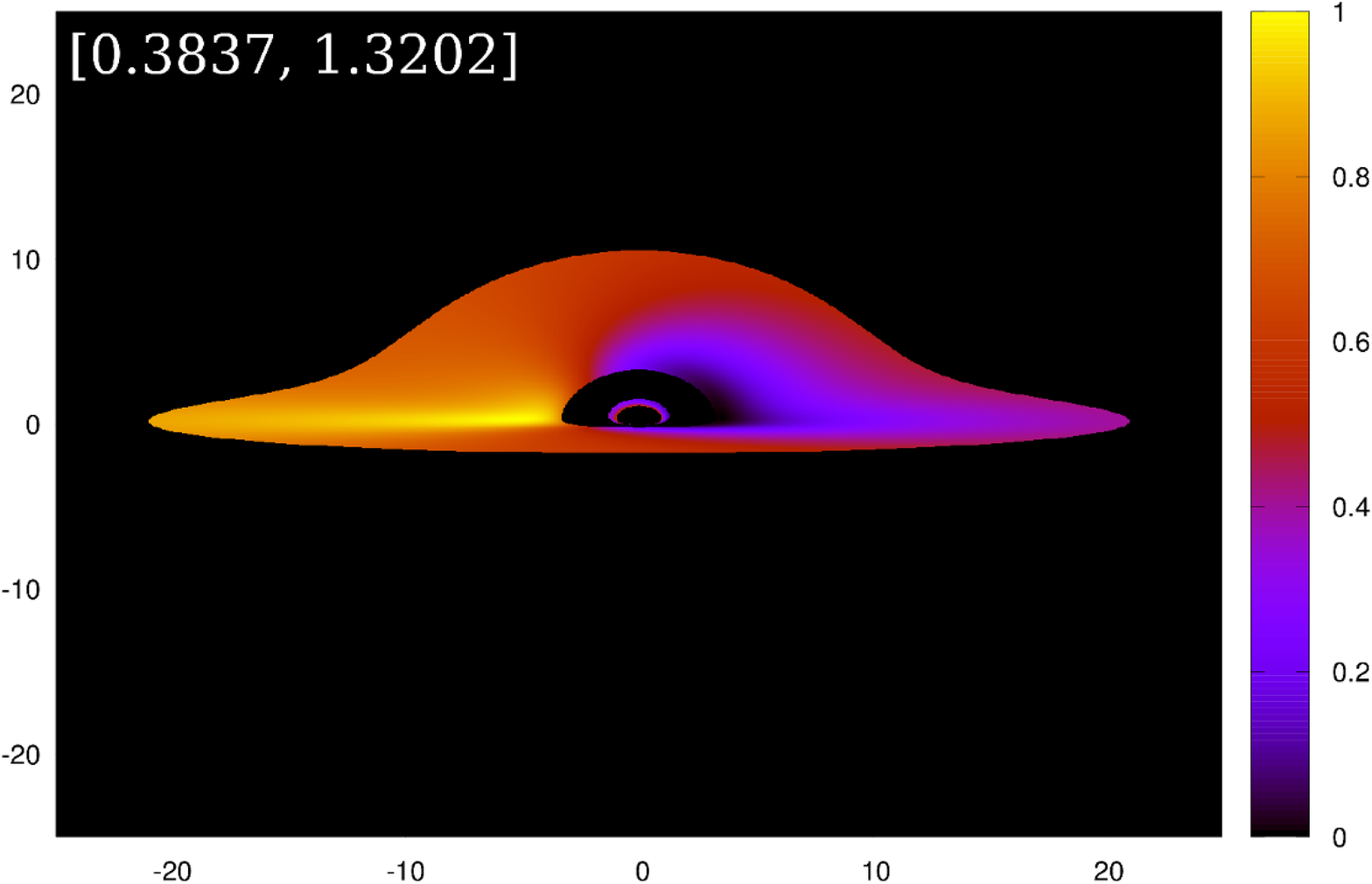}	
		\end{tabular}
		\caption{Frequency shift map of the Keplerian disc \emph{primary} image in the vicinity of RN naked singularity determined by charge parameter $Q=1.01$, $3/(2\sqrt{2})$, $1.118$, and $1.3$. The observer inclination angle is $85^\circ$.}
	\end{center}	
\end{figure}

We can see that the ghost images occur at all the considered inclination angles and all the considered RN naked singularity spacetimes. The direct ghost images increase with increasing charge parameter $Q$, being created by the low impact parameter photons coming from whole the Keplerian disc. These photons are reflected by the repulsive barrier at $r \sim 0$. Details of this phenomenon will be discussed later. Here we can see a clear qualitative difference in comparison to the direct ghost images created in the Bardeen no-horizon spacetimes. On the other hand, the direct ghost images in the RN naked singularity spacetimes are similar to those occuring in the Kerr naked singularity spacetimes, having also the same origin \cite{Stu-Sche:2010:CLAQG:}. 

The range of the frequency shift of the direct images of the Keplerian discs orbiting in the RN naked singularity spacetimes demonstrates similar behavior as those created in the Bardeen no-horizon spacetimes. 

\subsubsection{Secondary images}

Results of the construction of the standard and ghost secondary (indirect) images in the typical RN naked singularity spacetimes allowing for existence of the secondary images are presented in Figure 26-28. 

The dependence of the frequency range ($z_{max}$ -- $z_{min}$) of the secondary images of the Keplerian discs on the inclination angle and the charge parameter of the RN naked singularity spacetimes has again similar properties as the secondary images generated in the Bardeen no-horizon spacetimes. 
\begin{figure}[H]
	\begin{center}
		\begin{tabular}{cc}
			\includegraphics[scale=0.1]{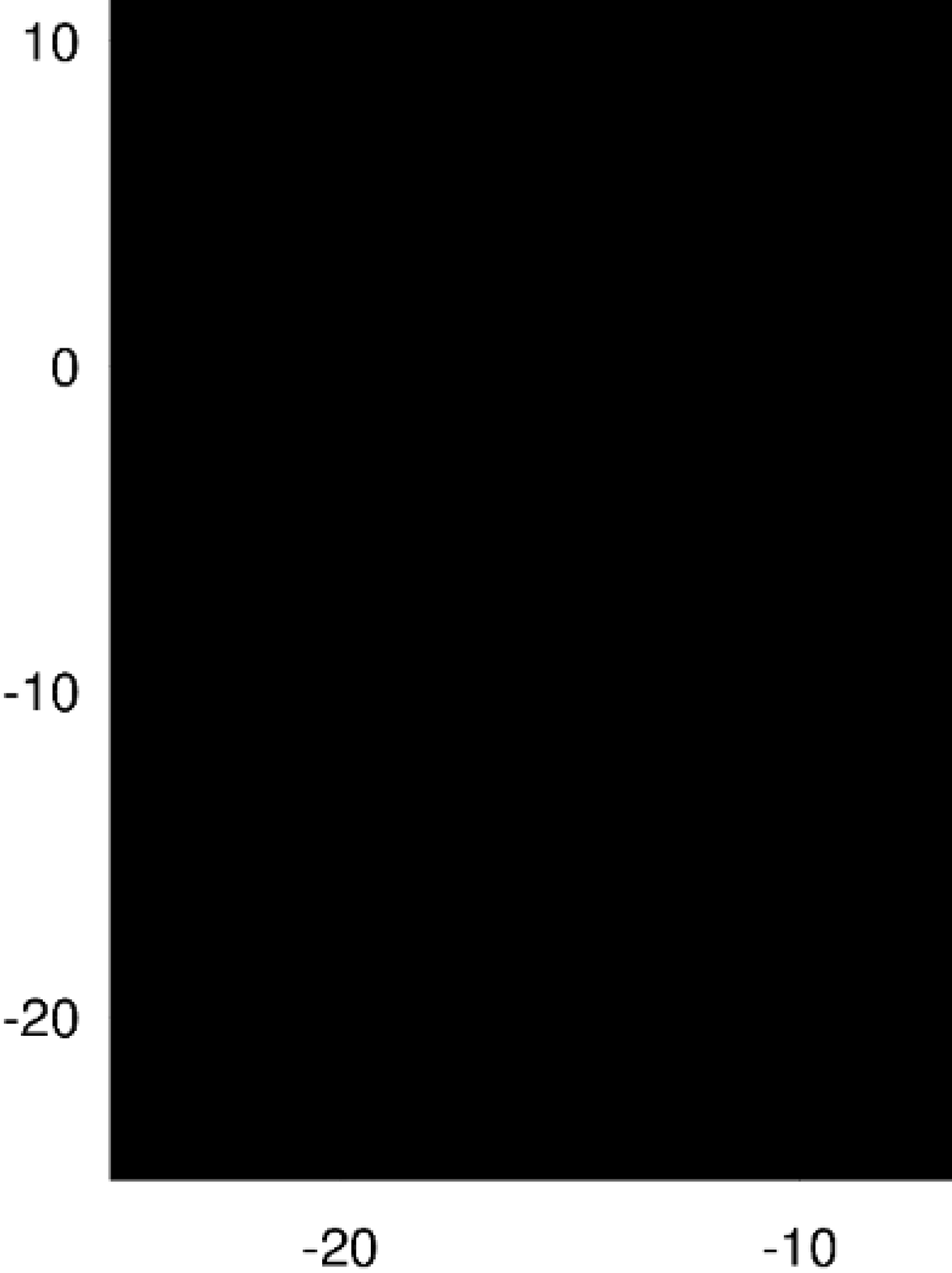}&\includegraphics[scale=0.1]{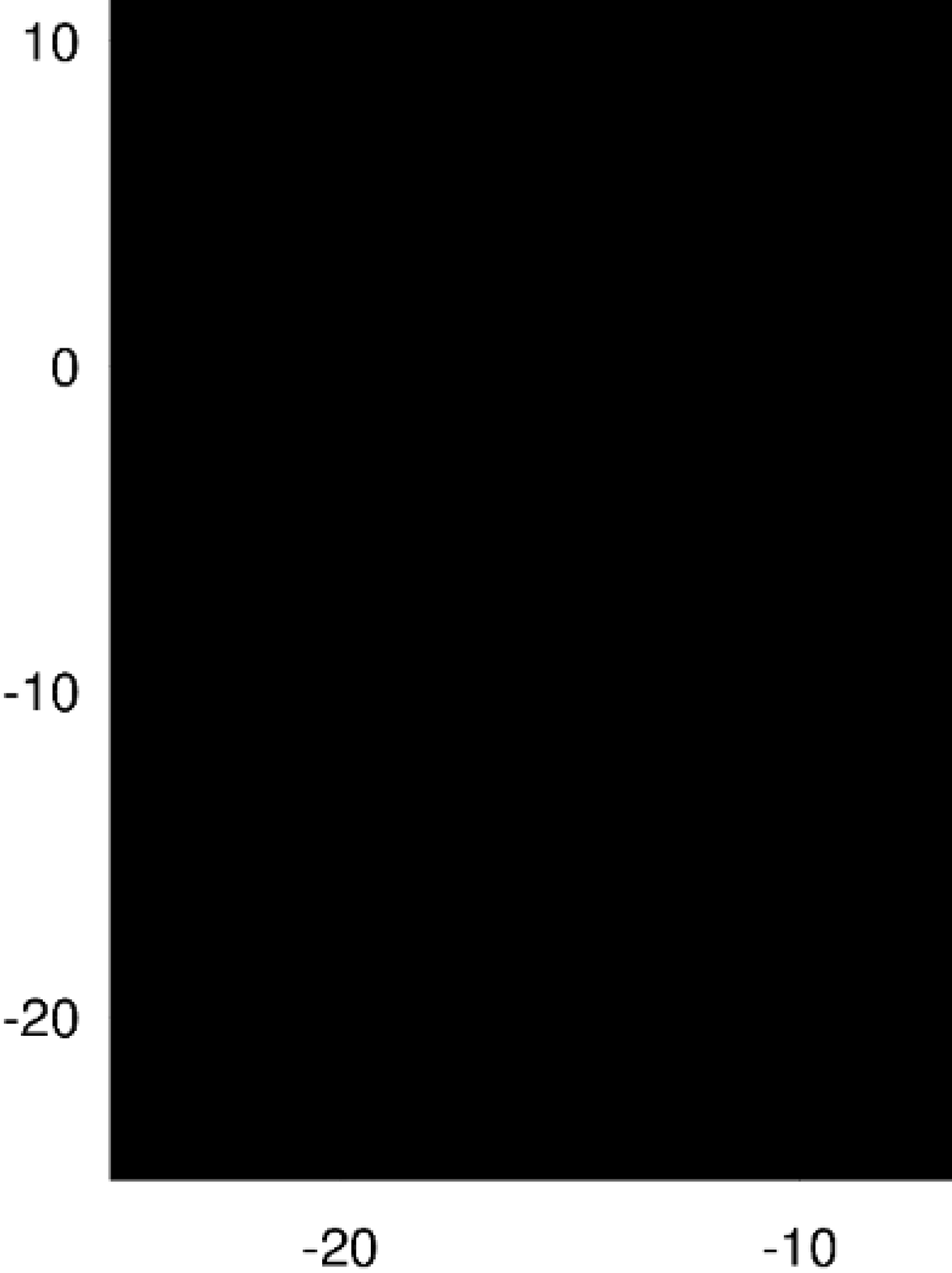}\\
			\includegraphics[scale=0.1]{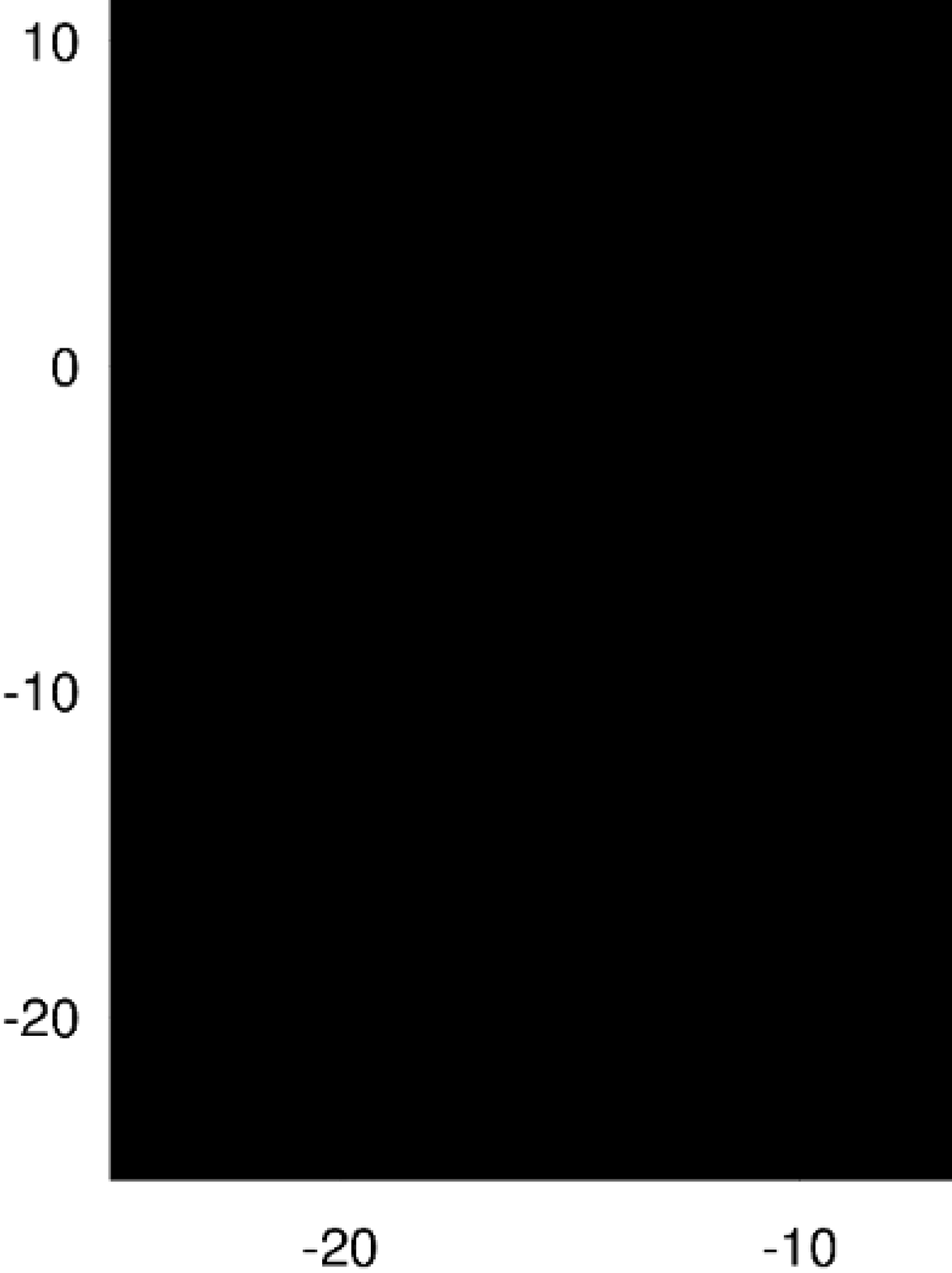}&\includegraphics[scale=0.1]{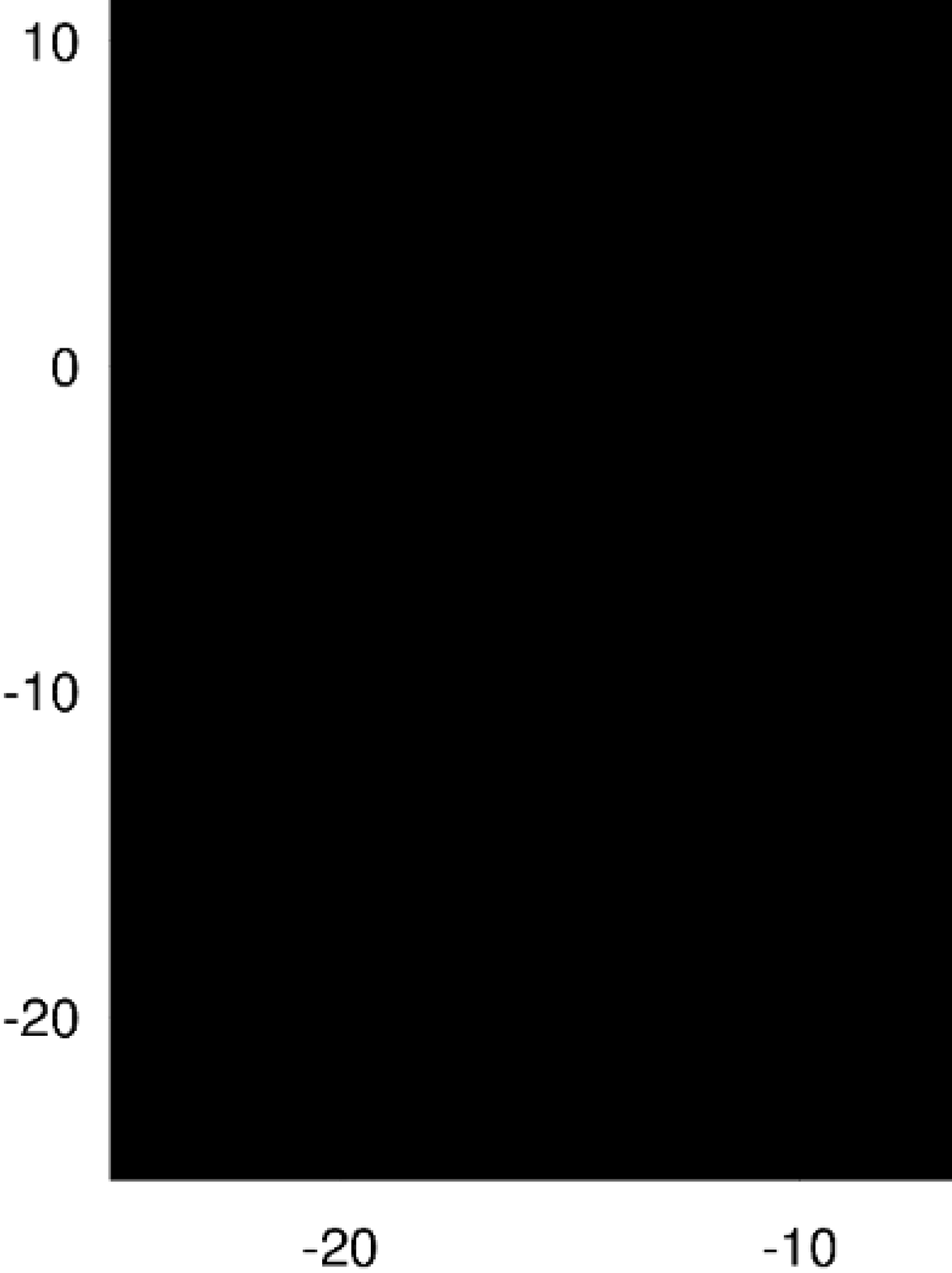}	
		\end{tabular}
		\caption{Frequency shift map of the Keplerian disc \emph{secondary} image in the vicinity of RN naked singularity determined by charge parameter $Q=1.01$, $3/(2\sqrt{2})$, $1.118$, and $1.3$. The observer inclination angle is $30^\circ$.}
	\end{center}	
\end{figure}

\begin{figure}[H]
	\begin{center}
		\begin{tabular}{cc}
			\includegraphics[scale=0.1]{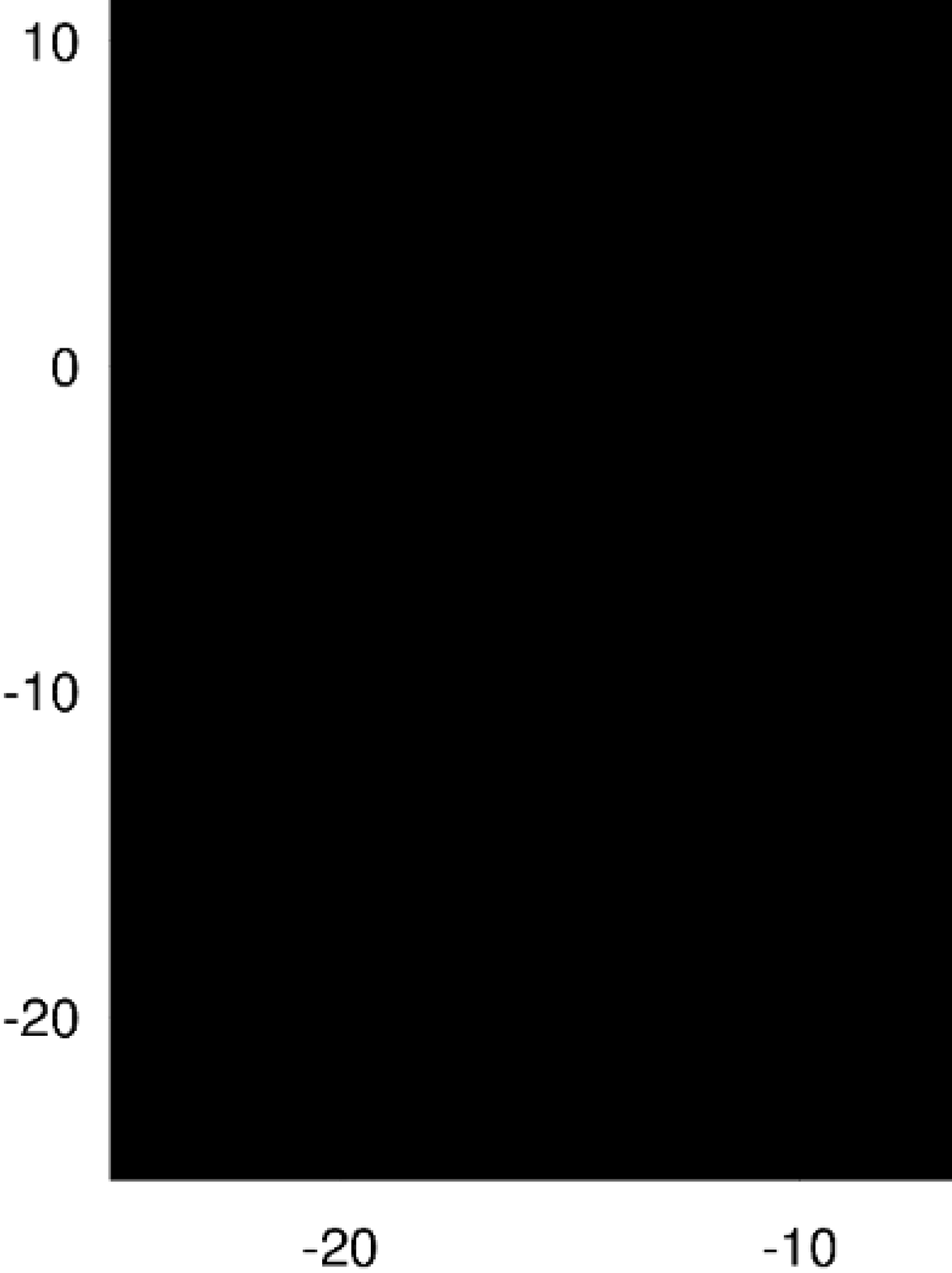}&\includegraphics[scale=0.1]{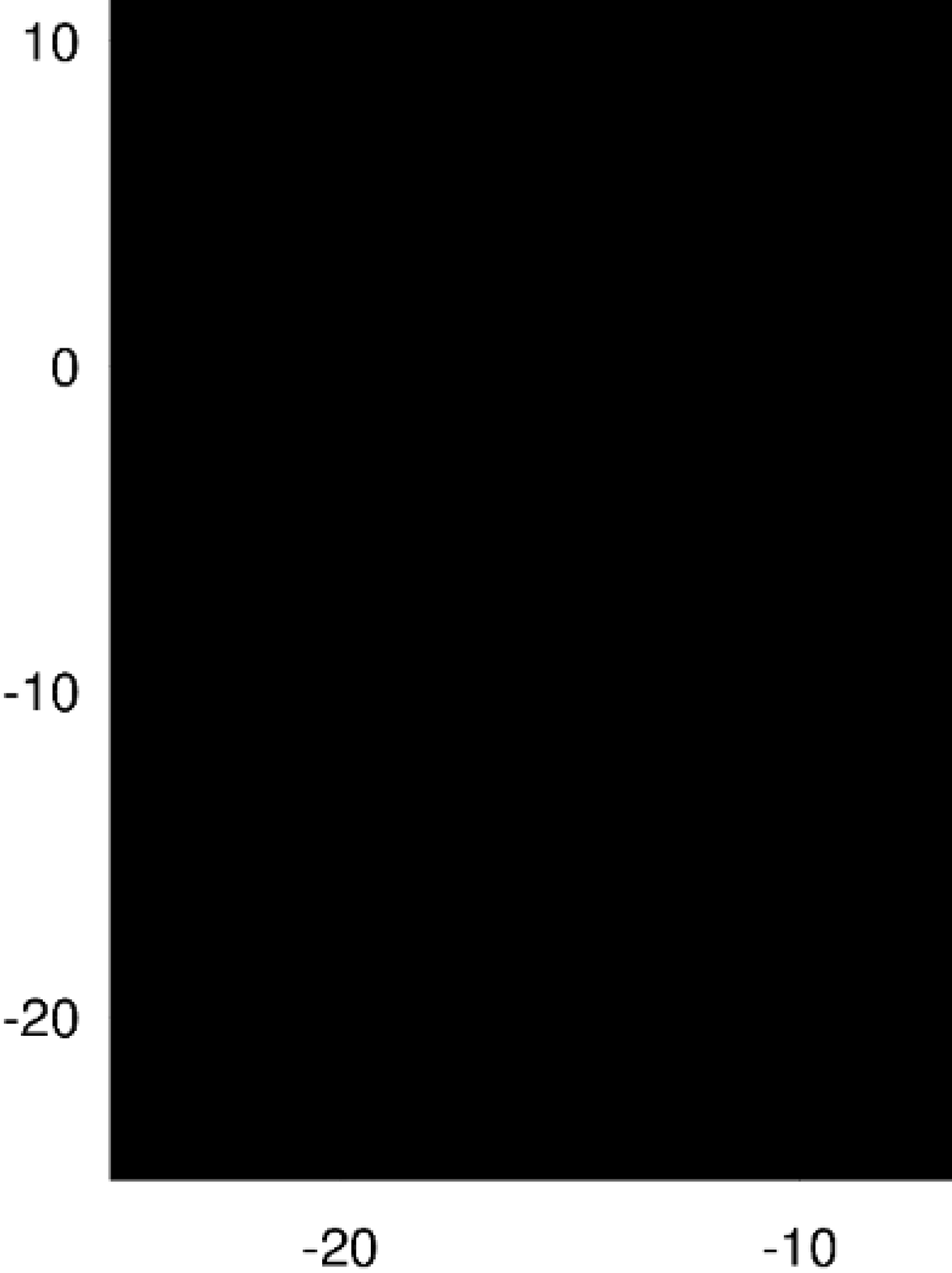}\\
			\includegraphics[scale=0.1]{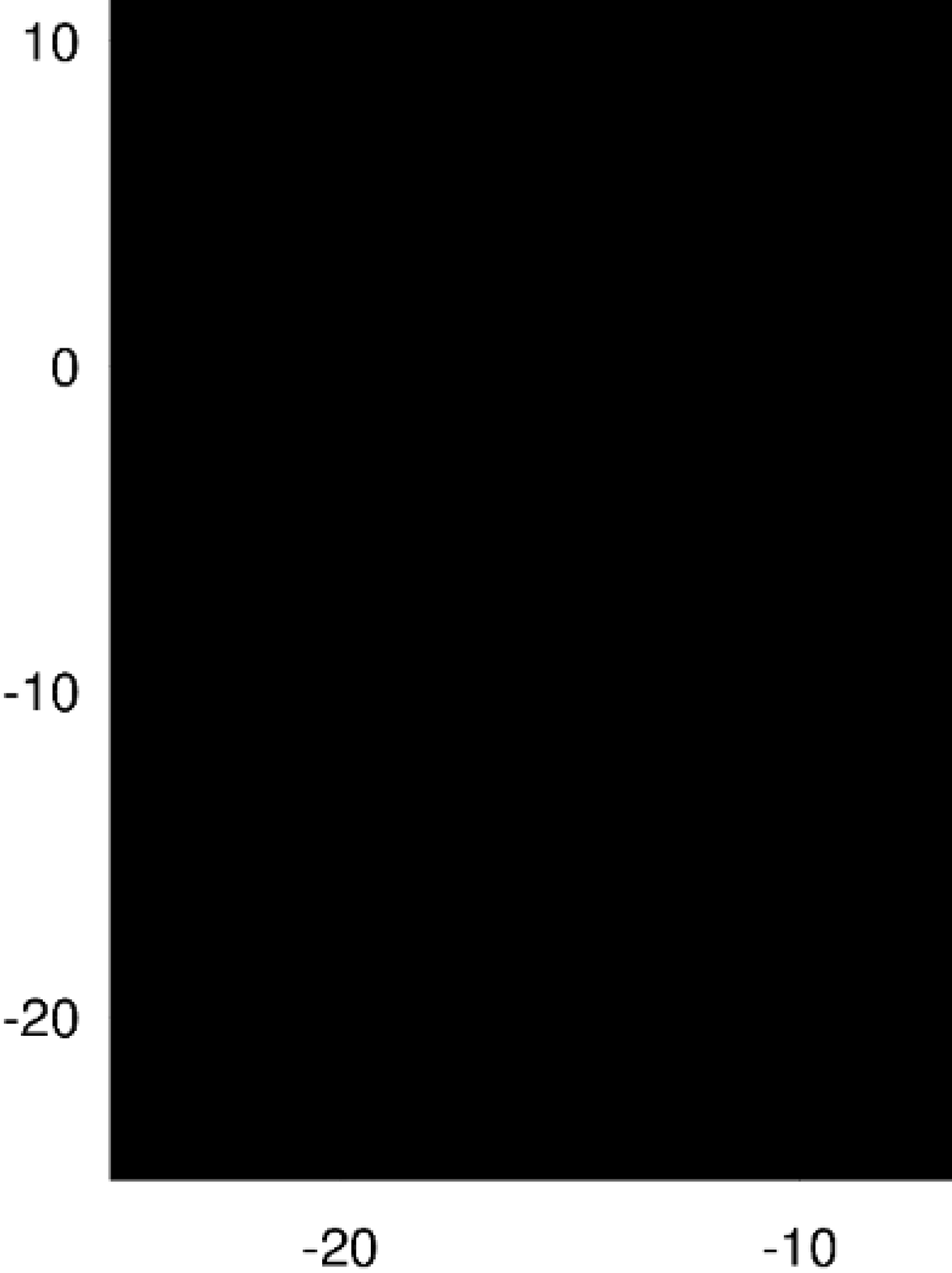}&\includegraphics[scale=0.1]{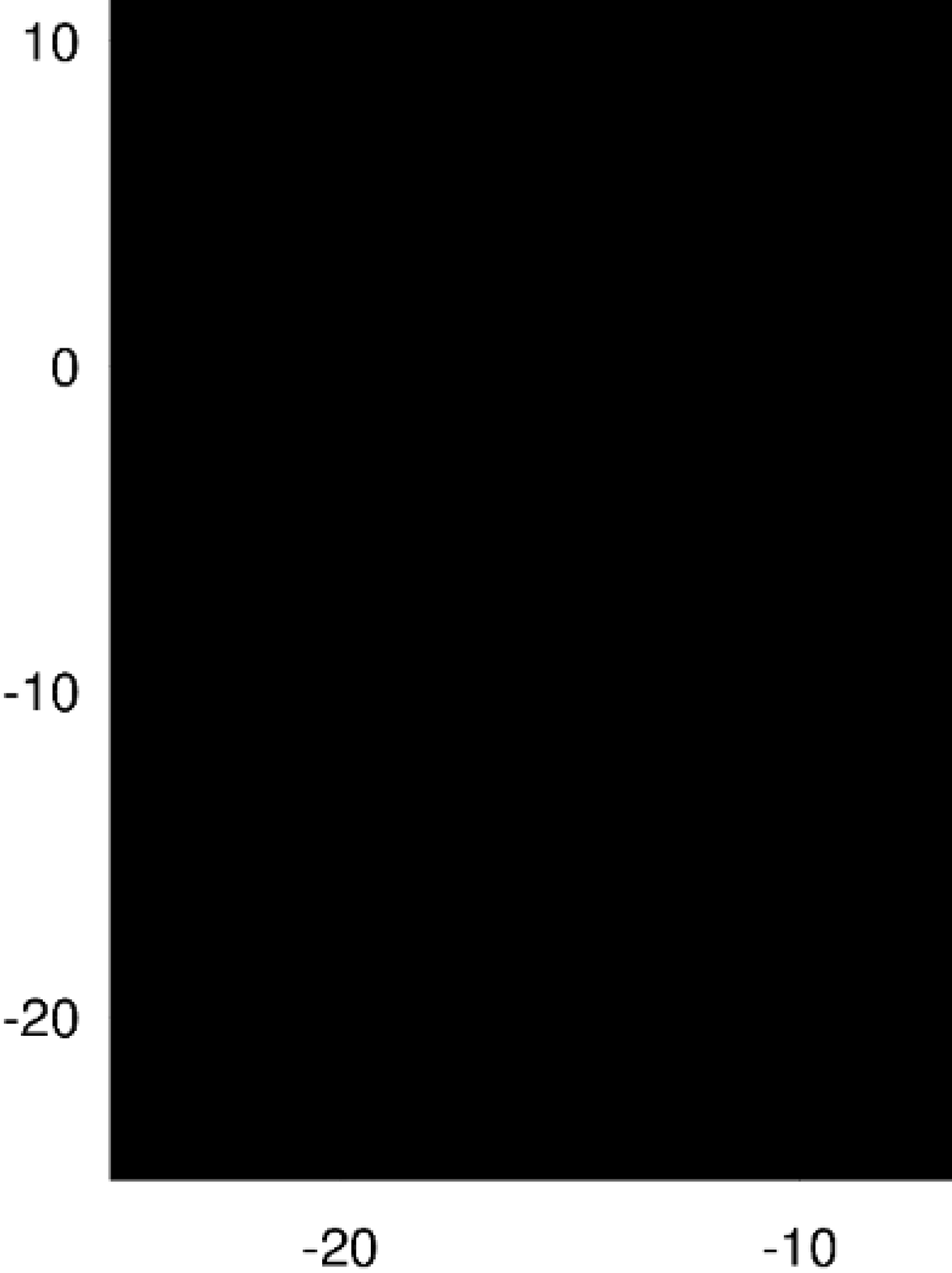}	
		\end{tabular}
		\caption{Frequency shift map of the Keplerian disc \emph{secondary} image in the vicinity of RN naked singularity determined by charge parameter $Q=1.01$, $3/(2\sqrt{2})$, $1.118$, and $1.3$. The observer inclination angle is $60^\circ$.}
	\end{center}	
\end{figure}

\begin{figure}[H]
	\begin{center}
		\begin{tabular}{cc}
			\includegraphics[scale=0.1]{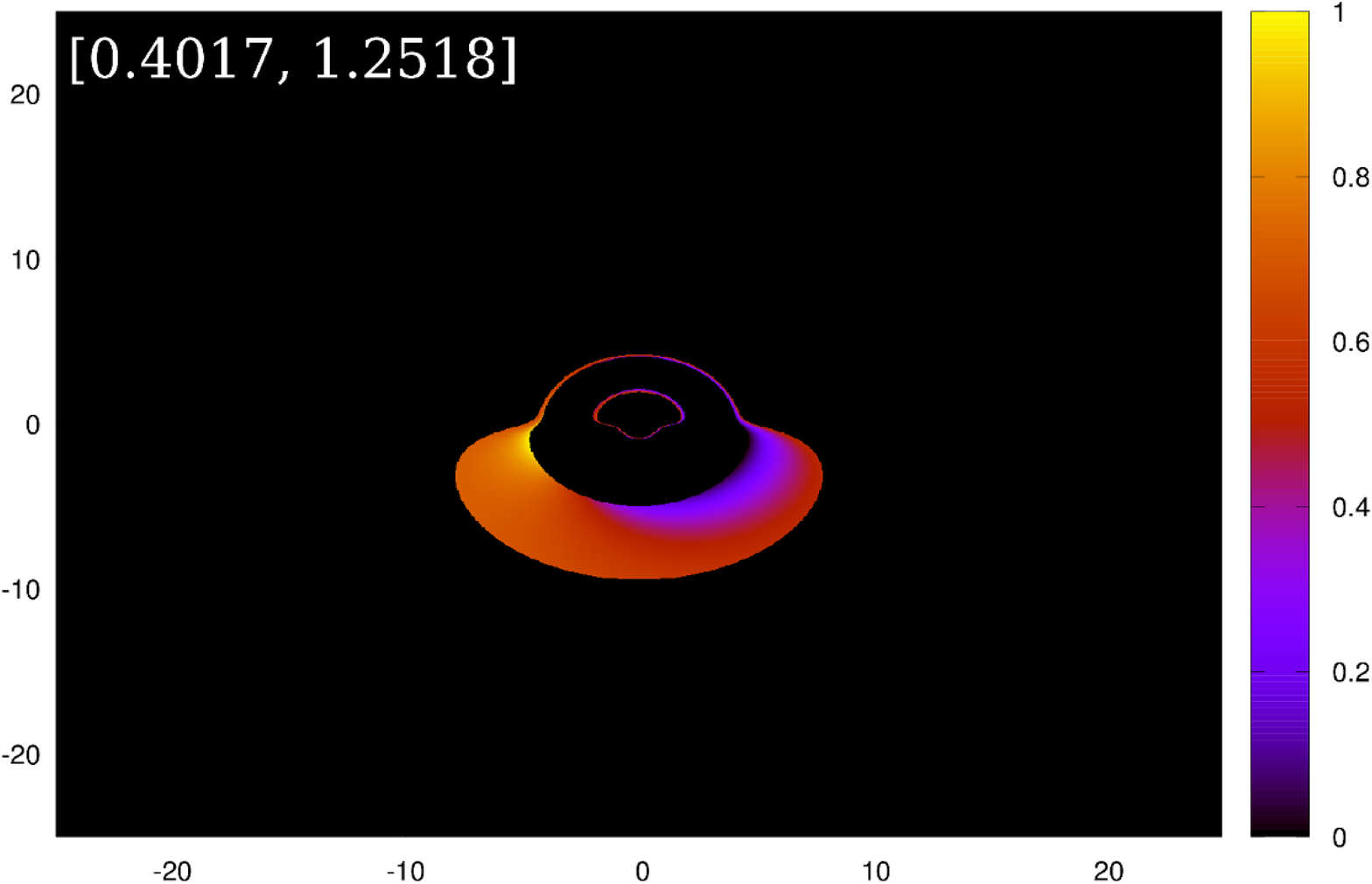}&\includegraphics[scale=0.2]{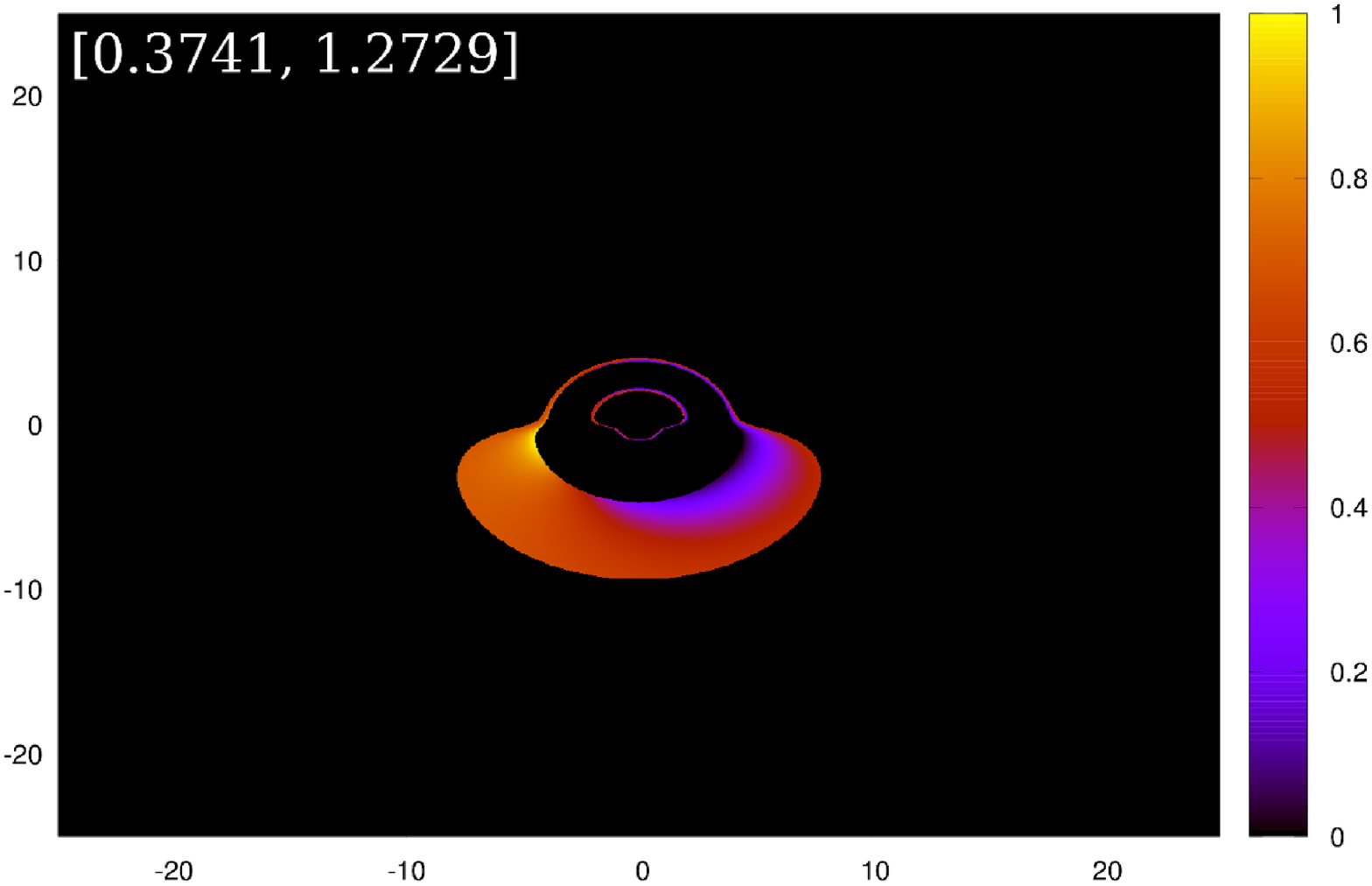}\\
			\includegraphics[scale=0.1]{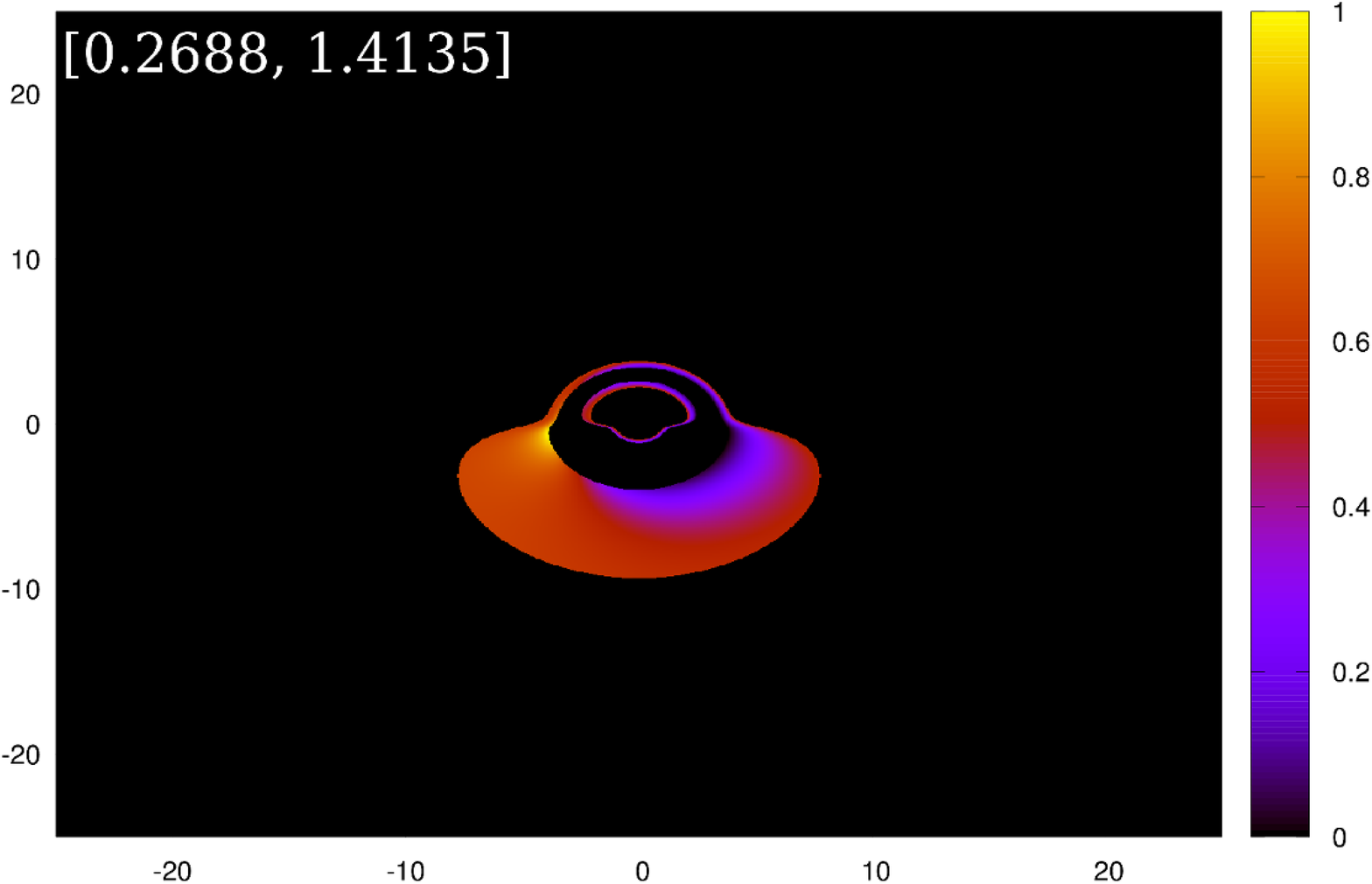}&\includegraphics[scale=0.1]{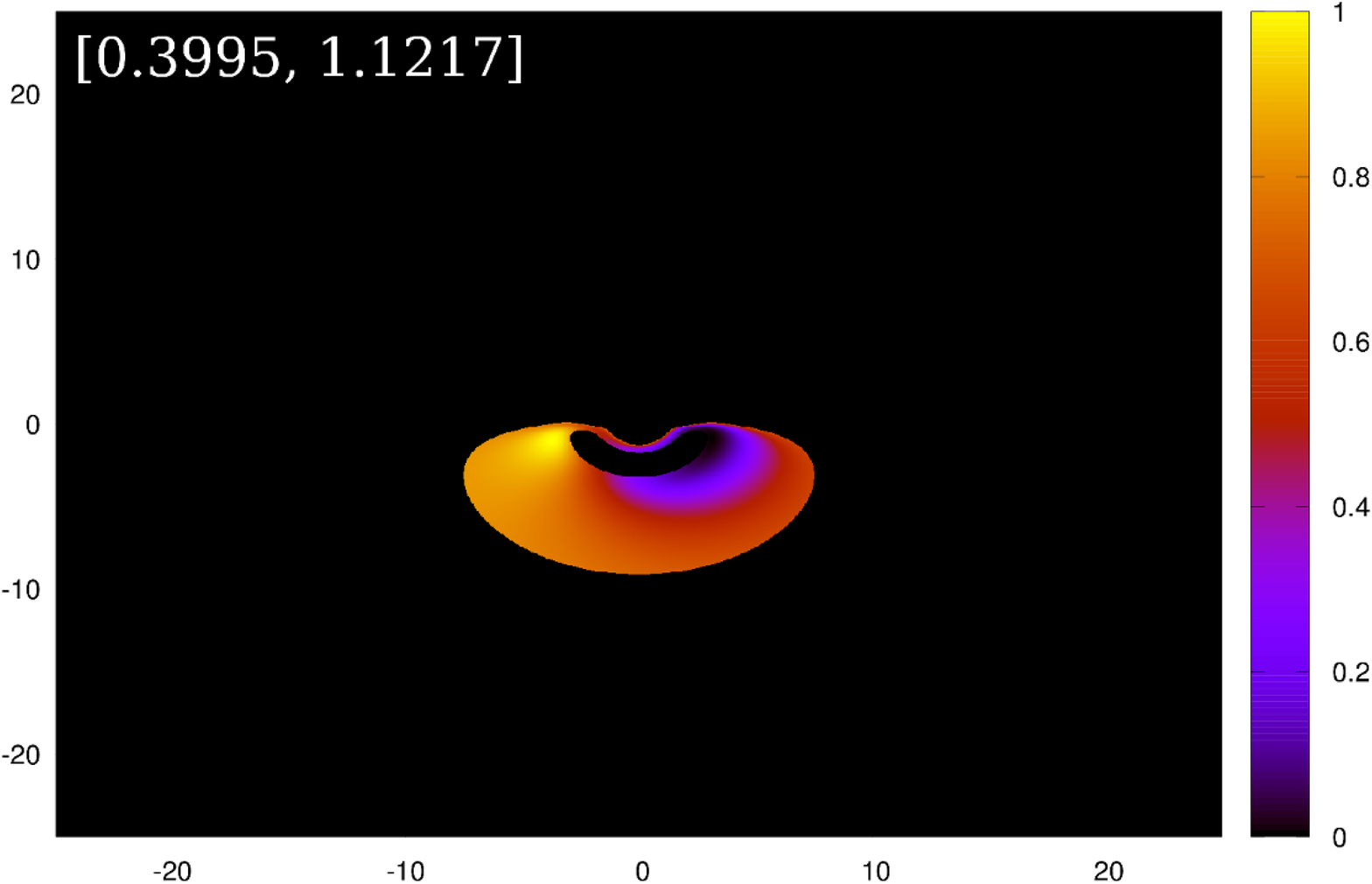}	
		\end{tabular}
		\caption{Frequency shift map of the Keplerian disc \emph{secondary} image in the vicinity of RN naked singularity determined by charge parameter $Q=1.01$, $3/(2\sqrt{2})$, $1.118$, and $1.3$. The observer inclination angle is $85^\circ$.}
	\end{center}	
\end{figure}

The indirect ghost images occur also for small inclination angles, as in the Bardeen no-horizon spacetimes. They are created by photons radiated with low impact parameters from the whole disc. These photons are reflected at $r \sim 0$ by the repulsive barrier. The secondary ghost images are oppositely shaped relative to the standard indirect images. For large enough charge parameter $Q$ (and for fixed inclination angle of the observer), the indirect images start to be deformed because of insufficient deflection of trajectories of photons coming from some part of the disc. Finally, for sufficiently large charge parameter, the indirect images disappear as no photon can be deflected enough to reach the observer at the given inclination angle. For the indirect images the situation in the RN naked singularity spacetimes is similar to those occuring in the Bardeen no-horizon spacetimes. Nevertheless, their origin could be different because of the different behavior of the lapse function at $r \sim 0$. Therefore, we shall study the mechanism of the secondary image vanishing also in the RN naked singularity spacetimes.

\subsection{Vanishing of the indirect disc image}

The secondary images are formed by photon rays satisfying conditions
\begin{eqnarray}
2 \pi &=& \psi_0+\left|\int^{u_t}_{u_o}+\int^{u_t}_{u_e}\right|\textrm{ for }l>0,\\
\pi &=& \psi_0-\left|\int^{u_t}_{u_o}+\int^{u_t}_{u_e}\right|\textrm{ for }l<0
\end{eqnarray}
where $\psi_0=\pi/2-\theta_o$. The secondary images vanish when neither of the above conditions is satisfied. For representative values of the observer inclination angle $\theta_o$, we have calculated the corresponding critical value of electric charge parameter $Q_{vanish}$. The secondary image then vanishes when $Q>Q_{vanish}$. The resulting values of the critical charge parameter are presented in the Table 5. 

\begin{table}[H]
	\begin{center}
		\caption{The values of electric charge parameter $Q_{vanish}$ corresponding to moment when secondary image vanishes calculated for particular value of observer inclination $\theta_o$ and emitter radial coordinate $r_e=20$. }
		\begin{tabular}{|c|cccccccc|}
			\hline
			$\theta_o$ & $10^\circ$ & $20^\circ$& $30^\circ$&$40^\circ$ & $50^\circ$ &$60^\circ$ & $70^\circ$ &$80^\circ$\\
			$Q_{vanish}$& $1.298$ & $1.443$ & $1.498$ & $1.568$ & $1.656$ & $1.771$ & $1.926$ & $2.138$\\
			\hline
		\end{tabular}
	\end{center}
\end{table} 

\subsection{Origin of direct ghost images in the RN naked singularity spacetimes}

The primary images of the Keplerian discs orbiting around the RN naked singuarities contain the primary ghost images -- see Figure 29 where the Keplerian disc image is constructed for the RN spacetime with electric charge parameter $Q=1.5$, and the distant observer is located at the inclination angle  $\theta_o=85^\circ$. In this case the direct ghost image is of different kind than in the regular Bardeen no-horizon spacetimes, resembling the ghost images discovered for imaging of the Keplerian discs orbiting Kerr naked singularities \cite{Stu-Sche:2010:CLAQG:}. 

\begin{figure}[H]
	\begin{center}
		\includegraphics[scale=0.5]{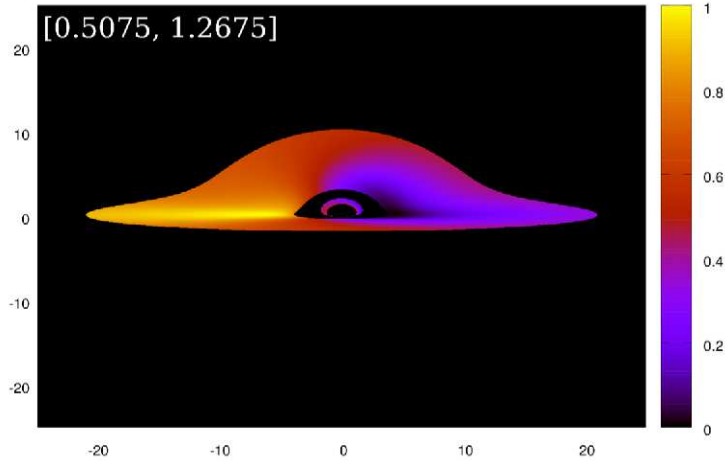}
		\caption{Primary image of Keplerian disk formed in the vicinity of RN naked singularity with electric charge parameter $Q=1.5$.}
	\end{center}
\end{figure}

In order to understand the origin of the direct ghost image in the RN naked singularity spacetimes, and the reason why it does not have the same structure as in the case of regular no-horizon spacetimes, we have constructed series of photon trajectories related to six representative values of the photon impact parameter $l=(10^{-6},0.5,1.0, 1.5,2.0,2.5)$ (we take the advantage of spherical symmetry of the RN spacetime and consider equatorial null geodesics in our analysis). The resulting photon trajectories are presented in Figure 30 representing an interesting interplay of the gravitational attraction of the source, and the repulsive phenomena related to the angular momentum of the photon and the repulsion near the centre of the RN spacetime. Due to this interplay, a typical deflection due to the attractive gravity occurs for large impact parameters, demonstrating deflection increasing with the impact parameter decreasing ($l = 2, 2.5$ in Figure 30). Then nearly straight trajectory occurs for intermediate values of the impact parameter ($l = 1.5$ in Figure 30). However, for small values of the impact parameter, direct repulsive effect enters the play, and the trajectories are reflected withour orbiting around the spacetime centre, demonstrating the limit of $\psi_{max} \to 0$ for $l \to 0$ ($l = 10^{-6}, 0.5, 1$ in Figure 30). We give the maximal deflection angle in dependence on the impact parameter, $\psi_{max}(l,Q=1.2)$ in Figure 31.

\begin{figure}[H]
	\begin{center}
		\includegraphics[scale=0.7]{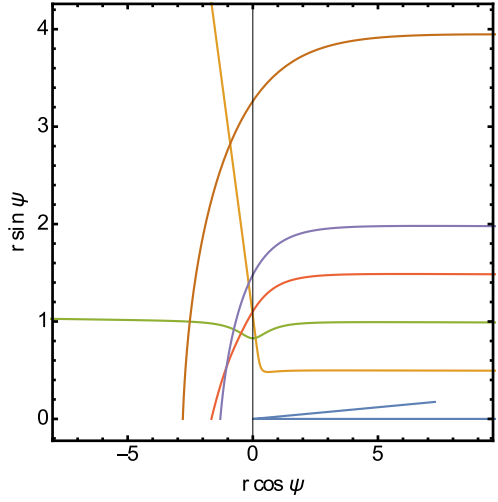}
		\caption{The series of representative trajectories of photons with the impact parameters $l=(10^{-6},0.5,1.0, 1.5,2.0,2.5)$ is constructed to demonstrate the origin of the direct ghost images in the RN naked singularity spacetimes. In this particular case the charge parameter of RN spacetime is $Q=1.2$}
	\end{center}
\end{figure} 

For an observer located at a given inclination angle $\theta_{o}$, there is a repulsive boundary determined by the specific value of the impact parameter $l_s=l_{s}(Q,\theta_o)$, separating the repulsive region from the standard attractive region (see Figure 31). The value of the separating impact parameter $l_{s}(Q,\theta_o)$ is implicitly determined by the formula
\begin{equation}
\psi_0(\theta_0) - 2\int_0^{u_t(l,Q)}\frac{l\diff u}{\sqrt{1-\tilde{f}(u)l^2 u^2}}=0.
\end{equation} 

\begin{figure}[H]
	\begin{center}
		\includegraphics[scale=0.9]{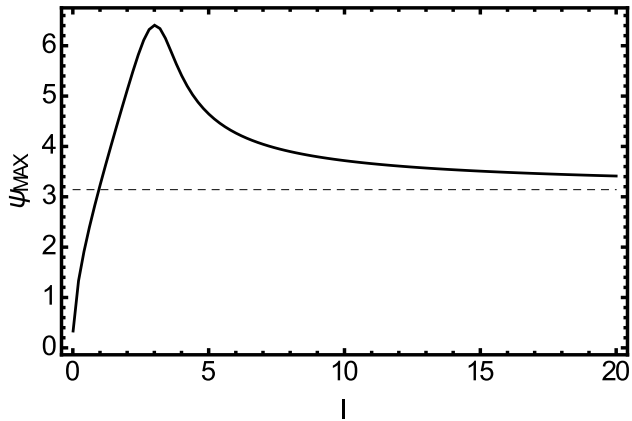}
		\caption{Plot of the value of $\phi_{max}$ for photon traveling from the asymptotically flat region ($r \to \infty$ back to this region given as a function of the impact parameter $l$ for fixed electric charge parameter $Q=1.2$ of the spacetime.}
	\end{center}
\end{figure}

For the RN spacetime with the electric charge parameter $Q=1.5$, the corresponding value of the separating impact parameter reads $l_S=1.39974$. Having determined the impact parameter $l_S$, we can  construct the source position function $u_{e}(l,Q)$, given implicitly by the relation 
\begin{equation}
\psi_0(\theta_0)-\left(\int_{0}^{u_t}\frac{l\diff u}{\sqrt{1-\tilde{f}(u)l^2 u^2}} + \int_{u_e(l)}^{u_t}\frac{l\diff u}{\sqrt{1-\tilde{f}(u)l^2 u^2}}\right)=0.\label{ue_formula}
\end{equation} 
The resulting character of the position function $u_{e}(l,Q,\theta_{o})$, defined for the impact parameter interval $l\in[l_{S},l_{max}]$, is presented in Figure 32, where in our calculations we put $l_{max}=20$. The region between the horizontal black lines determines the disk extension. The intersections with the blue line determine the extensions of the ghost and ordinary primary images. Here we have obtained for the ghost image the interval of impact parameter $l\in[1.45997,2.14393]$, and for the ordinary image the impact parameter interval $l\in[3.21631, 10.2152]$ (on the 2D observer screen those intervals determine the extension of the images). 

\begin{figure}[H]
	\begin{center}
		\includegraphics[scale=0.9]{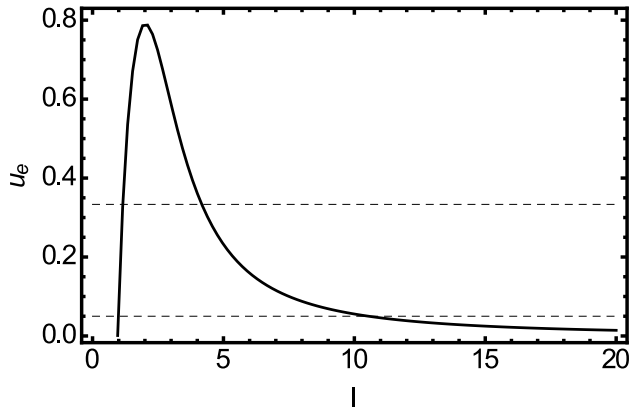}
		\caption{The position function $u_e=u_e(l;Q,\theta_{o})$ determined by formula (\ref{ue_formula}) is given for the electric charge parameter $Q=1.2$ of the spacetime and the inclination angle of the observer $\theta_{o}=85^{\circ}$.}
	\end{center}
\end{figure}

One can see that there is a minimal radius for which a photon with turning point can reach the infinity. This is consequence of the existence of the repulsive region that is the reason why we have topologically different direct ghost images in the RN naked singularity and the regular no-horizon spacetimes. 

The origin of the direct ghost image is also clearly seen from the plot of the position function $u_e(l;Q,\theta_{o})$ in Figure 32. For a given value of the emitter radius $r_e=1/u_e$, we have two values of the impact parameter $l_1$ and $l_2$, corresponding to two different regions on the observer screen.  

For the sake of illustration of the ghost image origin, we pick a representative value of the emitter reciprocal radius $u_e=0.25$, and trace the photon trajectories with impact parameter values $l_1=1.82219$ and $l_2=4.14839$ -- see Figure 33. 

\begin{figure}[H]
	\begin{center}
		\includegraphics[scale=0.7]{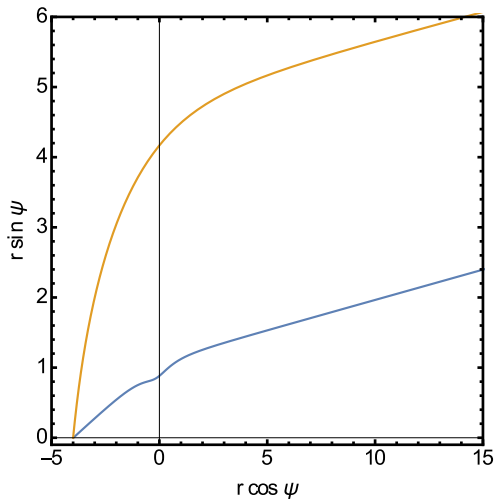}
		\caption{The trajectories of photons forming the ghost and ordinary primary image of the emitter at $r_e=1/0.25=4.0$. The RN spacetime electric charge parameter is, again, $Q=1.2$. }
	\end{center}
\end{figure}

\subsection{Complete images of Keplerian discs in the Reissner-N\"{o}rdstr\"{o}m naked singularity spacetimes}

For comparison with the complete images of Keplerian discs in the Bardeen spacetimes we give the complete image for the Keplerian discs orbiting in the Reissner-N\"{o}rdstr\"{o}m naked singularity spacetimes. The images are constructed in the same cases as for the Bardeen no-horizon spacetimes. 

\subsubsection{Complete images in the naked singularity spacetimes admitting circular photon orbits}

We first give in Figure 34 the overall complete image of the Keplerian disk formed in the Reissner-N\"{o}rdstr\"{o}m naked singularity spacetime with the specific electric charge parameter $Q=1.01$ allowing for existence of photon circular orbits. To visualize it we follow the approach of Luminet where the colors represent the bolometric flux of radiation coming from the disk \cite{Lum:1979:AAP}. We use arbitrary units being interested in relative luminosity of the higher-order images; for physical definition see \cite{Lum:1979:AAP}.  Again, an infinite number of higher-order images has to be generated in vicinity of the unstable photon circular orbit. 

\begin{figure}[H]
	\begin{center}
		\includegraphics[scale=0.6]{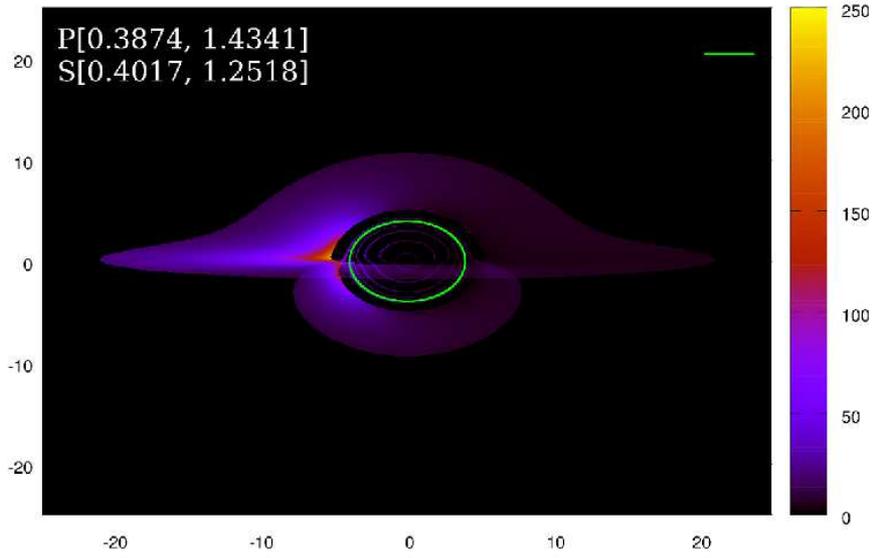}
		\caption{Complete image of Keplerian disc formed in the Reissner-N\"{o}rdstr\"{o}m spacetime with the electric charge parameter $Q=1.01$ allowing for existence of the circular photon orbits. The disc spans from $r_{in}=r_{ISCO}$ to $r_{out}=20$. The observer inclination angle reads $\theta_o=85^\circ$. The green circle determines the unstable photon circular orbit. The frequency shift range is given for both the primary (P) and secondary (S) images. The colors represent the bolometric flux.}
	\end{center}
\end{figure}

\begin{figure}[H]
	\begin{center}
		\begin{tabular}{c}
			\includegraphics[scale=0.42]{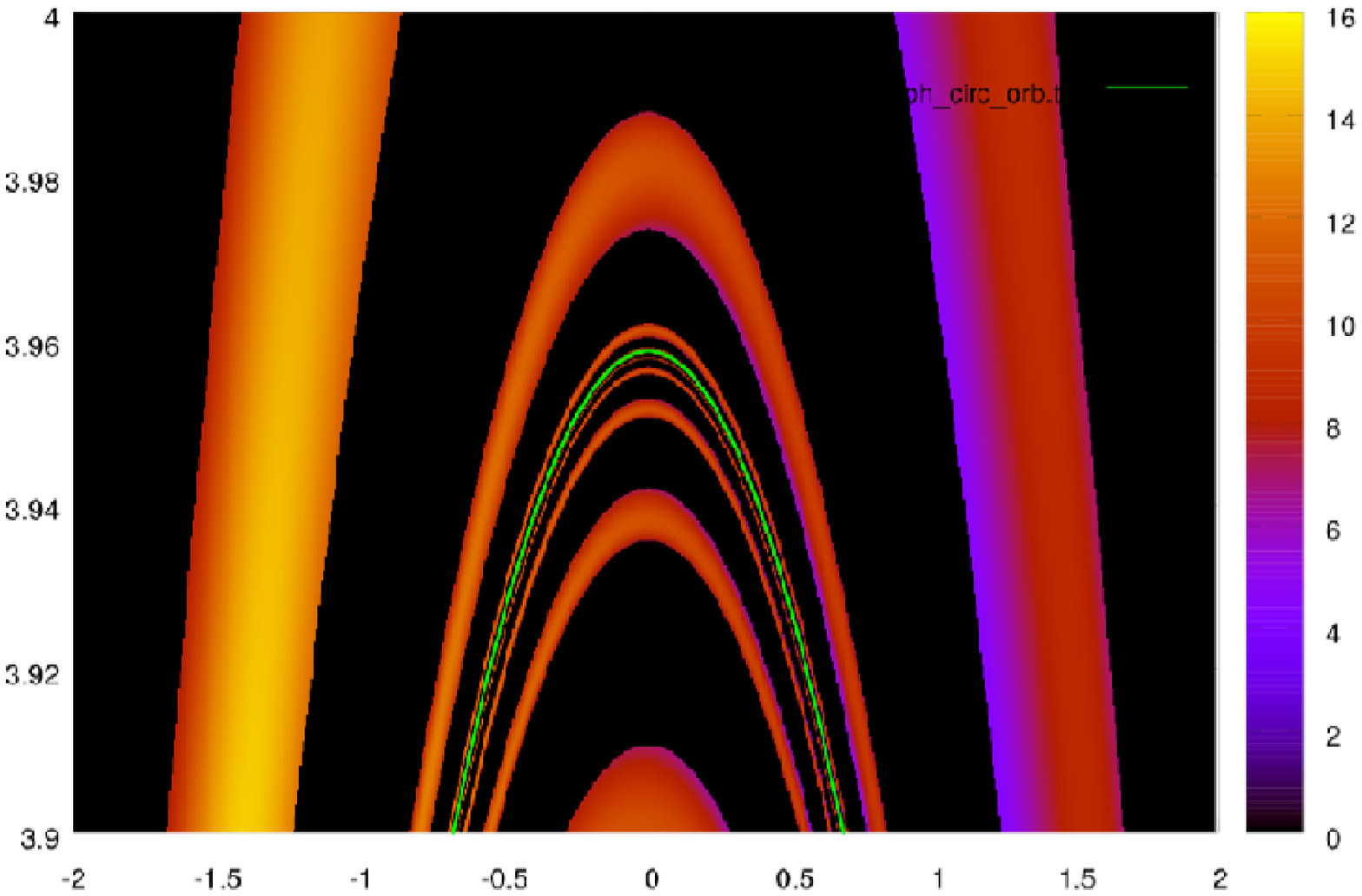}\\
			\includegraphics[scale=0.42]{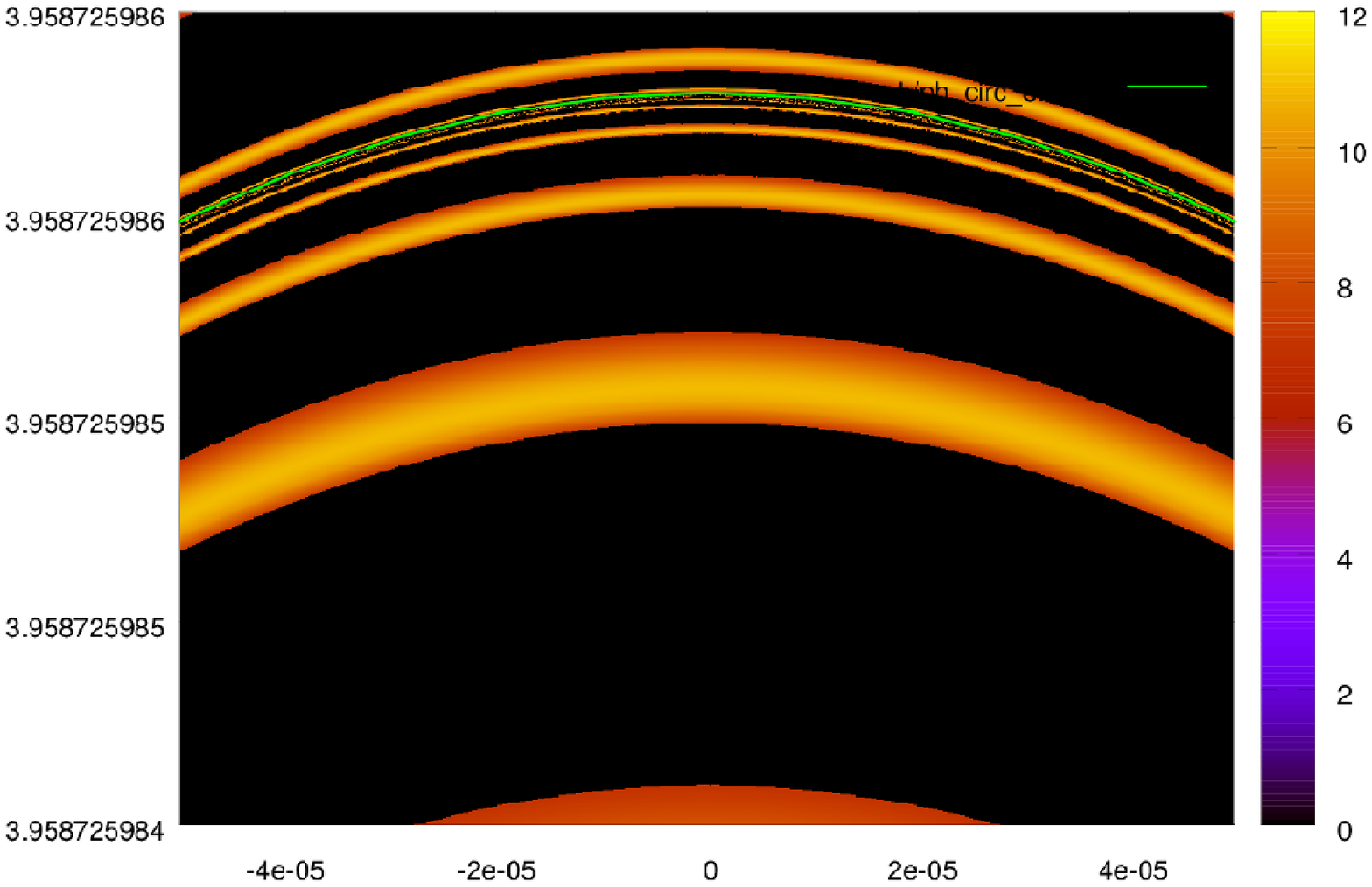}
		\end{tabular}
		\caption{The enlarged regions of the Keplerian disc complete image in close vicinity of unstable  circular photon orbit (green line) in the Reissner-No\"{o}rdstr\"{o}m spacetime with the electric charge parameter $Q=1.01$. The disc extension is the same as in the previous figure. The observer inclination angle is again $85^\circ$. The colors represent the bolometric flux.}
	\end{center}
\end{figure}

In Figure 35 we present enlarged region of the complete image located close to the unstable photon circular orbit with impact parameter $l=l_{ph(u)}(Q)$ demonstrating in similarity to the Bardeen case an asymmetry in distribution of the $n>3$-order images in the region inside ($l<l_{ph(u)}(g)$) and outside ($l>l_{ph(u)}(g)$) the photon circular orbit. The structure is quite identical to the case of Keplerian disc image in the Bardeen no-horizon spacetimes admitting existence of photon circular orbits. This is given by the fact that in both spacetimes the effective potential of the photon motion is essentially of the same character in vicinity of the unstable photon circular orbit.  

\subsubsection{Complete image in the naked singularity spacetime without the circular photon orbit}

\begin{figure}[H]
	\begin{center}
		\includegraphics[scale=0.22]{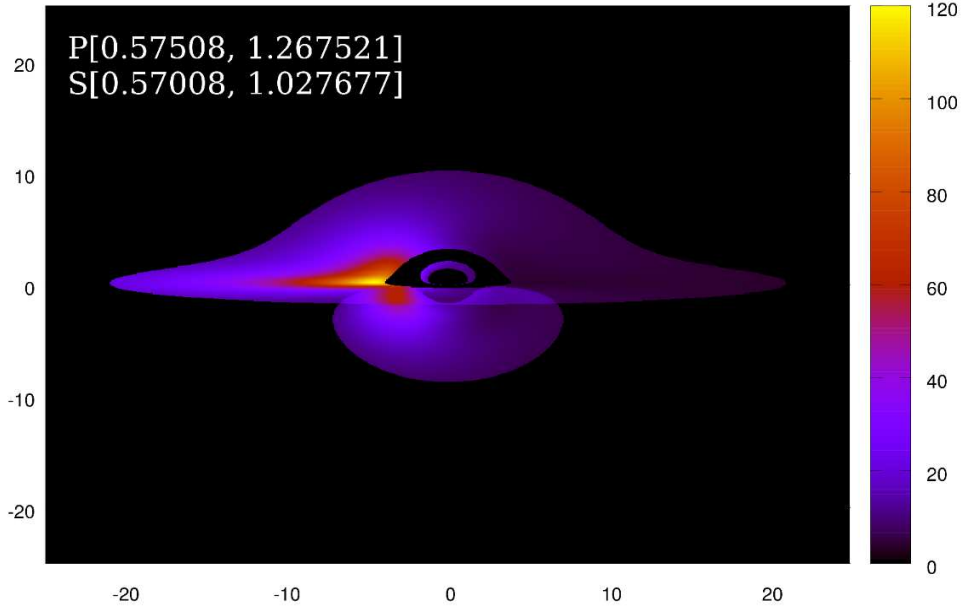}
		\caption{Complete image of the Keplerian disc orbiting in the Reissner-N\"{o}rdstr\"{o}m spacetime with the charge parameter $Q=1.5$. In this particular case there are no photon circular orbits. The disc spans from $r_{in}=r_{\Omega max}$ to $r_{out}=20$. The observer inclination angle is chosen $\theta_o=85^\circ$. In this case no higher order images can occur. The direct ghost image is the clearly different from the one of the Bardeen no-horizon spacetime. The frequency shift range is given for both the primary (P) and secondary (S) images. The colors represent the bolometric flux. }
	\end{center}
\end{figure}

For completeness we also consider the case of the Reissner-N\"{o}rdstr\"{o}m naked singularity spacetime where no photon circular orbit is admitted. The corresponding complete Keplerian disk image is illustrated in Figure 31. When compared with the Bardeen case  we immediately see the difference which  is in the shape of ghost image given by the different nature of Bardeen and RN spacetimes in the close vicinity of their center.  

\section{Conclusions}

We have maped the dependence of the deflection angle of the photon trajectories on the impact parameter of photons in the Bardeen no-horizon spacetimes, demonstrating existence of maximal deflection for trajectory with a critical impact parameter, and decreasing of the deflection angle for the impact parameter decreasing under the critical value. This special property is related to the existence of near-flat region of the regular no-horizon Bardeen spacetimes at their centre. 

Then we have used the extraordinary properties of the deflection angle dependence on the impact parameter in order to construct the so called ghost images of the innermost regions of the Keplerian discs and related them to their standard direct images. We determined both the ghost and standard direct images including also the frequency shift distribution across these images. We have shown that the ghostly imaged regions of the Keplerian discs are strongly limited and located around the line of sight of the distant observers just behind the coordinate origin. The ghost images arise only for large inclination angles of the distant observer. In the case of the Keplerian discs their occurrence is restricted to the no-horizon spacetimes allowing only for existence of stable circular geodesics and no unstable circular geodesics. The frequency range of the radiation related to the ghost images, along with those of the standard direct images, enables in principle determination of the charge parameter $g/m$ of the Bardeen no-horizon spacetime if the method discussed in \cite{Stu-Sche:2010:CLAQG:} could be applied. We have shown that in the RN naked singularity spacetimes, the ghost images occur in both direct and indirect images. They are of different origin as those created in the regular no-horizon spacetimes, being related to the repulsion effects occuring at $r \sim 0$ that is not present in the regular spacetimes. The ghost images in the RN spacetimes are thus of the same origin as those discovered in the Kerr naked singularity spacetimes \cite{Stu-Sche:2010:CLAQG:}. 

We conclude that the ghost images are clear signatures of the regular no-horizon spacetimes as no such phenomena could occur in black hole or naked singularity spacetimes. We expect that such images could be discovered by the new observational system GRAVITY enabling extremely precise angular resolution that has to start work at near future \cite{Gil-etal:SPIE:2012}. The assumed precision of the measurements made by GRAVITY can be as high as $\sim 10 \mu as$, enabling thus in the case of the supermassive black hole (or some alternate object as a superspinar \cite{Stu-Sche:2010:CLAQG:}, or a boson star \cite{Lid-Mad:IJMPD:1992}) in the Galaxy source Sgr A$^{*}$, to test predictions of alternative gravity theories, e.g., for braneworld black holes \cite{Sche-Stu:2008:IJMPD:,Boz-Man:ApJ:2012}. The precision $\sim 10 \mu as$ is high enough to distinguish the direct ghost images, if the inclination angle of the observer is high enough, and the expected extension of the standard direct image of the Keplerian disc is $\sim 50 \mu as$ in the Sgr A$^{*}$ central object as given in \cite{Sche-Stu:2008:IJMPD:}. 

Other interesting phenomena related to the ghost images could be expected in the profiled spectral lines generated in the innermost regions of the Keplerian discs or toroidal configurations, or in relation to the weak lensing of distant objects by supermassive no-horizon spacetimes. On the other hand, cross section of scattering of particles on such no-horizon spacetimes could give another interesting information that could be relevant, if such a strong gravity microscopic no-horizon objects will be created in the LHC experiments at CERN. 

The ghost images and related phenomena are not confined to the Bardeen spacetimes only, but they could be considered to be typical effects in all the no-horizon regular spacetimes with the central region having a near-flat geometry. 

\section*{Acknowledgements}
Z.S. and J.S. acknowledge the Albert Einstein Centre for Gravitation and Astrophysics supported by the~Czech Science Foundation Grant No. 14-37086G.


\begin{thebibliography}{99}



\bibitem{Ama-Eir:2012:PHYSR4:}
{{Amarilla}, L. and {Eiroa}, E.~F.},
\newblock{Shadow of a rotating braneworld black hole},
\newblock{\em Phys. Rev. D}, 85, 064019, 2012

\bibitem{AyB-Gar:1998:PhysRevLet:}
\newblock{{Ay{\'o}n-Beato}, E. and {Garc{\'{\i}}a}, A.},
\newblock{Regular Black Hole in General Relativity Coupled to Nonlinear Electrodynamics},
\newblock{\em Phys. Rev. Lett.}, 80, p.5056-5059, 1998

\bibitem{AyB-Gar:1999:PhysLetB:}
\newblock{{Ayon-Beato}, E.},
\newblock{New regular black hole solution from nonlinear electrodynamics},
\newblock{\em Phys. Lett. B}, 464, p.25--29, 1999

\bibitem{AyB-Gar:1999:GenRelGrav:}
\newblock{{Ayen-Beato}, E. and {Garcia}, A.},
\newblock{Non-Singular Charged Black Hole Solution for Non-Linear Source},
\newblock{\em Gen. Rel. and Grav.}, 31, 629, 1999

\bibitem{AyB-Gar:2000:PhysLetB:}
\newblock{{Ay{\'o}n-Beato}, E. and {Garc{\'{\i}}a}, A.},
\newblock{The Bardeen model as a nonlinear magnetic monopole},
\newblock{\em Phys. Lett. B}, 493, p.149--152, 2000

\bibitem{Mus:2014:PHYSR4:}
\newblock{{Azreg-A{\"i}nou}, M.},
\newblock{Generating rotating regular black hole solutions without complexification},
\newblock{\em Phys. Rev. D}, 90, 064041, 2014


\bibitem{Bam-Mod:2013:PhysLet:}
\newblock{{Bambi}, C. and {Modesto}, L.},
\newblock{Rotating regular black holes},
\newblock{\em Phys. Lett. B}, 721, p.329-334, 2013

\bibitem{Bao-Stu:1992:ApJ:}
\newblock{{Bao}, G. and {Stuchl{\'{\i}}k}, Z.},
\newblock{Accretion disk self-eclipse - X-ray light curve and emission line},
\newblock{\em Astrophys. J.}, 400, p.163-169, 1992


\bibitem{Bar:1968:GR5Tbilisi:}
\newblock{Bardeen, J.},
presented at GR5, Tbilisi, U.S.S.R., and published in the conference proceedings in the U.S.S.R., 1968.

\bibitem{Bar:1973:BlaHol:}
\newblock{{Bardeen}, J.~M.},
\newblock{Rapidly rotating stars, disks, and black holes.},
\newblock{\em Black Holes (Les Astres Occlus), editors: {{Dewitt}, C. and {Dewitt}, B.~S.} }, p.241-289, 1973

\bibitem{Boz-Man:ApJ:2012}
{{Bozza}, V. and {Mancini}, L.},
\newblock{ Observing Gravitational Lensing Effects by Sgr A* with GRAVITY},
\newblock{\em Astrophys. J.}, 753, 56, 2012

\bibitem{Car:1973:BlaHol:}
{{Carter}, B.},
\newblock{Black hole equilibrium states.},
\newblock{\em In Black Holes (Les Astres Occlus), editted by {{Dewitt}, C. and {Dewitt}, B.~S.}}, p.57-214, 1973

\bibitem{Cun-Bar:1975:ApJ:}
\newblock{{Cunningham}, C.~T.  and {Bardeen}, J.~M.},
\newblock{The optical appearance of a star orbiting an extreme kerr black hole},
\newblock {\em The Astrophysical Journal}, 183:237--264, 1973.

\bibitem{deF:1974:ASTRA:}
{{de Felice}, F.},
\newblock{\em Repulsive Phenomena and Energy Emission in the Field of a Naked Singularity}.
\newblock{Astron. \& Astrophys.}, 34, p.15, 1974

\bibitem{Doe-etal:2009:APJ:}
{{Doeleman}, S.~S., {Fish}, V.~L., {Broderick}, A.~E., {Loeb}, A., and {Rogers}, A.~E.~E.},
\newblock{Detecting Flaring Structures in Sagittarius A* with High-Frequency VLBI},
\newblock{\em Astrophys. Jour.}, 695, p.59-74, 2009



\bibitem{Eir-Rom-Tor:2002:PHYSR4:}
{{Eiroa}, E.~F. and {Romero}, G.~E. and {Torres}, D.~F.},
\newblock{Reissner-Nordstr{\"o}m black hole lensing},
\newblock{\em Phys. Rev. D}, 66, 024010, 2002

\bibitem{Eir-Sen:2013:PHYSR4:}
{{Eiroa}, E.~F. and {Sendra}, C.~M.},
\newblock{Regular phantom black hole gravitational lensing},
\newblock{\em Phys. Rev. D}, 88, 103007, 2013

\bibitem{Eir-Sen:2014:EJPC:}
 {{Eiroa}, E.~F. and {Sendra}, C.~M.},
 \newblock{Strong deflection lensing by charged black holes in scalar-tensor gravity},
\newblock{\em Euro. Phys. Jour. C}, 74, 3171, 2014

\bibitem{Fan-etal:1997:PASJ:}
\newblock{{Fanton}, C., {Calvani}, M., {de Felice}, F., and {Cadez}, A.},
\newblock{Detecting Accretion Disks in Active Galactic Nuclei},
\newblock{\em Pub. of the Astro. Soc. of Jap.}, 49, p.159-169, 1997

\bibitem{Fer-etal:2012:ExpA:}
 {{Feroci}, M., {Stella}, L., {van der Klis}, M., {Courvoisier}, T.~J.-L. , 
   	{Hernanz}, M., {Hudec}, R., {Santangelo}, A., {Walton}, D.,  
   	{Zdziarski}, A.,  {Barret}, D.,  {Belloni}, T.,  {Braga}, J.,  
   	{Brandt}, S.,  {Budtz-J{\o}rgensen}, C.,  {Campana}, S.,  
   	{den Herder}, J.-W.,  {Huovelin}, J.,  {Israel}, G.~L.,  
   	{Pohl}, M.,  {Ray}, P.,  {Vacchi}, A.,  {Zane}, S.,  
   	{Argan}, A.,  {Attin{\`a}}, P.,  {Bertuccio}, G.,  {Bozzo}, E.,  
   	{Campana}, R.,  {Chakrabarty}, D.,  {Costa}, E.,  {De Rosa}, A.,  
   	{Del Monte}, E.,  {Di Cosimo}, S.,  {Donnarumma}, I.,  
   	{Evangelista}, Y.,  {Haas}, D.,  {Jonker}, P.,  {Korpela}, S.,  
   	{Labanti}, C.,  {Malcovati}, P.,  {Mignani}, R.,  {Muleri}, F.,  
   	{Rapisarda}, M.,  {Rashevsky}, A.,  {Rea}, N.,  {Rubini}, A.,  
   	{Tenzer}, C.,  {Wilson-Hodge}, C.,  {Winter}, B.,  {Wood}, K.,  
   	{Zampa}, G.,  {Zampa}, N.,  {Abramowicz}, M.~A.,  {Alpar}, M.~A.,  
   	{Altamirano}, D.,  {Alvarez}, J.~M.,  {Amati}, L.,  {Amoros}, C.,  
   	{Antonelli}, L.~A.,  {Artigue}, R.,  {Azzarello}, P.,  
   	{Bachetti}, M.,  {Baldazzi}, G.,  {Barbera}, M.,  {Barbieri}, C.,  
   	{Basa}, S.,  {Baykal}, A.,  {Belmont}, R.,  {Boirin}, L.,  
   	{Bonvicini}, V.,  {Burderi}, L.,  {Bursa}, M.,  {Cabanac}, C.,  
   	{Cackett}, E.,  {Caliandro}, G.~A.,  {Casella}, P.,  {Chaty}, S.,  
   	{Chenevez}, J.,  {Coe}, M.~J.,  {Collura}, A.,  {Corongiu}, A.,  
   	{Covino}, S.,  {Cusumano}, G.,  {D'Amico}, F.,  {Dall'Osso}, S.,  
   	{De Martino}, D.,  {De Paris}, G.,  {Di Persio}, G.,  
   	{Di Salvo}, T.,  {Done}, C.,  {Dov{\v c}iak}, M.,  {Drago}, A.,  
   	{Ertan}, U.,  {Fabiani}, S.,  {Falanga}, M.,  {Fender}, R.,  
   	{Ferrando}, P.,  {Della Monica Ferreira}, D.,  {Fraser}, G.,  
   	{Frontera}, F.,  {Fuschino}, F.,  {Galvez}, J.~L.,  {Gandhi}, P.,  
   	{Giommi}, P.,  {Godet}, O.,  {G{\"o}{\v g}{\"u}{\c s}}, E.,  
   	{Goldwurm}, A.,  {G{\"o}tz}, D.,  {Grassi}, M.,  {Guttridge}, P.,  
   	{Hakala}, P.,  {Henri}, G.,  {Hermsen}, W.,  {Horak}, J.,  
   	{Hornstrup}, A.,  {in't Zand}, J.~J.~M.,  {Isern}, J.,  
   	{Kalemci}, E.,  {Kanbach}, G.,  {Karas}, V.,  {Kataria}, D.,  
   	{Kennedy}, T.,  {Klochkov}, D.,  {Klu{\'z}niak}, W.,  
   	{Kokkotas}, K.,  {Kreykenbohm}, I.,  {Krolik}, J.,  {Kuiper}, L.,  
   	{Kuvvetli}, I.,  {Kylafis}, N.,  {Lattimer}, J.~M.,  {Lazzarotto}, F.,  
   	{Leahy}, D.,  {Lebrun}, F.,  {Lin}, D.,  {Lund}, N.,  
   	{Maccarone}, T.,  {Malzac}, J.,  {Marisaldi}, M.,  {Martindale}, A.,  
   	{Mastropietro}, M.,  {McClintock}, J.,  {McHardy}, I.,  
   	{Mendez}, M.,  {Mereghetti}, S.,  {Miller}, M.~C.,  {Mineo}, T.,  
   	{Morelli}, E.,  {Morsink}, S.,  {Motch}, C.,  {Motta}, S.,  
   	{Mu{\~n}oz-Darias}, T.,  {Naletto}, G.,  {Neustroev}, V.,  
   	{Nevalainen}, J.,  {Olive}, J.~F.,  {Orio}, M.,  {Orlandini}, M.,  
   	{Orleanski}, P.,  {Ozel}, F.,  {Pacciani}, L.,  {Paltani}, S.,  
   	{Papadakis}, I.,  {Papitto}, A.,  {Patruno}, A.,  {Pellizzoni}, A.,  
   	{Petr{\'a}{\v c}ek}, V.,  {Petri}, J.,  {Petrucci}, P.~O.,  
   	{Phlips}, B.,  {Picolli}, L.,  {Possenti}, A.,  {Psaltis}, D.,  
   	{Rambaud}, D.,  {Reig}, P.,  {Remillard}, R.,  {Rodriguez}, J.,  
   	{Romano}, P.,  {Romanova}, M.,  {Schanz}, T.,  {Schmid}, C.,  
   	{Segreto}, A.,  {Shearer}, A.,  {Smith}, A.,  {Smith}, P.~J.,  
   	{Soffitta}, P.,  {Stergioulas}, N.,  {Stolarski}, M.,  
   	{Stuchlik}, Z.,  {Tiengo}, A.,  {Torres}, D.,  {T{\"o}r{\"o}k}, G.,  
   	{Turolla}, R.,  {Uttley}, P.,  {Vaughan}, S.,  {Vercellone}, S.,  
   	{Waters}, R.,  {Watts}, A.,  {Wawrzaszek}, R.,  {Webb}, N.,  
   	{Wilms}, J.,  {Zampieri}, L.,  {Zezas}, A. and {Ziolkowski}, J.
   },
   \newblock{The Large Observatory for X-ray Timing (LOFT)},
   \newblock{\em Exper. Astron.}, 34, 415--444, 2012
   


\bibitem{Gar-Hac-Kun-Lam:2013:arXiv:1306.2549:}
\newblock{{Garcia}, A., {Hackmann}, E., {Kunz}, J., {L{\"a}mmerzahl}, C., {Macias}, A.},
\newblock{\em Motion of test particles in a regular black hole space--time},
\newblock{\em ArXiv e-prints}, gr-qc:1306.2549, 2013


\bibitem{Gil-etal:SPIE:2012}
{{Gillessen}, S., {Lippa}, M., {Eisenhauer}, F., {Pfuhl}, O., 
   	{Haug}, M., {Kellner}, S., {Ott}, T., {Wieprecht}, E., 
   	{Sturm}, E., {Hau{\ss}mann}, F., {Kister}, C.~F., {Moch}, D. and 
   	{Thiel}, M.},
\newblock{GRAVITY: metrology},
\newblock{In \em Society of Photo-Optical Instrumentation Engineers (SPIE) Conference Series}, 8445, 84451O, 2012

\bibitem{Haw-Elli:1973:LargeScaleStructure:}
\newblock{{Hawking}, S.~W. and {Ellis}, G.~F.~R.},
\newblock{The large-scale structure of space-time},
\newblock{Cambridge University Press}, 1973

\bibitem{Hay:2006:PhysRevLet:}
\newblock{{Hayward}, S.~A.},
\newblock{Formation and Evaporation of Nonsingular Black Holes},
\newblock{\em Phys. Rev. Lett.}, 96, 031103, 2006

\bibitem{Hor:2009:PHYSRL:}
{{Ho{\v r}ava}, P.},
\newblock{\em Spectral Dimension of the Universe in Quantum Gravity at a Lifshitz Point},
\newblock{Phys. Rev. Lett.}, 102, 16, 2009

\bibitem{Hor:2009:PHYSR4:}
{{Ho{\v r}ava}, P.},
\newblock{\em Quantum gravity at a Lifshitz point},
\newblock{Phys. Rev. D}, 79, 8, 2009


\bibitem{Keh-Sfe:2009:PhysLetB:}
\newblock{{Kehagias}, A. and {Sfetsos}, K.},
\newblock{The black hole and FRW geometries of non-relativistic gravity},
\newblock{\em Phys. Lett. B}, 678,p.123-126, 2009

\bibitem{Kra:2011:CLAQG:}
{{Kraniotis}, G.~V.},
\newblock{Precise analytic treatment of Kerr and Kerr-(anti) de Sitter black holes as gravitational lenses},
\newblock{\em Class. and Quant. Grav.}, 28, 085021, 2011

\bibitem{Kra:2014:GRG:}
{{Kraniotis}, G.~V.},
\newblock{Gravitational lensing and frame dragging of light in the Kerr-Newman and the Kerr-Newman (anti) de Sitter black hole spacetimes},
\newblock{Gen. Rel. and Grav.}, 46, 1818, 2014


\bibitem{Lid-Mad:IJMPD:1992}
{{Liddle}, A.~R. and {Madsen}, M.~S.},
\newblock{The Structure and Formation of Boson Stars},
\newblock{\em Int. Jour. of Mod. Phys. D}, 1, p.101-143, 1992

\bibitem{Lum:1979:AAP}
{{Luminet}, J.-P.},
\newblock{Image of a spherical black hole with thin accretion disk},
\newblock{Astron. and Astrophys.}, 75, 228--235, 1979

\bibitem{Mis-Tho-Whe:1973:Gravitation:}
{{Misner}, C.~W., {Thorne}, K.~S., and {Wheeler}, J.~A.},
\newblock {Gravitation},
\newblock{\em San Francisco: W.H.~Freeman and Co.}, 1973

\bibitem{Mod-Nic:2010:PHYSR4:}
\newblock{{Modesto}, L. and {Nicolini}, P.},
\newblock{Charged rotating noncommutative black holes},
\newblock{\em Phys. Rev. D}, 82, 104035, 2010

\bibitem{Nev-Saa:2014:arXiv:1402.2694:}
\newblock{{Neves}, J.~C.~S. and {Saa}, A.},
\newblock{Regular rotating black holes and the weak energy condition},
\newblock{\em Phys. Lett. B}, 734, p.44--48, 2014

\bibitem{Nov-Tho:1973:BlaHol:}
\newblock{{Novikov}, I.~D. and {Thorne}, K.~S.},
\newblock{Astrophysics of black holes},
\newblock{\em Black Holes (Les Astres Occlus), editors: {{Dewitt}, C. and {Dewitt}, B.~S.}}, p.343-450, 1973

\bibitem{Nun:2010:PHYSR4:}
{{Bin-Nun}, A.~Y.},
\newblock{Relativistic images in Randall-Sundrum II braneworld lensing},
\newblock{\em Phys. Rev. D}, 81, 123011, 2010

\bibitem{Pag-Tho:1974:ApJ:}
\newblock{{Page}, D.~N. and {Thorne}, K.~S.},
\newblock{Disk-Accretion onto a Black Hole. Time-Averaged Structure of Accretion Disk},
\newblock{\em Astrophys. J.}, 191, p.499-506, 1974

\bibitem{Pat-Jos:2011:CLAQG:}
{{Patil}, M. and {Joshi}, P.},
\newblock{\em Kerr naked singularities as particle accelerators},
\newblock{Class. and Quant. Grav.}, 28, 23, 2011

\bibitem{Pat-Jos:2012:PHYSR4:}
\newblock{{Patil}, M. and {Joshi}, P.~S.},
 \newblock{Ultrahigh energy particle collisions in a regular spacetime without black holes or naked singularities},
 \newblock{\em Phys. Rev. D}, 86, 044040, 2012 

\bibitem{Pug-Que-Ruf:2011:PHYSR4:}
\newblock{{Pugliese}, D. and {Quevedo}, H. and {Ruffini}, R.},
\newblock{Motion of charged test particles in Reissner-Nordstr{\"o}m spacetime},
\newblock{\em Phys. Rev. D}, 83, 104052, 2011 

\bibitem{Pug-Que-Ruf:2011b:PHYSR4:}
 {{Pugliese}, D. and {Quevedo}, H. and {Ruffini}, R.},
 \newblock{Equatorial circular motion in Kerr spacetime},
\newblock{\em Phys. Rev. D}, 84, 044030, 2011

\bibitem{Rau-Bla:1994:ApJ:}
{{Rauch}, K.~P. and {Blandford}, R.~D.},
\newblock{Optical caustics in a kerr spacetime and the origin of rapid X-ray variability in active galactic nuclei},
\newblock{\em Astrophys. Jour.}, 421, 46--68, 1994


\bibitem{Sche-Stu:2008:IJMPD:}
\newblock{{Schee}, J. and {Stuchl{\'{\i}}k}, Z.},
\newblock{Optical Phenomena in the Field of Braneworld Kerr Black Holes},
\newblock{\em Int. Jour. of Mod. Phys. D}, 18, p.983--1024, 2009

\bibitem{Sche-Stu:2008:GenRelGrav:}
\newblock{{Schee}, J. and {Stuchl{\'{\i}}k}, Z.},
\newblock{Profiles of emission lines generated by rings orbiting braneworld Kerr black holes},
\newblock{\em Gen. Rel. and Grav.}, 41, p.1795-1818, 2009

\bibitem{Sche-Stu:2013:JCAP:}
\newblock{{Schee}, J. and {Stuchl{\'{\i}}k}, Z.},
\newblock{Profiled spectral lines generated in the field of Kerr superspinars},
\newblock{\em Journ. of Cosmolog. and Astropart. Phys.}, 4, 005, 2013

\bibitem{Ser:2009:PHYSRL:}
{{Sereno}, M.},
\newblock{Role of {$\Lambda$} in the Cosmological Lens Equation},
\newblock{\em Phys. Rev. Lett.}, 102, 021301, 2009

\bibitem{Stu:1980:BAC:}
{{Stuchl{\'{\i}}k}, Z.},
\newblock{\em Equatorial circular orbits and the motion of the shell of dust in the field of a rotating naked singularity}.
\newblock{Bull. of the Astronom. Inst. of Czechoslovakia}, 31, p.129--144, 1980

\bibitem{Stu:1981:BAC:}
{{Stuchl{\'{\i}}k}, Z.},
\newblock{\em Evolution of Kerr naked singularities}.
\newblock{Bull. of the Astron.Inst. of Czech.}, 32, p.68--72, 1981


\bibitem{Stu-Cal:1991:GenRelGrav:}
\newblock{{Stuchlik}, Z. and {Calvani}, M.},
\newblock{Null geodesics in black hole metrics with non-zero cosmological constant},
\newblock{\em Gen. Rel. and Grav.}, 23, p.507-519, 1991


\bibitem{Stu-Hle:2000:CLAQG:}
{{Stuchl{\'{\i}}k}, Z. and {Hled{\'{\i}}k}, S.}
\newblock{Equatorial photon motion in the Kerr-Newman spacetimes with a non-zero cosmological constant},
\newblock{\em Class. and Quant. Grav.}, 17,  p.4541-4576, 2000

\bibitem{Stu-Hle:2002:ActaPhysSlov:}
\newblock{{Stuchl\'{i}k, Z.} and  {Hled\'{i}k, S.}},
\newblock{Properties of the Reissner-Nordstrom spacetimes with a nonzero cosmological constant}
\newblock{\em Acta Phys. Slovac.} 52, 363, 2002 

\bibitem{Stu-Hle-Tru:2011:CLAQG:}
Stuchl{\'{i}}k, Z., Hled{\'{i}}k, S. and Truparov{\'{a}}, K., 
\newblock{\em Evolution of kerr superspinars due to accretion counterrotating
  keplerian discs}.
\newblock{Class. and Quant. Grav.}, 28, 155017, 2011


\bibitem{Stu-Sche:2010:CLAQG:}
\newblock{{Stuchl{\'{\i}}k}, Z. and {Schee}, J.},
\newblock{Appearance of Keplerian discs orbiting Kerr superspinars},
\newblock{\em Class. and Quant. Grav.}, 27, 215017, 2010

\bibitem{Stu-Sche:2012a:CLAQG:}
{{Stuchl{\'{\i}}k}, Z. and {Schee}, J.},
\newblock{\em Observational phenomena related to primordial Kerr superspinars},
\newblock{Class. and Quant. Grav.}, 29, 065002, 2012

\bibitem{Stu-Sche:2012b:CLAQG:}
{{Stuchl{\'{\i}}k}, Z. and {Schee}, J.},
\newblock{\em Counter-rotating Keplerian discs around Kerr superspinars}",
\newblock{Class. and Quant. Grav.}, 29, 025008, 2012

\bibitem{Stu-Sche:2013:CLAQG:}
\newblock{{Stuchl{\'{\i}}k}, Z. and {Schee}, J.},
\newblock{Ultra-high-energy collisions in the superspinning Kerr geometry},
\newblock{\em Class. and Quant. Grav.}, 30, 075012, 2013

\bibitem{Stu-Sche:2014:CLAQG:}
\newblock{{Stuchl{\'{\i}}k}, Z. and {Schee}, J.},
\newblock{Optical effects related to Keplerian discs orbiting Kehagias-Sfetsos naked singularities},
\newblock{\em Class. and Quant. Grav.}, 31, 195013, 2014

\bibitem{Stu-Sche-Abd:2014:PHYSR4:}
\newblock{{Stuchl{\'{\i}}k}, Z., {Schee}, J. and {Abdujabbarov}, A.},
\newblock{Ultra-high-energy collisions of particles in the field of near-extreme Kehagias-Sfetsos naked singularities and their appearance to distant observers},
\newblock{\em Phys. Rev. D}, 89, 104048, 2014 

\bibitem{Stu-Sche:2015:IJMPD:}
\newblock{{Stuchlik}, Z. and {Schee}, J.},
\newblock{Circular geodesic of Bardeen and Ayon-Beato-Garcia regular black-hole and no-horizon spacetimes}
\newblock{Accepted in Int. Jour. of Mod. Phys. D}, 2015

\bibitem{Tak-Har:2010:CLAQG:}
{{Takahashi}, R. and {Harada}, T.},
\newblock {\em Observational testability of a Kerr bound in the x-ray spectrum of black hole candidates}.
\newblock{Class. and Quant. Grav.}, 27, 7, 2010



\bibitem{Tos-Abd-Ahm-Stu:2014:PHYSR4:}
\newblock{{Toshmatov}, B., {Ahmedov}, B., {Abdujabbarov}, A.,and 
	{Stuchl{\'{\i}}k}, Z.},
\newblock{Rotating regular black hole solution},
\newblock{\em Phys. Rev. D}, 89, 104017, 2014

\bibitem{Vir-Elli:2002:PHYSR4:}
{{Virbhadra}, K.~S. and {Ellis}, G.~F.},
\newblock{\em Gravitational lensing by naked singularities},
\newblock{Phys. Rev. D}, 65, 10, 2002

\bibitem{Vir-Kee:2008:PHYSR4:}
{{Virbhadra}, K.~S. and {Keeton}, C.~R.},
\newblock{Time delay and magnification centroid due to gravitational lensing by black holes and naked singularities},
\newblock{\em Phys. Rev. D}, 72, 124014, 2008

\bibitem{Vie-etal:2014:PHYSR4:}
\newblock{{Vieira}, R.~S.~S., {Schee}, J., {Klu{\'z}niak}, W., {Stuchl{\'{\i}}k}, Z., and {Abramowicz}, M.},
\newblock{Circular geodesics of naked singularities in the Kehagias-Sfetsos metric of Ho{\v r}ava's gravity}, 
\newblock{\em Phys. Rev. D}, 90, 024035, 2014
\end{thebibliography}
\end{document}